%% file: main.tex
\pdfoutput=1

\documentclass[a4paper,10pt]{article}
\usepackage{amsfonts,graphicx,multicol,shadow}
\usepackage{textcomp}
\usepackage{graphicx}
\usepackage{placeins}
\usepackage{array}
\usepackage{makecell} 
\usepackage[utf8]{inputenc}
\usepackage{pst-fill,pst-grad}
\usepackage{float}
\usepackage[matrix,arrow,curve]{xy}
\usepackage{pstricks} 
\usepackage{amsmath,amsfonts,verbatim,afterpage,euscript,mathrsfs,amssymb}
\usepackage{hyperref}
\usepackage{microtype} 
\usepackage{subcaption} 
\usepackage{authblk} 
\usepackage{cancel} 
\usepackage{soul}

\usepackage[numbers,sort&compress]{natbib}

\usepackage[linesnumbered,ruled,vlined]{algorithm2e}

\usepackage{amsthm}
\usepackage{ragged2e}
\usepackage{dsfont}
\usepackage{tikz,pgfplots}
\usepackage{comment}
\pgfplotsset{compat=1.13}
\usepgfplotslibrary{fillbetween}
\usepgfplotslibrary{groupplots}
\usetikzlibrary{calc}
\usepackage{enumitem}
\usetikzlibrary{arrows}

\input{my_macros}

\begin{document}

\sloppy 
\renewcommand\Affilfont{\footnotesize} 


\title{Physics-informed, boundary-constrained Gaussian process regression for the reconstruction of fluid flow fields}
\author[1,2]{Adrian Padilla-Segarra\thanks{Corresponding author: adrian.padilla\_segarra@onera.fr}}
\author[2]{Pascal Noble}
\author[2]{Olivier Roustant}
\author[3]{\'Eric Savin}
\affil[1]{ONERA, DTIS, FR 31000 Toulouse, France}
\affil[2]{Institute of Mathematics of Toulouse, INSA Toulouse, FR 31400 Toulouse, France}
\affil[3]{ONERA, DTIS, Universit\'e Paris-Saclay, FR 91120 Palaiseau, France}
\date{\today}

\maketitle \rule[0.5mm]{170mm}{0.3mm}


\begin{abstract}
Gaussian process regression techniques have been used in fluid mechanics for the reconstruction of flow fields from a reduction-of-dimension perspective. A main ingredient in this setting is the construction of adapted covariance functions, or kernels, to obtain such estimates. In this paper, we present a general method for constraining a prescribed Gaussian process on an arbitrary compact set. The kernel of the pre-defined process must be at least continuous and may include other information about the studied phenomenon. This general boundary-constraining framework can be implemented with high flexibility for a broad range of engineering applications. From this, we derive physics-informed kernels for simulating two-dimensional velocity fields of an incompressible (divergence-free) flow around aerodynamic profiles. These kernels allow to define Gaussian process priors satisfying the incompressibility condition and the prescribed boundary conditions along the profile in a continuous manner. We describe an adapted numerical method for the boundary-constraining procedure parameterized by a measure on the compact set. The relevance of the methodology and performances are illustrated by numerical simulations of flows around a cylinder and a NACA 0412 airfoil profile, for which no observation at the boundary is needed at all.\\

\noindent\textbf{Keywords:} Gaussian process regression, Physics-informed Gaussian processes, Uncertainty quantification, Computational fluid dynamics.
\end{abstract}

\tableofcontents


\section{Introduction}

\subsection*{Context and aim of the paper}

Complex physical phenomena such as fluid flows have been extensively studied from different perspectives, by using both deterministic and probabilistic approaches \cite{CHO73,CON08,DEL16,FLA08,HIR07,POP00}. The development of various techniques for measuring physical quantities in flows \cite{TAV24} such as velocity, vorticity, pressure, or Lagrangian particle motion \cite{Maas1993,Malik1993}, as well as the availability of high-precision numerical simulations, offer a variety of information from different nature for the description of fluid dynamics. In the last decade, there has been a strong interest in developing techniques that can be able to model these dynamics based on information from different sources; see for example \cite{SAR21,SYM17}. Nevertheless, the classical deterministic methods based on mesh construction (such as finite differences, finite volumes, or finite elements) are not necessarily adaptable for considering multiple-source numerical estimates and assimilating the available information if any. Uncertainty quantification (UQ) of the estimates becomes a technical challenge that must be addressed.
In the present paper, we are interested in developing a physics-informed Gaussian process (GP) approach for the reconstruction of physical fields of an incompressible fluid flow around an aerodynamic profile. We present a general methodology for constraining a prescribed GP on an arbitrary compact set, which corresponds to the boundary of the profile in the applications. The pre-defined covariance function may include other physical information and is only required to be continuous on the compact set for performing the constraining procedure. In particular, we use this framework for the reconstruction of two-dimensional (2D) incompressible flows around a cylinder and a NACA aerodynamic profile. The enforcement of a slip boundary condition is done continuously for all points of the profile boundary. The obtained estimates must predict the physical fields at any unobserved spatial position based on a set of discrete direct observations of the field. Moreover, they have to simultaneously satisfy continuous-type conditions in agreement with the physics of the flow.

\subsection*{Contributions}

Our main contributions are the description and implementation of a physics-informed GP framework for reconstruction of flows past an aerodynamic profile that integrates multiple physics-constraining techniques simultaneously. In particular, we choose a general constraining method for the profile slip condition that modifies any GP with a continuous base covariance kernel to satisfy a linear scalar condition on arbitrary compact sets. We develop an adapted version compatible with GP derivatives needed for velocity and vorticity reconstructions. Furthermore, we outline and implement a numerical method for performing flow reconstructions under non-trivial geometries, such as a NACA airfoil. The combination of these elements and their general applicability are the main novelties of our work, which are absent in other boundary-constraining methods with more restrictive constructive approaches.

To achieve this, we firstly describe in a general context the spectral GP constraining method (initially developed in \cite{Gauthier2012}). This presentation is intended such that the proposed approach can be used in other engineering applications involving boundary constraints under general geometries.
We also describe a numerical method for solving the associated eigenvalue problem, which is parameterized by a probability measure defined on the compact set. We study two choices for the definition of this measure.
Next, we define our GP modelling framework by imposing incompressibility and a slip condition on the profile boundary to \emph{a priori} distributions of the physical fields. 
This is done by modelling the scalar stream function of the fluid flow as a boundary-constrained GP (BCGP), so that its velocity counterpart is a physics-informed prior that already fulfills the foregoing conditions in a continuous sense at all positions. Namely, these constraints are imposed over a continuous set rather than discretely at specific observation points. Ultimately, these defined priors can be used in Gaussian process regression (GPR) based on pointwise observations of the fluid flow fields.
The main advantage of this approach is that there is no need to save some part of the budget of the design of experiments for the inclusion of the physical constraints as observations since they are already included in the \textit{a priori} distributions. Furthermore, this procedure is a relevant mesh-free approach describing the fluid dynamics that can be directly merged within Lagrangian-based estimates \cite{Owhadi2023}. This framework can prevent wrong predictions of Lagrangian particles moving inside the aerodynamic profile, which may occur when using discrete observations of the boundary condition instead. Moreover, an important aspect of using GPs for the reconstruction of velocity and vorticity fields is the fact that we do not need to solve a pressure equation in order to get estimates of these fields.

\newpage

\subsection*{Related works}

\paragraph{Constrained GPs.} In order to completely define a GP, a mean function and a covariance function (or a kernel as defined in the dedicated literature) are required. These functions define the mathematical properties of the prior estimate, thus yielding an automatic satisfaction of the problem constraints if desired. The nature of these constraints can be set by physical specifications of the modelling problem, still no unique rule-of-thumb exists for incorporating them in the mean and/or kernel functions. Specific techniques to learn them from the data have been proposed in \emph{e.g.} \cite{Akian2022,Owhadi2019,Hamzi2021}. Divergence-free random vector fields have been considered in \cite{Scheuerer2012} and used for Lagrangian flow particle simulation in \cite{Owhadi2023}. Additionally, interests in constraining GP priors to boundary conditions on the boundary of domains with different geometries have also emerged \cite{Dalton2024,Gulian2022,Solin2019}. In \cite{Gulian2022}, the BCGPs are built from a particular structure based on Fourier expansions, whereas \cite{Solin2019} presents a spectral framework based on eigenfunctions of the Laplacian that is applicable to stationary base kernels. Other methods, such as \cite{Dalton2024}, define BCGP kernels from multiplication of the base kernel with a symmetric function satisfying the boundary conditions.
Our chosen method, developed firstly by \cite{Gauthier2012}, modifies any given GP with a continuous kernel to satisfy a null condition (also known as homogeneous Dirichlet condition in the context of partial differential equations (PDEs)) over any compact subset of its domain. We provide an associated numerical method for scalability of this procedure under different choices of the probability measure. To our knowledge, this is the first time that such a general boundary-constraining method (adapted for computing kernel derivatives) is applied to an arbitrary continuous kernel (not only stationary) and under a complex definition of the arbitrary constraining set, such as the boundary of a NACA airfoil.

\paragraph{Physics-informed learning models.} From a broader perspective, various methods of machine learning for image analyses for example have been adapted to fluid mechanics and data assimilation in this context; see \emph{e.g.} \cite{BRU20,BUZ23,DUR19,LIN23}. Recent advances in deep learning architectures include physics informed neural networks (PINNs) \cite{Raissi2019}, graph neural networks \cite{GNN09}, Fourier neural operators \cite{LI21FNO}, operator learning algorithms such as DeepONets \cite{LU21}, Kolmogorov–Arnold networks \cite{liu2025KAN}, or foundation models \cite{herde2024poseidon} among others. They have shown promising performances on a variety of fluid dynamics problems, the main difficulties being to handle the long-term stability and accuracy required for complex turbulent regimes \cite{TOS24,WAN24}. Besides, kernel methods such as GPR offer a flexible framework for estimating physical variables by directly combining information of different types \cite{BAC23,Kanagawa2018,SCH01}. GPR is based on a well-defined mathematical setting that includes interpolation representation formulas for estimates computation, as well as counterparts for quantifying uncertainty \cite{RAS05}. While the classic approach of using GPR techniques is data-driven, some of the recently proposed methods are purely physics-driven. It has been adapted for developing numerical methods for the approximation of solutions of PDEs \cite{Chen2021,Henderson2023}, through a collocation design of the PDE constraints. These schemes are analogous to those based on PINNs \cite{Raissi2019}. However, the GPR setting allows to control some of the mathematical properties of the estimates, such as linear constraints imposed to the prior distributions of the underlying GPs \cite{Ginsbourger2016}, and even the regularity of its sample paths \cite{Henderson2024}. In this way, these approaches can incorporate problem conditions from a continuous perspective rather than using discrete observation points. Hence GPR can also be of interest for the definition and implementation of data assimilation workflows in fluid dynamics.

\subsection*{Outline of the paper}

\begin{figure}[H]
    \centering
    \includegraphics[width=0.92\textwidth]{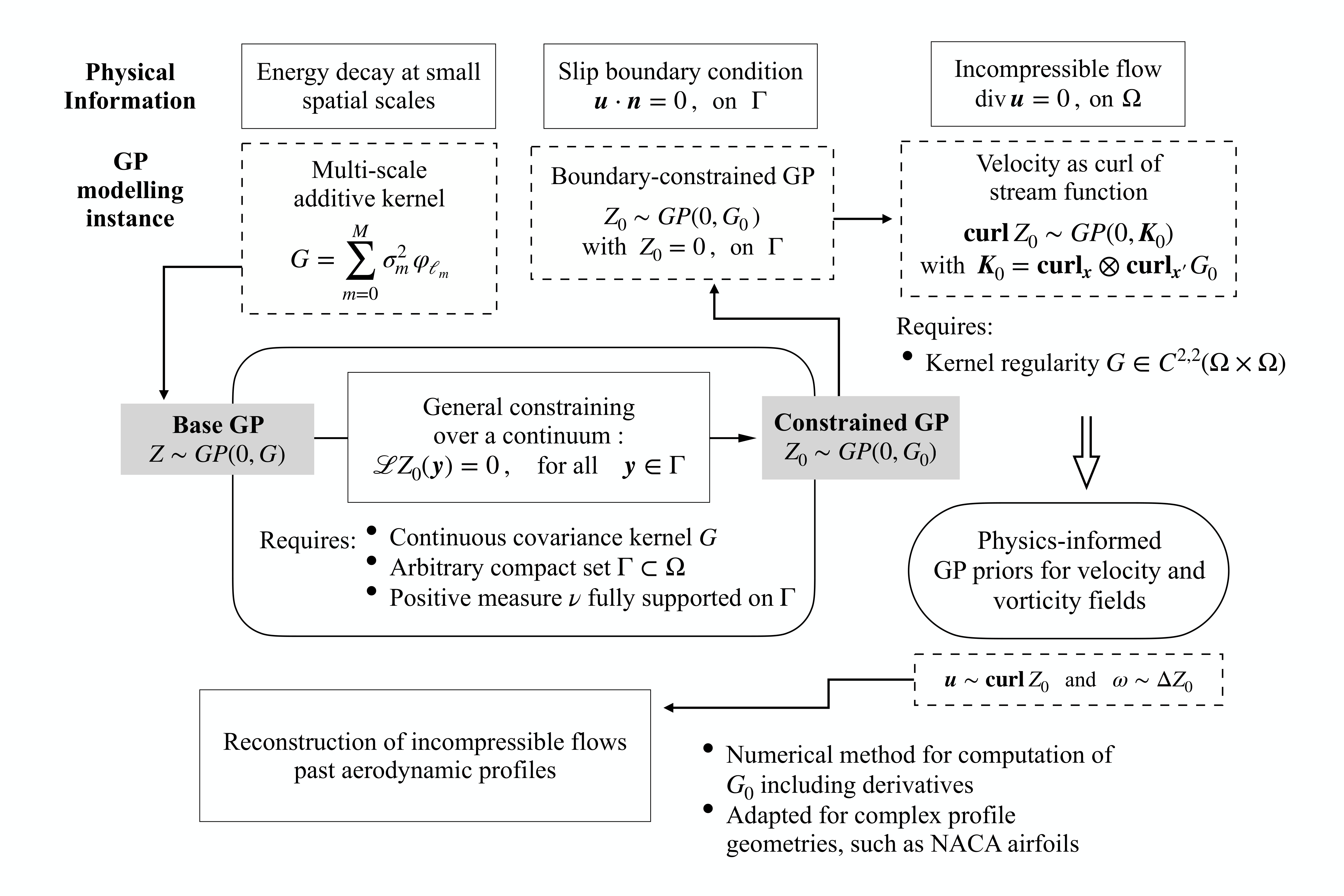}
    \caption{
    Summary of the methodology to enforce physical constraints on GP priors for velocity and vorticity fields. Three physical constraints can be considered simultaneously: incompressibility, slip boundary condition on profile, and energy decay at small spatial scales. The base GP must have a continuous kernel for boundary constraining via a general spectral method over an arbitrary compact set (see Prop.\ref{Prop_1} in \sref{sec:GPconstrained}). To enforce incompressibility, the GP kernel must be at least $\C^{2,2}$ differentiable (\sref{sec:PIGP}). A numerical method to compute the constrained kernel and its derivatives is described in Algorithms \ref{algo_spectral_factor} and \ref{algo_BCGP_derivatives}. Applications are performed for the reconstruction of 2D flows past a cylinder (\sref{sec:cylinder}) and at the leading edge of a NACA airfoil (\sref{sec:NACA}).}
    \label{fig_summary}
\end{figure}

The rest of the paper is organized as follows. The mathematical framework of GPR is briefly recalled in \sref{sec:math}. Then the main result used in our methodology for constructing a BCGP over a compact subset of its domain is outlined in \sref{sec:GPconstrained}. In particular, a numerical algorithm is described for the computation of the derivatives of boundary-constrained kernels for general computational geometries. In \sref{sec:PIK} we show how kernels in GPR can be informed by the physics they aim to approximate, with a particular focus on 2D incompressible fluid flows described by Navier-Stokes equations. A couple of numerical examples are then presented in \sref{sec:Aero} to illustrate the proposed approach. Here we consider standard profiles in aerodynamic applications, namely a cylinder and a NACA airfoil. Some conclusions and perspectives are finally drawn in \sref{sec:Conclusions}. A summary of the framework is presented in \fref{fig_summary}.

\subsection*{Notations}\label{sec:notations}

For a bounded subset $B\subset\R^n$, the $n$-dimensional volume of $B$ is denoted by $\abs{B}$. For a discrete set $A$, its cardinality is denoted by $\#(A)$. $\Mat_{m,n}(\R)$ stands for the set of $m\times n$ real matrices, with $\Mat_{n}(\R) = \Mat_{n,n}(\R)$. $\Trace{\boldsymbol A}$ stands for the trace of the square matrix ${\boldsymbol A}$. The usual Euclidean norm of a vector ${\boldsymbol a}$ in $\R^n$ is denoted by $\norm{{\boldsymbol a}}$, and the scalar product of two vectors ${\boldsymbol a}$ and ${\boldsymbol b}$ in $\R^n$ is denoted by ${\boldsymbol a}\cdot{\boldsymbol b}$. Their tensor product is denoted by ${\boldsymbol a}\otimes{\boldsymbol b}$, such that $({\boldsymbol a}\otimes{\boldsymbol b}){\boldsymbol c}=({\boldsymbol b}\cdot{\boldsymbol c}){\boldsymbol a}$ for any ${\boldsymbol c}\in\R^n$. $\Nabla$ and $\curl$ stand for the usual gradient and curl vector operators, respectively, which are also denoted by $\smash{\Nabla_\x}$ and $\smash{\curl_\x}$, respectively, to make explicit that they are computed with respect to the coordinates $\x$ if necessary. We denote by ${\boldsymbol a}^\perp$ the counterclockwise orthogonal vector to ${\boldsymbol a}\in\R^2$, such that ${\boldsymbol a}\cdot{\boldsymbol a}^\perp={\boldsymbol a}^\perp\cdot{\boldsymbol a}=0$. For $\x=(x_1,x_2)\in\R^2$ we have $\curl_\x = \smash{\Nabla_\x^\perp} = \smash{(-\partial_{x_2},\partial_{x_1})^\itr}$, where $\itr$ stands for the transpose operator. For $\Nderiv,r\in\N$ and $A,B\subset \R^d$, $\C^{\Nderiv,r}(A\times B)$ stands for the set of differentiable functions over $A\times B$ with continuous derivatives up to the order $\Nderiv$ in the first argument and to the order $r$ in the second argument.
For real-valued, second-order random variables $\ZGP$ and $\ZGP'$ defined on a probability space $(X,{\mathcal F},P)$, their mathematical expectations, or averages, are denoted by $\esp\braces{\ZGP}$ and $\esp\braces{\ZGP'}$, respectively, their covariance is denoted by $\Cov\braces{\ZGP,\ZGP'}=\esp\braces{(\ZGP-\esp\braces{\ZGP})(\ZGP'-\esp\braces{\ZGP'})}$, and their variances are denoted by $\Var\braces{\ZGP}=\Cov\braces{\ZGP,\ZGP}$ and $\Var\braces{\ZGP'}=\Cov\braces{\ZGP',\ZGP'}$, respectively. A Gaussian process $\{\ZGP(\x),\x\in\DomG\}\sim\GP(m,\G)$ indexed on some set $\DomG$ is defined as the collection of Gaussian random variables $\ZGP(\x)$ such that $\esp\braces{\ZGP(\x)}=m(\x)$ and $\Var\braces{\ZGP(\x),\ZGP(\x)}=\G(\x,\x)$, where $\x\mapsto m(\x)$ is their mean function and $\x,\xp\mapsto\G(\x,\xp)$ is their covariance function. Here we note $\ZGP\sim\law$ if the random variable $\ZGP$ is distributed according to the probability law $\law$. The classical Gaussian probability density function on $\R$ is $f_{\Normal}(z;m,\sigma^2)=\frac{1}{\sqrt{2\pi\sigma^2}}\exp[-\demi(\frac{z-m}{\sigma})^2]$.

\section{Reminder on Gaussian Process Regression}\label{sec:math}

We first recall the general setting of GPR from a probabilistic and functional approach. We refer to \cite{BAC23,OWH19,PAU16}, for example, for a general introduction on reproducing kernel Hilbert spaces (RKHS) and kernel methods. Let us consider a domain $\DomG\subset\R^d$ where $d$ is a positive integer. Given a scalar-valued positive semi-definite symmetric function, or kernel $\G : \DomG\times\DomG\to\R$, denote by $\HG$ the RKHS associated to $\G$, with induced inner product $\inner{\cdot,\cdot}_\G$. By definition, we have the following properties:
\begin{enumerate}[label=(\roman*)]
    \item $\G(\cdot,\x)\in\HG$ for all $\x\in\DomG$;
    \item The reproducing property is satisfied, \textit{i.e.} $h(\x) = \inner{h,\G(\cdot,\x)}_\G$ for all $h\in\HG$ and $\x\in\DomG$.
\end{enumerate}

We will denote as $\ZGP =\{\ZGP(\x),\x\in\DomG\} \sim\GP\parenth{0,\G}$ the zero-mean GP indexed on $\x\in\DomG$, such that its covariance function is the kernel $\G$. This function defines the structure of the sample paths of the GP in a certain extent, even though these sample paths do not necessarily belong to the RKHS associated to the kernel \cite{Kanagawa2018}.

Now, consider a set of $N$ points $\X = \smash{\{\x_i\}_{i=1}^N}$ inside $\DomG$, and a vector $\Y = \smash{(y_1,\ldots,y_N)^\itr}$ of $N$ fixed real-valued observations of $\ZGP$ at these points. A generic interpolation formula to estimate the value of $\ZGP(\x)$ at any unobserved $\x\in\DomG$ conditioned on the information that $\ZGP(\x_i) = y_i$ for all $i=1,\ldots, N$ (denoted by $\ZGP(\X)=\Y$) is given by the representer theorem \cite{OWH19}:
\begin{equation}\label{eq_2a}
\esp\braces{\ZGP(\x)\ |\ \ZGP(\X)=\Y} = \G(\x,\X)\G(\X,\X)^{-1}\Y\,. 
\end{equation}
Here $\G(\X,\X)$ is the Gram matrix with elements $[\G(\X,\X)]_{ij}=\G(\x_i,\x_j)$ and $\G(\x,\X)$ is the row vector with elements $\G(\x,\X)_i=\G(\x,\x_i)$. Moreover, the counterpart for quantifying the covariance of the random variables $\ZGP(\x)$ and $\ZGP(\xp)$ associated to the locations $\x$ and $\xp$ in the domain $\DomG$, conditioned on $\ZGP(\X)=\Y$, reads:
\begin{equation}\label{eq_2o}
    \Cov\braces{\ZGP(\x),\ZGP(\xp)\ |\ \ZGP(\X)=\Y} = \G(\x,\xp) - \G(\x,\X) \G(\X,\X)^{-1}\G(\X,\xp)\,.
\end{equation}

For a kernel $\G$ that is regular enough, we can access to its pointwise derivatives and thus obtain estimates based on derivative information of the underlying target or phenomena. In this case, considering $M$ data points $\Q=\smash{\braces{\q_i}_{i=1}^{M}}$ in $\DomG$ for the observation of derivatives, we can estimate $\ZGP(\x)$ for any $\x\in\DomG$, conditioned on the information $\Li_i\ZGP(\q_i) = v_i$, $i=1,\dots,M$, for the fixed real values in the vector $\V = \smash{(v_1,\dots,v_M)^\itr}$ and the corresponding linear differential operators $\Li_i$:
\begin{equation}\label{eq_2b}
    \esp\braces{\ZGP(\x)\ |\ \Li_i\ZGP(\q_i)=v_i,\ 1\leq i\leq M} = {\Li^\prime} \G(\x,\Q)\left(\Li\otimes\Li^\prime\G(\Q,\Q)\right)^{-1}\V\,.
\end{equation}
Here $\smash{\Li\otimes\Li^\prime\G(\Q,\Q)}$ is the Gram matrix of entries $\Li_i\otimes\Li_j^\prime\G(\q_i,\q_j)$ for $i,j=1,\dots,M$, and $\smash{\Li_i}$, $\smash{\Li_j^\prime}$ refer to the differential operators acting on the first and second entries of $\G$, respectively. The order of these differential operators is limited by the regularity of $G$. The representation identity of \eref{eq_2b} remains valid if we replace the differential operators $\Li_i$ with any other linear operators acting on the GP \cite{Ginsbourger2016} since formally any linear evaluation of a GP remains a GP. The covariance counterpart for $\x,\xp\in\DomG$ reads:
\begin{equation}\label{eq_2p}
    \Cov\braces{\ZGP(\x),\ZGP(\xp)\ |\ \Li_i\ZGP(\q_i) = v_i,\ 1\leq i\leq M} = \G(\x,\xp) -  \Li^\prime\G(\x,\Q)  \left(\Li\otimes\Li^\prime\G(\Q,\Q)\right)^{-1}  \Li\G(\Q,\xp) \,.
\end{equation}

The functions of $\x\in\DomG$ from \eqref{eq_2a} and \eqref{eq_2b} are thus linear combinations of functions of the form $\G(\cdot,\x_i)$ and $\Li_i\G(\cdot,\q_i)$. Actually they can be interpreted as minimizers of an optimization problem in an \emph{adhoc} RKHS with a regression loss function defined from the discrete observations; see \emph{e.g.} \cite{OWH19}. Also, ridge regression versions of these representation formulas, which account for problem regularization, can be defined \cite{Micchelli2004}. Furthermore, estimates based on mixed information, from direct observations and derivative observations simultaneously, can be established in the same way as formulas \eqref{eq_2a} and \eqref{eq_2b} by the representer theorem. However, so far, all the interpolation conditions are presented in a discrete way. We are interested in the imposition of continuous-type conditions on GPs, as described in the next section.

\section{Constraining a GP over a continuum}\label{sec:GPconstrained}

In this section we show how to constrain a GP over a continuum by carefully crafting its covariance kernel. We start by considering an arbitrary compact set within the domain $\DomG$ in \sref{sec:GPcompact}, before we turn to the more particular case of a parameterized compact manifold in \sref{sec:BCGP}. We also show how to approximate it numerically in \sref{sec:Num}. 

\subsection{Constraints over a compact set}\label{sec:GPcompact}

For a compact domain $\DomG\subset\R^d$ and arbitrary scalar GP $\ZGP =\{\ZGP(\x),\x\in\DomG\} \sim\GP\parenth{0,\G}$ with covariance continuous kernel $\G:\DomG\times\DomG\rightarrow\R$, we now detail how to modify $\G$ such that the resulting zero-mean GP satisfies homogeneous Dirichlet boundary conditions on a compact set $\curveD$ inside $\DomG$. For a non-homogeneous condition, the resulting constrained kernel can be associated to a GP with a mean function $m$ satisfying the desired Dirichlet boundary condition. The results presented in this section are based on \cite{Gauthier2012}.

Firstly, recall the Karhunen-Lo\`eve expansion (KLE) of $\ZGP$ with respect to a particular measure $\measureD$ over the compact set $\curveD$:
\begin{equation}\label{eq_2h}
    \ZGP(\x)=\sum_{n=0}^{+\infty}\sqrt{\lameigenD_n}\,\eigenphiD_n(\x)\,\xiD_{n}\,,\quad\text{for } \x\in\curveD\,,   
\end{equation}
where $\xiD_1,\xiD_2,\ldots\sim\Normal(0,1)$ are independent, identically distributed (i.i.d.) normal random variables, and the pairs $(\lameigenD_n,\eigenphiD_n)$, $n\in\N$, are positive eigenvalues and $L^2(\curveD,\measureD)$-normalized eigenfunctions associated to the Hilbert-Schmidt operator $\Top_\curveD : L^2(\curveD,\measureD)\to L^2(\curveD,\measureD)$ defined by:
\begin{equation}\label{eq_2q}
    \Top_\curveD[h](\x) = \int_\curveD\G(\x,\y)h(\y)\dd\measureD(\y)\,,\quad \text{for }h\in L^2(\curveD,\measureD)\,,\,\x\in\curveD\,.
\end{equation}
The measure $\measureD$ can be chosen freely as any positive Borel measure fully supported on $\curveD$. This ensures that the operator $\Top_\curveD$ is positive, self-adjoint and compact. Its eigenvalues are then real, positive, and can be ordered in a decreasing sequence converging to $0$.

Secondly, we can extend the definition of these eigenfunctions, or features, to the whole initial domain, \textit{i.e.} we consider the lifting
$\eigenphiExt{n}$ of $\smash{\eigenphiD_n}$ to $\DomG$ for each $n\in\N$:
\begin{equation}\label{eq_2t}
    \eigenphiExt{n}(\x) = \frac{1}{\lameigenD_n}\int_{\curveD}\G(\x,\y)\eigenphiD_n(\y) \dd\measureD(\y)\,,\quad\text{for }\x\in\DomG\, .
\end{equation}
Indeed, by definition of the KLE \eqref{eq_2h}, the associated eigenfunctions of $\Top_\curveD$ verify:
\begin{equation}\label{eq_2s}
\int_{\curveD}\G(\x,\y)\eigenphiD_n(\y) \dd\measureD(\y) = \lameigenD_n \eigenphiD_n(\x) \,,\quad\x\in\curveD\,,
\end{equation}
and thus we have $\smash{\eigenphiExt{n}(\x)} = \smash{\eigenphiD_n(\x)}$ for all $\x\in\curveD$ and $n \in \N$.

With these elements, we are able to construct a GP satisfying the homogeneous Dirichlet condition on the compact set $\curveD$ as presented in the following proposition.

\begin{proposition}\label{Prop_1}
    Consider an arbitrary scalar Gaussian process $\ZGP =\{\ZGP(\x),\x\in\DomG\} \sim\GP\parenth{0,\G}$, with continuous covariance kernel $\G:\DomG\times\DomG\rightarrow\R$, and a compact subset $\curveD$ of its domain $\DomG$.
    Consider also a positive Borel measure $\measureD$ fully supported on $\curveD$.
    Let $\lameigenD_n, \eigenphiD_n$ and $\xiD_n$, $n\in\N$, come from the Karhunen-Lo\`eve expansion of $\ZGP$ localized on $\curveD$ with respect to $\measureD$, and $\eigenphiExt{n}$ be the lifting of $\eigenphiD_n$ to $\DomG$ as defined in \eref{eq_2t}. Then $\ZGPobstacle =\braces{\ZGPobstacle(\x),\x\in\DomG }$ defined as: 
    \begin{equation}\label{eq_2u}
        \ZGPobstacle(\x) = \ZGP(\x) - \sum_{n=0}^{+\infty} \sqrt{\lameigenD_n}\,\eigenphiExt{n}(\x)\,\xiD_n \,, \quad\x\in\DomG\,,
    \end{equation}
    is a centered Gaussian process that satisfies an homogeneous Dirichlet boundary condition on $\curveD$, i.e.:
    \begin{equation}\label{eq_2z}
        \ZGP_0(\x) = 0\,, \quad \forall \x\in\curveD\,,
    \end{equation}
    in the mean-square sense. Moreover, its associated kernel reads:
    \begin{equation}\label{eq_scalar_BCGP_kernel}
        \Gobstacle(\x,\xp) = \G(\x,\xp) - \sum_{n = 0}^{+\infty} \lameigenD_n\,\eigenphiExt{n}(\x)\,\eigenphiExt{n}(\xp)\,, \quad \forall \x,\xp\in\DomG\,,
    \end{equation}
    and is such that $\Gobstacle(\cdot,\xp) = 0$ over $\curveD$ for all $\xp\in\DomG$ uniformly.
\end{proposition}
\noindent The result is proved in \cite{Gauthier2012} from a theoretical point of view. For the sake of completeness, we provide here a simpler sketch of the proof, focusing on the main ideas rather than technical details.\\

\begin{proof_sketch}
    For each integer $N\in\N$, define:
    \begin{equation*}
    Y_N(\x) = \sum_{n=0}^{N} \sqrt{\lameigenD_n}\,\eigenphiExt{n}(\x)\,\xiD_n\,, \quad \x \in \DomG\,.
    \end{equation*}
    The main point is that the random variable $Y_N(\x)$ is the orthogonal projection of $\ZGP(\x)$ onto $\text{span}(\xiD_0, \dots, \xiD_N)$:
    \begin{equation} \label{eq:orthoProj}
       Y_N(\x) = \esp\braces{ \ZGP(\x)\ |\ \xiD_0, \dots, \xiD_N }. 
    \end{equation}
    Indeed, the conditional expectation has the form $\esp\braces{ \ZGP(\x)\ \vert\ \xiD_0, \dots, \xiD_N } = \sum_{n=0}^N \alpha_n(\x)\,\xiD_n$ where $\alpha_n(\x)$ are real numbers. By definition of the orthogonal projection, we must have for all $\x\in\DomG$ and $n=0,\ldots,N$,
    $$ \alpha_n(\x) = \Cov\braces{\ZGP(\x), \xiD_n}\,.$$
    Now, notice that by the $L^2(\curveD,\measureD)$-orthogonality of the eigenfunctions $\smash{\eigenphiD_n}$, we have that for all $n\in\N$, the random variables associated to the expansion \eqref{eq_2h} over $\curveD$ read:
    \begin{equation*}
    \xiD_n = \frac{1}{\sqrt{\lameigenD_n}}\int_\curveD \ZGP(\y)\eigenphiD_n(\y)\dd\measureD(\y)\,,
    \end{equation*}
    which leads to: 
    \begin{equation*}
    \begin{split}
        \alpha_n(\x) &= \frac{1}{\sqrt{\lameigenD_n}}\int_\curveD G(\x,\y)\eigenphiD_n(\y)\dd\measureD(\y) \\
    &= \sqrt{\lameigenD_n}\,\eigenphiExt{n}(\x)
        \end{split}
    \end{equation*}
    and proves \eref{eq:orthoProj}.
    
    As a corollary, we obtain that $\smash{\lim_{N}} Y_N(\x)$ is a well-defined Gaussian random variable. Indeed, as $\smash{Y_N(\x)}$ is a centered Gaussian random variable, it is sufficient to prove that $\Var\braces{Y_N(\x)}$ converges when $N$ tends to infinity. Now, $\smash{\Var\braces{Y_N(\x)}} = \smash{\sum_{n=0}^N \lameigenD_n \eigenphiExt{n}^2(\x)}$, 
    is an increasing sequence in $N$ bounded by $\smash{\Var\braces{\ZGP(\x)}}$ due to \eqref{eq:orthoProj}, and thus converges. By similar arguments, it can be proved that $\ZGPobstacle$ is a GP.
    
    Now, the intuition behind the definition of $\ZGPobstacle(\x)$ is clear: it represents the difference between $\ZGP(\x)$ and its orthogonal projection onto $\text{span}\{\xiD_n,\, \forall n \in \mathbb{N} \} = \text{span}\{\ZGP(\y),\, \forall \y\in \curveD \}.$ 
    Clearly, we expect that:
    \begin{equation}\label{eq_2y}
        \ZGPobstacle(\x) = 0\,, \quad \forall \x \in \curveD\,.
    \end{equation}
    This is immediately verified since, if $\x \in \curveD$, the lifting 
    $\eigenphiExt{n}$ coincides with $\eigenphiD_n$, implying that the
    term $\sum_{n=0}^{+\infty} \sqrt{\lameigenD_n}\,\eigenphiExt{n}(\x)\,\xiD_n$ coincides with the KLE of $\ZGP$ at $\x$.
    
    We can subsequently compute the covariance function associated to $\ZGPobstacle$. 
    To that end, using the interpretation of $Y_N(\xp)$ in terms of the orthogonal projection \eqref{eq:orthoProj}, we have:
    $$ \Cov\braces{Y_N(\x), \ZGP(\xp) - Y_N(\xp)} = 0\,, \quad \x,\,\xp \in \DomG\,. $$
    This implies that, for all $\x,\,\xp \in \DomG$,
    \begin{equation*}
    \begin{split}
    \Cov\braces{\ZGP(\x) - Y_N(\x), \ZGP(\xp) - Y_N(\xp)} &= \Cov\braces{\ZGP(\x), \ZGP(\xp) - Y_N(\xp)}\\
    &= G(\x, \xp) - \sum_{n=0}^N \sqrt{\lameigenD_n}\, \eigenphiExt{n}(\xp)\,\Cov\{\ZGP(\x), \xiD_n\}\,.
    \end{split}
    \end{equation*}
    Now, we have obtained above, from the expression of $\alpha_n(\x)$, that:
    $$
    \Cov\braces{\ZGP(\x),\xiD_n} = \sqrt{\lameigenD_n}\,\eigenphiExt{n}(\x)\,, \qquad \x\in\DomG\,,\,n\in\N\,.
    $$
    Then, for any $\x,\,\xp\in\DomG$, we obtain:
    $$ \Cov\braces{\ZGP(\x) - Y_N(\x), \ZGP(\xp) - Y_N(\xp)} 
    = G(\x, \xp) - \sum_{n=0}^{N}\lameigenD_n\,\eigenphiExt{n}(\x)\,\eigenphiExt{n}(\xp)\,.
    $$
    Thus, taking the limit when $N$ tends to infinity, we get:
    \begin{align}
    G_0(\x, \xp) & = G(\x,\xp) - \sum_{n=0}^{+\infty}\lameigenD_n\,\eigenphiExt{n}(\x)\,\eigenphiExt{n}(\xp) \nonumber \\
    & =\G(\x,\xp) - \sum_{n=0}^{+\infty}\frac{1}{\lameigenD_n}\parenth{\int_\curveD \G(\x,\y)\eigenphiD_n(\y)\dd\measureD(\y)}\parenth{\int_\curveD \G(\xp,\y)\eigenphiD_n(\y)\dd\measureD(\y)}\,.\label{eq_2i}
    \end{align}
    Finally, due to \eref{eq_2y}, if $\x \in \curveD$ and $\xp\in\DomG$ (not necessarily in $\curveD$), we notice that:
    \begin{equation*}
        \Gobstacle(\x,\xp) = \Cov\braces{\ZGPobstacle(\x),\ZGPobstacle(\xp)} = 0\,.
    \end{equation*}

    \end{proof_sketch}

\begin{remark}
    The convergence of the series in \eref{eq_scalar_BCGP_kernel} is uniform, as a consequence of the uniform convergence of the Mercer decomposition of $\G$ over $\curveD$ \cite[Sect. 4.5]{Steinwart2008}. On the other hand, the convergence of the KLE of $\ZGP$ on the compact set $\curveD$ (and thus of the lifting counterparts on the whole domain $\DomG$) can be assured in the mean square sense. It can be established almost surely if we further assume a higher regularity degree of $\G$. However, the main idea is that the condition \eqref{eq_2z} is satisfied over a continuum, instead of a discretization of the compact set $\curveD$, while preserving the convergence properties of the KLE on the localized set.
\end{remark}

\subsection{Constraints over parameterized sets}\label{sec:BCGP}

Starting from the general framework for constraining a pre-defined GP over a compact subset $\curveD$ outlined in the previous section, we now specialize it to the situation where $\curveD$ is a parameterized compact manifold. The flexibility of choice of the probability measure $\measureD$ over $\curveD$ associated to the KLE of \eref{eq_2h} allow us to consider different geometries of $\curveD$. However, the numerical computation of GPR estimates based on a BCGP kernel can be influenced by a particular choice of the measure $\measureD$.

The following corollaries present two specific choices of $\measureD$ that simplify \eref{eq_2i}, and are helpful for computing derivatives of this kernel in \sref{sec:PIK}. In the first scenario, the geometry of $\curveD$ has no other restriction than to be compact in $\DomG$. In the second scenario, we suppose $\curveD$ to be a Lipschitz $(d-1)$-manifold in $\R^d$.

\begin{corollary}[Uniform measure on $\gammadom$]\label{Coro_1}
    In the context of Proposition \ref{Prop_1}, consider $\curveD$ as a compact manifold contained in $\DomG$ such that we can define a parameterization $\boldgamma$ with domain $\gammadom\subset\R^p$ (for an integer $p$) that is $(\gammadom,\curveD)$-measurable and such that $\curveD = \{\boldgamma(\s) = (\gamma_1(\s),\dots,\gamma_d(\s))^\itr,\ \s\in\gammadom\}$. Then, by choosing an adequate measure $\measureD$ for the Karhunen-Lo\`eve expansion of $\ZGP$ over $\curveD$, we have: 
    \begin{equation}\label{eq_uniform_gamma_measure}
        \eigenphiExt{n}(\x) = \frac{1}{\lameigenD_n} \int_\gammadom \G(\x,\boldgamma(\s)) \eigenphiD_n(\boldgamma(\s)) \frac{\dd\mesLebesgue(\s)}{\abs{\gammadom}}\, , \quad \text{ for all }\, n\in\N\, \text{ and }\, \x\in\DomG\,,
    \end{equation}
    where $\mesLebesgue$ is the Lebesgue measure on $\gammadom$.
\end{corollary}

\noindent We shall actually choose $\measureD$ as the normalized pushforward measure $\gamma_*\mesLebesgue / \abs{\gammadom}$ of the Lebesgue measure $\mesLebesgue$ over $\gammadom$ by the transformation $\boldgamma$. Therefore, the definition of the integral in \eref{eq_2t} will be reduced to:
\begin{equation*}
    \int_\curveD \G(\x,\y) \eigenphiD_n(\y) \frac{\dd\boldgamma_*\eta(\y)}{\abs{\gammadom}} = \int_\gammadom \G(\x,\boldgamma(\s)) \eigenphiD_n(\boldgamma(\s)) \frac{\dd\eta(\s)}{\abs{\gammadom}}\,,
\end{equation*}
for all $n\in\N$ and $\x\in\DomG$, by definition of the pushforward measure (see Theorem 3.6.1 in \cite{Bogachev2007}).

 \begin{corollary}[Surface measure on $\curveD$]\label{Coro_2}
    In the context of Proposition \ref{Prop_1}, consider $\curveD$ as a compact Lipschitz $(d-1)$-manifold contained in $\DomG$ such that we can define an injective Lipschitz parameterization $\boldgamma$ with domain $\gammadom\subset\R^p$ (for an integer $p$) such that $\curveD = \{\boldgamma(\s) = (\gamma_1(\s),\ldots,\gamma_d(\s))^\itr,\ \s\in\gammadom\}$. Then, by choosing an adequate measure $\measureD$ for the Karhunen-Lo\`eve expansion of $\ZGP$ over $\curveD$, we have: 
    \begin{equation}\label{eq_surface_measure}
        \eigenphiExt{n}(\x) = \frac{1}{\lameigenD_n} \int_\gammadom \G(\x,\boldgamma(\s)) \eigenphiD_n(\boldgamma(\s)) \sqrt{\det (J_\boldgamma(\s)^\itr J_\boldgamma(\s))}\ \frac{\dd\mesLebesgue(\s)}{\Hauss^{d-1}(\curveD)}\, , \quad \text{ for all }\, n\in\N\, \text{ and }\, \x\in\DomG\,,
    \end{equation}
    where $\mesLebesgue$ is the Lebesgue measure on $\gammadom$, and 
    $J_\boldgamma(\s) \in \Mat_{d,p}(\R)$ is the associated Jacobian matrix of $\boldgamma$ at $\s\in\gammadom$, a.e.
\end{corollary}

\noindent We actually choose $\measureD$ as the surface measure $\Hauss^{d-1}$ on $\R^d$ normalized by $\Hauss^{d-1}(\curveD)$, the $(p-1)$-dimensional measure of the manifold. The Lipschitz assumption on the curve $\curveD$ implies that the Jacobian of $\boldgamma$ is defined $a.e.$ on $\gammadom$. Then, by the change of variables formula on submanifolds from $\R^p$ to $\R^d$ (see Section 3.3 in \cite{Evans2018}), the integral in \eref{eq_2t} reads:
\begin{equation}\label{eq_integral_HausMeasure}
    \int_\curveD \G(\x,\y) \eigenphiD_n(\y) \frac{\dd\Hauss^{d-1}(\y)}{\Hauss^{d-1}(\curveD)} = \int_\gammadom \G(\x,\boldgamma(\s)) \eigenphiD_n(\boldgamma(\s)) \sqrt{\det (J_\boldgamma(\s)^\itr J_\boldgamma(\s))}\ \frac{\dd\mesLebesgue(\s)}{\Hauss^{d-1}(\curveD)}\,,
\end{equation}
for all $n\in\N$ and $\x\in\DomG$.  We can observe from this equation that under this choice of measure $\measureD$, the definition of $\eigenphiExt{n}$ is independent of the choice of the parameterization $\boldgamma$ of $\curveD$. In the particular case $p = 1$, the integral in \eref{eq_integral_HausMeasure} can be expressed in terms of the arc-length measure:
\begin{equation*}
\int_\curveD \G(\x,\y) \eigenphiD_n(\y) \frac{\dd\Hauss^{d-1}(\y)}{\Hauss^{d-1}(\curveD)} = \int_\gammadom \G(\x,\boldgamma(s)) \eigenphiD_n(\boldgamma(s)) \norm{\boldgamma'(s)} \frac{\dd\mesLebesgue(s)}{\Hauss^{d-1}(\curveD)}\,,
\end{equation*}
where $\boldgamma'(s)=\frac{\dd\boldgamma}{\dd s}$.

Either formula \eqref{eq_uniform_gamma_measure} or \eqref{eq_surface_measure} can then be plugged into \eqref{eq_2i}.
In what follows, a BCGP will be a zero-mean GP with a covariance function given in the form of \eref{eq_2i} under the choice of the measure $\measureD$ for the KLE along the parameterized set $\curveD$ as in one of the Corollaries \ref{Coro_1} and \ref{Coro_2}.

\subsection{A numerical method for approximating constrained kernels}\label{sec:Num}

In this section, we describe a numerical method to compute kernels for BCGPs, as defined by \eref{eq_2i} in the case where we consider a curve $\curveD\subset\R^d$ with a Lipschitz parameterization $\boldgamma : \gammadom \rightarrow \curveD$ and $\gammadom\subset\R^p$. For simplicity, we consider $p = 1$ and $\gammadom = [0,2\pi]$. As before, we denote by $\ZGPobstacle$ the associated BCGP, which is computed from an arbitrary base GP $\ZGP$ over the bounded domain $\DomG$, with continuous kernel $\G$. Since this continuous setting is described in terms of a spectral expansion of the base kernel, we perform a truncation of this spectrum. However, it is worth to remark that this numerical procedure does not represent a discrete enforcement of the boundary constraint \eqref{eq_2z} at collocation points, but it is instead an approximation of the geometry of the compact set verifying \eqref{eq_2z} at all points of the geometry.

We outline how the boundary-constrained kernel $\Gobstacle$ obtained by \eref{eq_2i} is computed through a carefully crafted approximation in terms of mode truncation and numerical integration. In order to completely define $\Gobstacle$, we seek to approximate the spectral pairs $\smash{(\lameigenD_n,\eigenphiD_n)}$ by discrete counterparts. By definition, these pairs are obtained from the integral representation with respect to the base kernel $\G$:
\begin{equation}
    \int_{\curveD}\G(\x,\y)\eigenphiD_n(\y)\,\dd\measureD(\y) = \lameigenD_n\eigenphiD_n(\x)\,,\,\text{ for } \x\in\curveD\,,\, n\in\N\,,
\end{equation}
and with respect to the measure $\measureD$ on $\curveD$. Our choices for $\measureD$ are detailed in Corollaries \ref{Coro_1} and \ref{Coro_2}. For the uniformity of notations, we introduce a weight $h$ defined on $\gammadom$ such that the spectral problem to approximate is:
\begin{equation}\label{eq_3f}
    \int_{\gammadom}\G(\x,\boldgamma(s))\eigenphiD_n(\boldgamma(s)) h(s)\, \dd\mesLebesgue(s) = \lameigenD_n\eigenphiD_n(\x)\,,\,\text{ for } \x\in\curveD\,,\, n\in\N\,,
\end{equation}
with $h \equiv 1/2\pi$ whenever the measure $\measureD$ is chosen as that of \cref{Coro_1}, and $h(s) = \norm{\boldgamma'(s)}/ \Hauss^{d-1}(\curveD)$, $s\in\gammadom$, whenever it is chosen as that of \cref{Coro_2}. In any case, we have that $h\geq 0$.

Hence, we start with a discretization of $\gammadom$ with $\Nint + 1$ nodes $\smash{s_i}$ such that $0=s_0 < s_1 <\dots < s_\Nint = 2\pi$ and define $\smash{h_{i}} = \smash{h(s_{i})(s_{i+1}-s_i)}$ for $i=0,\dots, \Nint - 1$ such that $\weightMat=\diag(h_0,h_1,\ldots,h_{\Nint-1})\in \Mat_{\Nint}(\R)$ is the diagonal matrix of weights of the $\Nint$ associated intervals in $\gammadom$. We consider a truncation parameter $\Neigen$ for the spectral modes in the KLE of $\ZGPobstacle$ (\ref{eq_2h}), corresponding to the largest $J+1$ eigenvalues. Then, for each $n = 0,\ldots, J$ and $\Nint$ large enough, we consider the discrete counterparts of \eqref{eq_3f} by replacing the pairs $\smash{(\lameigenD_n,\eigenphiD_n\circ\boldgamma)} \in \smash{\R^+\times L^2(\gammadom)}$ by their discrete versions $\smash{ ( \lameigenD^{(\Nint)}_n,{\boldsymbol\eigenphiD}^{(\Nint)}_n ) }\in\smash{\R^+\times\R^{\Nint + 1}}$. A Nystr\"om quadrature approximation of these integral relations (see for example \cite{Atkinson1997}), for each $n = 0,\ldots, J$, computed at the nodes $s_i\in\gammadom$ yields:
\begin{equation}
    \sum_{j = 0}^{\Nint-1} \G(\boldgamma(s_i),\boldgamma(s_j))\eigenphiD^{(\Nint)}_{n,j} h_{j} = \lameigenD^{(\Nint)}_n\eigenphiD^{(\Nint)}_{n,i}\,,
\end{equation}
for $\smash{\eigenphiD^{(\Nint)}_{n,i}}=\smash{({\boldsymbol\eigenphiD}^{(\Nint)}_n)_i}$ being the $i$-th component of $\smash{{\boldsymbol\eigenphiD}^{(\Nint)}_n}$. In matrix form, the generalized eigenvalue problem is:
\begin{equation}\label{eq_2l}
    \Ggram \weightMat \EigenphiMat = \EigenphiMat\LameigenMat\,,
\end{equation}
where $\Ggram$ is the symmetric matrix with elements $[\Ggram]_{ij} = \G\parenth{\boldgamma(s_i),\boldgamma(s_j)}$, $i,j = 0,\ldots,\Nint-1$; $\LameigenMat = \smash{\diag(\lameigenD^{(\Nint)}_0,\lameigenD^{(\Nint)}_1,\ldots,\lameigenD^{(\Nint)}_{\Nint-1})}\in\smash{\Mat_\Nint(\R^+)}$ is the diagonal matrix of eigenvalues; and $\EigenphiMat = \smash{ ( \boldsymbol{\eigenphiD}^{(\Nint)}_0,\ldots\boldsymbol{\eigenphiD}^{(\Nint)}_{\Nint-1} ) } \in\Mat_\Nint(\R)$ is the matrix of orthogonal eigenvectors. By using the square root of $\weightMat$, we can transform \eref{eq_2l} to the classical eigenvalue problem:
\begin{equation}\label{eq_eigenproblem}
    \GgramModif \EigenphiMatModif = \EigenphiMatModif\LameigenMat\,,
\end{equation}
with the symmetric matrix $\GgramModif = \weightMat^{\demi}\Ggram\weightMat^{\demi}$, and $\EigenphiMatModif = \weightMat^{\demi} \EigenphiMat$. Notice that the final approximation of the eigenfunctions $\boldsymbol{\eigenphiD}_n$ must be normalized in the sense of $L^2(\curveD, \measureD)$.

In practice, given an arbitrary continuous base kernel $\G$, we must consider a spectral truncation of $\Gobstacle$. In order to quantify the convergence of the spectral series, we define the spectral accuracy over $\curveD$ as:
\begin{equation}\label{eq_epsSpec}
    \epsSpec(\Nint,\Neigen) = 1-\frac{1}{\Trace\GgramModif} \sum_{n=0}^\Neigen \lameigenD^{(\Nint)}_n\,,\quad \Neigen\leq\Nint\,.
\end{equation}
Whenever $\epsSpec(\Nint,\Neigen)$ approaches to zero, the accuracy of the retained truncation of the sequences of eigenfunctions and eigenvalues for explaining the operator spectrum is higher. 
Therefore, we fix a threshold $\delta > 0$ such that, for a given value of $\Nint$, we choose $\Neigen$ as:
\begin{equation}\label{eq_Neigen_truncation}
    \Neigen = \min \braces{\, \Neigen^*\in\N\ :\ \epsSpec(\Nint,\Neigen^*) \leq \delta\, }\,.
\end{equation}
We redefine the eigenfunctions and eigenvalues matrices considering this truncation, which yields $\EigenphiMat,\EigenphiMatModif\in\Mat_{\Nint,\Neigen + 1}(\R)$, and $\LameigenMat\in\Mat_{\Neigen + 1}(\R)$, and correspond to the selected $\Neigen + 1$ eigenvalues in a decreasing order.

Then, an approximation of $\Gobstacle$ from \eref{eq_2i} defined in terms of $\Nint$ integration intervals and $J + 1$ spectral modes reads:

\begin{equation}\label{eq_discrete_kernel_prev}
    \Gobsdist{\Nint,\Neigen}(\x,\xp) = \G(\x,\xp)  - \smash{\sum_{n=0}^{\Neigen} \parenth{ \frac{1}{\sqrt{\lameigenD^{(\Nint)}_n} } \eigenphiExt{n}^{(\Nint)}(\x) } \parenth{ \frac{1}{\sqrt{\lameigenD^{(\Nint)}_n} } \eigenphiExt{n}^{(\Nint)}(\xp) }},
\end{equation}
where:
\begin{equation}\label{eq_discrete_liftings}
\eigenphiExt{n}^{(\Nint)} (\x) = \sum_{j=0}^{\Nint-1} \G\parenth{\x,\boldgamma(s_j)}\eigenphiD_{n,j}^{(\Nint)} h_{j} = \G(\x,\Xobstacleset) \weightMat \boldsymbol{\eigenphiD}_n^{(\Nint)} \,,
\end{equation}
for each $n = 0,\ldots,\Neigen$ and $\x,\xp\in\DomG$, with $\Xobstacleset = \smash{\{\boldgamma(s_j)\}_{j=0}^{\Nint-1}}$. Then \eref{eq_discrete_kernel_prev} in compact form reads:
\begin{equation}\label{eq_discrete_kernel}
    \Gobsdist{\Nint,\Neigen}(\x,\xp) = \G(\x,\xp) - \parenth{\G(\x,\Xobstacleset) \SpecFactor } \parenth{ \G(\xp,\Xobstacleset) \SpecFactor }^\itr\,,
\end{equation}
with:
\begin{equation}\label{eq_spectral_factor_mat}
    \SpecFactor = \weightMat^{\demi}\EigenphiMatModif\LameigenMat^{-\demi} \in\Mat_{\Nint,\Neigen + 1}(\R) \,,
\end{equation}
as the spectral factor matrix. It can be computed offline in order to reduce the computational complexity of evaluation of a BCGP kernel. The methodology is summarized in Algorithm \ref{algo_spectral_factor}.

\begin{algorithm}

    \caption{Computation of a BCGP kernel spectrum}\label{algo_spectral_factor}
    \KwIn{Base kernel $\G$ with fixed hyperparameters, number of integration intervals $\Nint$, profile boundary nodes $\Xobstacleset = \smash{\{\Xobstacle{k}\}_{k=0}^{\Nint-1}}$, measure weights $\weightMat$, spectral accuracy bound $\deltaSpec$}
    
    \KwOut{Spectral factor matrix $\SpecFactor$}
    
    Compute the modified Gram matrix $\GgramModif = \weightMat^{\demi}\smash{\G(\Xobstacleset,\Xobstacleset)} \weightMat^{\demi}$,

    Solve the eigenvalue problem \eqref{eq_eigenproblem} to obtain the ordered pairs $\smash{(\lameigenD^{(\Nint)}_n,\, \weightMat^{\demi}\eigenphiD^{(\Nint)}_n)_{n=0}^{\Nint-1}}$\;

    Compute the minimal number of spectral modes $\Neigen + 1$ satisfying the condition \eqref{eq_Neigen_truncation} given $\Nint$ and $\deltaSpec$\;

    Normalize the selected eigenvectors $\eigenphiD^{(\Nint)}_n$ in the sense of the $L^2(\curveD,\measureD)$ norm\;

    Compute the spectral factor matrix $\SpecFactor$ from \eref{eq_spectral_factor_mat}.

\end{algorithm}

\subsection{Benefits and limitations of constraining over a continuum}

From a numerical perspective, the computation of $\smash{\Gobsdist{\Nint,\Neigen}}$ is simplified by considering the matrix-form expression \eqref{eq_discrete_kernel}, which takes into account the choice of the measure $\measureD$, as in Corollaries \ref{Coro_1} and \ref{Coro_2} for instance. Moreover, the eigenvalue problem \eqref{eq_2l} can be solved as an offline procedure only once in order to define the kernel $\Gobstacle$, as explained in Algorithm \ref{algo_spectral_factor} (and subsequently in Algorithm \ref{algo_BCGP_derivatives}). Therefore, this step of BCGP construction needs not be included at each GPR estimation. The computational cost for the spectral factor in the offline procedure, up to $\Neigen+1$ modes, is of order $\text{O}\smash{(\Nint^2(\Neigen + 1))}$, which is negligible compared to the computational cost of online GPR regression (with or without BCGP) described in \tref{tab_computational_cost_comparison} given a number of observations $N$ in the regime $N\gg\Neigen$. Furthermore, if we need to compute kernel matrices of the form $\Gobstacle(\X,\Y)$, where $\X$ and $\Y$ are two subsets of points of $\DomG$, we notice that the spectral term in the right hand side of \eref{eq_discrete_kernel} depends on the kernel evaluation on $\X$ and $\Y$ separately. If for example $\X = \Y$, the covariance contribution has to be computed only once in practice.

When performing regression by means of formulas \eqref{eq_2a} and \eqref{eq_2b}, the most expensive computational task is the evaluation and inversion of the corresponding Gram matrices. For instance, given a set of $N$ observations $\Y = \smash{(y_1,\ldots,y_N)^\itr} \in \R^N$ corresponding to the locations $\X = \smash{\braces{\x_i}_{i=1}^N}$ inside $\DomG$, consider the representation formula:
\begin{equation}\label{eq_3g}
    \esp\braces{\ZGPobstacle(\x)\ |\ \ZGPobstacle(\X)=\Y} = \Gobstacle(\x,\X)\Gobstacle(\X,\X)^{-1}\Y\,,
\end{equation}
for all new observations $\x\in\DomG$ using the BCGP $\ZGPobstacle \sim\GP(0,\Gobstacle)$. An alternative method for enforcing the condition \eqref{eq_2z} on the posterior mean estimate will be to discretize $\curveD$ in $\Nobstacle$ points $\Xobstacleset = \smash{\{\Xobstacle{i}\}_{i=1}^\Nobstacle}$ and set $\ZGP(\Xobstacle{i}) = 0$ for each $i=1,\ldots,\Nobstacle$:
\begin{equation}\label{eq_discrete_approach}
    \esp\braces{\ZGP(\x)\ |\ \ZGP(\X)=\Y\,,\, \ZGP(\Xobstacleset) = 0\, } = \G(\x,\tilde{\X})\G(\tilde{\X},\tilde{\X})^{-1}\tilde{\Y}\,,
\end{equation}
with $\tilde{\X} = \X \cup \Xobstacleset$ and $ \tilde{\Y} = \smash{(y_1,\ldots,y_N,0,\ldots,0)^\itr} \in \R^{N+\Nobstacle}$. This approach uses $N + \Nobstacle$ design points at each step, which directly increases the computational complexity of inverting the associated Gram matrix to the order $\smash{\text{O}((N + \Nobstacle)^3)}$. The amount of discrete points $\Nobstacle$ must be considerably increased to obtain a satisfactory accuracy of the Dirichlet condition along $\curveD$ on the posterior mean estimate. 

On the other hand, the inversion cost with the proposed method is of order $\text{O}\smash{(N^3)}$ since no observation points are required in the compact set $\curveD$. Even though the evaluation of the BCGP kernel $\Gobstacle$ is more expensive in comparison to that of $\G$ (see \tref{tab_computational_cost_comparison}), the computational complexity of this procedure is $\text{O}\smash{(N^2(\Neigen + 1) + N(\Neigen + 1)\Nint)}$, for $\Neigen + 1$ being the number of spectral modes and $\Nint + 1$ being the number of selected nodes in $\gammadom$ for the approximation of the eigenvalue problem. Accordingly, the total computational cost of using BCGPs is not larger than $\text{O}\smash{(N^3)}$ since in practice $J\ll N$. Furthermore, the main advantage of using BCGPs relies on the fact that the constraint \eqref{eq_2z} is satisfied continuously, where the approximation procedure is not a spatial discretization but a discretization of the spectrum defining the geometry of $\curveD$. Lastly, it is worth to mention that the spectrum decay of the localized GP $\ZGPobstacle$ is shaped by the geometry of the compact set and choice of variance, length-scales and other kernel hyperparameters. Therefore, a proper choice of hyperparameters can reduce the number of spectral modes needed to achieve some desired accuracy of the constraint. In \sref{sec:CrossValidation}, we present a strategy based on UQ coverage in order to tune the kernel hyperparameters for improving learning fit and computational costs.

\begin{table}
    \centering
    \begin{tabular}{
        |>{\centering\arraybackslash}m{4cm}
        |>{\centering\arraybackslash}m{6cm}
        |>{\centering\arraybackslash}m{6cm}|}
        \hline
        \textbf{Procedure} & \textbf{BCGP} using \eref{eq_3g} & $\curveD$ \textbf{discretization} using \eref{eq_discrete_approach} \\
        \hline
        \textbf{Evaluation of Gram matrix} & $ \approx 2N^2(\Neigen+1) + 2 N (\Neigen + 1) \Nint$ & $\approx \displaystyle \smash{\frac{1}{2}} (N+\Nobstacle)(N+\Nobstacle + 1)$ \\
        \hline
        \textbf{Inversion of Gram matrix} & $\approx \displaystyle\smash{\frac{1}{3}} N^3$ & $\approx \displaystyle\smash{\frac{1}{3}} (N+\Nobstacle)^3 $ \\
        \hline
    \end{tabular}
    \caption{Comparison of the online computational cost (flops) of two approaches for the enforcement of the Dirichlet boundary condition \eqref{eq_2z} on the posterior distribution from the base GP $\ZGP$. As a reference, we consider the complexity of a single evaluation of kernel $\G$ of order O$(1)$. The inversion of the Gram matrix is performed by the Cholesky decomposition.}
    \label{tab_computational_cost_comparison}
\end{table}

\section{Physics-informed GPR for incompressible flows in aerodynamics}\label{sec:PIK}

This section outlines the approach for constructing physics-informed GPR in the context of fluid dynamics. We particularly address incompressible fluid flows past an aerodynamic profile as described in \sref{sec:fluid}. In \sref{sec:PIGP} we focus on how the divergence-free condition pertaining to incompressible fluid flows can be imposed to GP prior definitions, while simultaneously constraining them with the condition imposed along the profile boundary by means of BCGPs. The overall GPR strategy we have developed is fully detailed in \sref{sec:GPR-Design}. Particular aspects for computing BCGP kernel derivatives as needed in this setting are discussed in \sref{sec:dBCGP}, while some considerations on the use of non-constrained kernels are diverted to \sref{sec:discrete_kernel} with the aim to compare this more classical approach to the proposed one.

\subsection{Fluid dynamics context} \label{sec:fluid}

Let us consider a fluid flow in a 2D configuration around a profile as sketched on \fref{fig_4d}. The spatial domain is a rectangular box around the profile denoted by $\DomG$ and the boundary of the profile is denoted by $\curveD$. Two cases will be addressed in the subsequent numerical examples, namely a cylinder and a NACA airfoil \cite{Ladson1996} (see the definition of NACA profiles in \sref{sec:profiles} below). In the cylinder case the boundary is regular, whereas in the airfoil case the defined boundary is piecewise $C^1$. The fluid motion of an incompressible viscous flow is described by the velocity field $\velocity$ over a time interval $[0,T]$, with the incompressibility condition:
\begin{equation}\label{eq_3b}
    \diver\velocity(\x,t) = 0,\quad \text{for all}\, (\x,t)\in\DomG\times[0,T]\, .
\end{equation}
Since a 2D flow is considered, the velocity is written in terms of a scalar velocity stream function $\vstream$ as $\velocity=\curl\,\vstream$. Moreover, we impose a slip boundary condition on the profile boundary, so that:
\begin{equation}\label{eq_3c}
    \velocity(\x,t)\cdot\normalD(\x) = \curl\,\vstream(x,t) \cdot\normalD(\x) = 0,\quad \text{for all}\, (\x,t)\in\curveD\times[0,T]\,,
\end{equation}
where $\smash{\normal(\x)}$ is the outward unit normal vector to $\curveD$ at $\x$. In this scenario, the gradient $\Nabla\vstream$ of the stream function is normal to $\curveD$, so that $\vstream$ is actually constant along this curve.

The external boundary $\partial\DomG$ of $\DomG$ excluding the boundary of the profile $\curveD$ is the rectangle from $0$ to $L_1$ in the horizontal (streamwise) direction and from $0$ to $L_2$ in the vertical one. The boundary conditions over $\partial\DomG=\partial\DomG_\text{in}\cup\partial\DomG_\text{out}\cup\partial\DomG_\text{w}$ are: 
\begin{enumerate}[label = (\roman*)]
    \item Inlet and outlet conditions: they are given by the fields $\uinlet$ and $\uoutlet$ such that for all $t\in[0,T]$,
    \begin{equation}\label{eq_3a}
    \velocity(\x,t) = \begin{cases}
    \uinlet(\x,t)\ , & \text{for }\x\in\partial\DomG_\text{in} = \braces{0}\times(0,L_2)\,, \\
    \uoutlet(\x,t)\ , & \text{for }\x\in\partial\DomG_\text{out} = \braces{L_1}\times(0,L_2)\,. \\
    \end{cases}
    \end{equation}
    \item Top and bottom boundary conditions: they are given by the velocity field $\uwall$ for all  $t\in [0,T]$:
    \begin{equation}\label{eq_3d}
        \velocity(\x,t) = \uwall(\x,t)\,\quad \text{ for }\, \x\in\partial\DomG_\text{w} = (0,L_1)\times\braces{0,L_2}\,.
    \end{equation}
\end{enumerate}
The definition of $\uinlet$ and $\uoutlet$ will typically come from flow data, whereas $\uwall$ arises either from physics or from flow data, depending on the specific configuration of the scenario.

\subsection{Divergence-free, boundary-constrained GPs} \label{sec:PIGP}

We now outline the GP framework to model flow fields while accounting for the divergence-free and boundary conditions continuously. A natural choice is to model the velocity stream function $\vstream$ as a scalar GP with a prior definition that includes the slip boundary condition on $\curveD$ and the divergence-free condition within $\DomG$.

More precisely, we consider an arbitrary GP $\ZGP$ indexed on the spatial domain $\DomG$ with a covariance kernel $\G$. Since the profile boundary $\curveD$ is a compact set inside $\DomG$, we model $\vstream$ by a BCGP defined from this base GP $\ZGP$. That is, we consider $\vstream = \ZGPobstacle \sim\GP(0,\Gobstacle)$ as the BCGP associated to $\ZGP$ through the procedure of Proposition \ref{Prop_1}, for $\Gobstacle$ being the constrained kernel defined in \eref{eq_2i}. It follows that the flow velocity can be modelled as the zero-mean vector-valued GP:
$$
\velocity = \curl\,\ZGPobstacle \sim \GP(0,\Kobstacle)\,,
$$
with a matrix-valued covariance function $\Kobstacle:\DomG\times\DomG\to\Mat_2(\R)$ defined as:
\begin{equation}
\Kobstacle(\x,\xp) =\Cov\braces{\curl\,\ZGPobstacle(\x),\curl\,\ZGPobstacle(\xp)} \\
= \curl_\x\otimes\curl_{\xp}\,\Gobstacle(\x,\xp)\,, \label{eq_2d}
\end{equation}
for $\x,\xp\in\DomG$, with $\curl_\x=\smash{\Nabla_\x^\perp}=\smash{(-\partial_{x_2},\partial_{x_1})^\itr}$ and $\curl_{\xp}=\smash{\Nabla_{\xp}^\perp}=\smash{(-\partial_{x^\prime_2},\partial_{x^\prime_1})^\itr}$ acting on the first and the second entries of $\Gobstacle$, respectively. This kernel is well-defined pointwise if we assume that the base kernel $\G$ has the required regularity \cite{Ginsbourger2016,Owhadi2023}. The uniform convergence of the series in \eref{eq_2i} is actually conserved considering the partial derivatives of $\Gobstacle$, as stated in the following corollary. The regularity of the base kernel $\G$ is thus inherited by the BCGP kernel $\Gobstacle$.

\begin{corollary}\label{Coro_kernel_derivatives}
    In the context of Proposition \ref{Prop_1}, let $\Nderiv\in\N$ and assume that the kernel $\G$ has regularity $\C^{\Nderiv,\Nderiv}(\DomG\times\DomG)$. 
    Then for a multi-index $\smash{\alpha\in\N^d}$ of order less or equal to $\Nderiv$, the Gaussian process
    \begin{equation}\label{eq_2aa}
        \partial_\x^\alpha\ZGPobstacle(\x) = \partial_\x^\alpha\ZGP(\x) - \sum_{n=0}^{+\infty} \sqrt{\lameigenD_n}\,\partial_\x^\alpha\eigenphiExt{n}(\x)\,\xiD_n \,, \quad\x\in\DomG\,,
    \end{equation}
    is well-defined in the mean square sense. Moreover, given another multi-index $\smash{\beta\in\N^d}$ of order less or equal to $\Nderiv$, the associated covariance kernel can be computed following \eref{eq_2i} as:
    \begin{equation}\label{eq_2v}
        \partial^\alpha_\x\otimes\partial^\beta_{\xp}\,\Gobstacle(\x,\xp) = \partial^\alpha_\x\otimes\partial^\beta_{\xp}\,\G(\x,\xp) - \sum_{n=0}^{+\infty}\frac{1}{\lameigenD_n}\parenth{\int_\curveD \partial_\x^\alpha\G(\x,\y)\eigenphiD_n(\y)\dd\measureD(\y)} \parenth{\int_\curveD \partial_{\xp}^\beta\G(\xp,\y)\eigenphiD_n(\y)\dd\measureD(\y)}\,,
    \end{equation}
    for $\x,\xp\in\DomG$, with uniform convergence of the series.
\end{corollary}

\begin{remark}
    We observe from expression \eqref{eq_2v} that the derivatives of the eigenfunctions $\phi_n$ themselves are not needed in order to compute the derivatives of $\Gobstacle$. Moreover, uniform convergence of \eref{eq_2aa} can also be obtained if the spectrum of the associated operator is uniformly bounded: $\smash{\sum_{n=0}^{+\infty} \sqrt{\lambda_n} \norm{\partial^\alpha_\x\eigenphiD_n}_{\infty}} < +\infty$. In practice, we are interested in the well-posedness of posterior estimates resulting from conditioning a GP prior on field observations.
    Therefore, we only require the base kernel to satisfy $\G\in\C^{2,2}(\DomG\times\DomG)$ to be able to compute estimates of the velocity and vorticity fields.
\end{remark}

We observe that in this case the GP $\velocity = \curl\,\ZGPobstacle$ satisfies for each $\x\in\DomG$:
\begin{equation}\label{eq_2e}
    \diver\curl\,\ZGPobstacle(\x) = \diver\Nabla^\perp\,\ZGPobstacle(\x)= O\,,
\end{equation}
where $O$ stands for the null Gaussian random variable. This property is also satisfied for the elements of the RKHS $\HKobstacle$ associated to $\Kobstacle$. In particular, for all $\Beta\in\R^2$ and $\xp\in\DomG$:
\begin{equation*}
    \Nabla_\x\cdot\Kobstacle(\cdot,\xp)\Beta = \Nabla_\x\cdot\Nabla^\perp_\x\otimes\Nabla^\perp_{\xp}\,\Gobstacle(\cdot,\xp) \Beta = 0\,.
\end{equation*}
Thus the realizations of $\velocity$ will be almost surely (a.s.) divergence-free vector fields defined over $\DomG\subset\R^2$. Moreover, it is also verified that:
\begin{equation}\label{eq_BCGP}
    \curl\,\ZGPobstacle(\x)\cdot\normalD(\x) = O,\quad \forall\x\in\curveD\, ,
\end{equation}
in a mean-square sense, since a Dirichlet condition along $\curveD$ holds for $\ZGPobstacle$ by the derivation presented previously in \sref{sec:GPcompact}; see \eref{eq_2y}. Therefore, under this choice of GP, the prior estimates of the velocity field in our GPR strategy automatically satisfy the physical conditions \eqref{eq_3b} and \eqref{eq_3c} in a continuous sense for all points. Since the condition is enforced only at the profile boundary $\curveD$, the values of the kernel at the interior of the obstacle are ignored, and no observation points are needed there. Merging this approach within time-evolving Lagrangian models guarantees that no particle penetrates the profile boundary since the slip condition is satisfied by the prior and posterior GPs. The smooth zero-continuation of the kernel towards the interior region has no numerical consequences since no observation points are required.

The next step is to define the structure of the base kernel $\G$, from which the boundary-constrained kernel $\Gobstacle$ is computed. Since $\G\in\C^{2,2}(\DomG\times\DomG)$ must hold, one possible choice is to define it as a standard anisotropic radial-basis function (RBF):
\begin{equation}\label{eq_RBF}
    \G(\x,\xp) = \sigma^2 \exp \brackets{ -\frac{ (x_1 - x'_1)^2 + \alpha^2 (x_2-x'_2)^2 }{2 \lcor^2}  },\quad \x,\xp\in\DomG\,,
\end{equation}
where $\sigma^2, \lcor$, and $\alpha$ are hyperparameters for variance, correlation length and anisotropy, respectively. However, the dynamics of the flow evolves at different spatial scales, and so a single lengthscale may not be sufficient to capture this behavior. In the next section, we explain how to account for multiple scales from physical arguments.

\subsection{Energy decay GPs} \label{sec:ernegy_decay_kernel}

In this work, we consider a power law for decay of the velocity energy spectrum $E(k)$ at wavenumber $k$. In a 2D isotropic scenario at small spatial scales, we have that $E(k)$ is (approximately) proportional to $k^{-3}$ \cite[Sect. 4.5]{Frisch1995}. This law corresponds to the behavior of the flow for high wavenumbers $k$ in the direct Richardson cascade \cite{Boffetta2012}. In this case, the expectation of the squared velocity increment from a point $\x\in\DomG$ to $\x + \h\in\DomG$ is approximately of order $\smash{\norm{\h}^2}$. This information can be included in the GP kernel definition through a multi-scale additive structure. We adapt the ideas proposed in \cite[Sect. 5]{Owhadi2023} to our non-periodic anisotropic scenario.

Instead of using solely \eqref{eq_RBF}, we rather define the base kernel $\G$ as an additive kernel with different orders of correlation lengths:
\begin{equation}\label{eq_kernel_additive}
    \G(\x,\xp) = \sum_{m=0}^M \sigma_m^2 \exp \brackets{ -\frac{ (x_1 - x'_1)^2 + \alpha_m^2 (x_2-x'_2)^2 }{2 \lcor_m^2}  },\quad \x,\xp\in\DomG\,,
\end{equation}
where $M + 1$ is the number of scales, and $\sigma_m^2, \lcor_m$, and $\alpha_m$ are the corresponding hyperparameters to be defined. We have now to set a reference structure and fit the resulting decay parameter, say $\gamma$, of the kernel modelling counterpart to follow the energy decay power law.

To fix this reference, we set the decay of the correlation lengths as $\smash{\lcor_m} = \smash{\lcor_0 /2^{3m}}$ for $m = 0,1,\ldots,M$ in order to capture the small spatial scales. Moreover since the power law is isotropic in these scales, we set the anisotropy parameters with a lower limit as $\smash{\alpha_m} = \smash{\max\parenth{1\,,\, 2^{3-m}}}$. This definition ensures that the horizontal and vertical dynamics are different in magnitude at the coarser scales, which is adapted to our flow scenario past an aerodynamic obstacle. It also satisfies the condition $\alpha_m\to 1$ in order to recover the isotropic power law at the smaller scales. Finally, we set a reference structure for the decay law of the velocity variance increments in terms of the decay parameter $\gamma\in\N$, such that $\smash{\sigma_m} = \smash{\sigma_0 / 2^{\gamma m}}$. Then, under this reference, we can fix $\gamma$ as a physics-informed parameter by fitting the 2D energy decay law.

To do this, we observe that the kernel \eqref{eq_kernel_additive} is stationary and thus:
\begin{align*}
    \esp\braces{\ \norm{\curl\,\ZGP(\x + \h) - \curl\,\ZGP(\x)  }^2\ } & = 2\,\curl_\x\cdot\curl_{\xp}\parenth{ \G(\x, \x) - \G(\x + \h, \x) } \\
        & = 2 \sum_{m=0}^M \frac{\sigma_m^2}{\lcor_m^2} \brackets{ 1 + \alpha_m^2 + \exp\parenth{- \frac{h_1^2 + \alpha_m^2 h_2^2}{2\lcor_m^2}}\parenth{ \frac{ h_1^2 + (\alpha_m^2 h_2)^2 }{\lcor_m^2} - 1 - \alpha_m^2 } }\,,
\end{align*}
where the sum is dominated by the modes for which $\norm{\h}^2 > 2\lcor_m^2$. For an increment in the direction of $\h$ of order $\norm{\h} \simeq 2^{-q}$ with $ 3 < q\leq M$, this yields similarly to \cite{Owhadi2023}:
$$
    \esp\braces{\ \norm{\curl\,\ZGP(\x + \h) - \curl\,\ZGP(\x)  }^2\ } \propto \sum_{m=q/3}^M \frac{\sigma_m^2}{\lcor_m^2} = O(2^{2q(1-\gamma/3)}) = O(\norm{\h}^{-2\parenth{1-\gamma/3}})\,.
$$
We obtain that $\gamma = 6$ must hold in order to fit the power law of the squared velocity increment of order $O(\norm{h}^2)$ when $\norm{\h}\rightarrow 0$. This behavior is then included in the prior and posterior distribution of the physical fields. The remaining hyperparameters $\lcor_0$ and $\sigma_0$, corresponding to the correlation length and variance at the initial coarse scale, are flexible and can be tuned using validation datasets.

The reasonings of this section can be further extended to consider more accurate power laws for velocity and/or vorticity increments, such as the Kraichnan correction for scale-free power laws \cite{Boffetta2012}. Moreover, an alternative strategy to include energy decay in velocity and vorticity kernels can be followed considering invariant-measure-informed kernels \cite{Hamzi2025}.

\subsection{GPR observation design} \label{sec:GPR-Design}

Once the physics-informed GP prior distribution of the velocity field $\velocity$ has been defined, the observation design used to perform GPR is established, with the aim to obtain estimates of $\velocity$ as the mean of the posterior distribution of the conditioned GP. The general GPR design approach is displayed on \fref{fig_4d}. Here we spot the observation points at the inlet and outlet boundaries of the computational domain $\DomG$ used for estimates, the observation points in the interior of $\DomG$, and the observation points for the prescribed physical boundary conditions at the top and bottom boundaries.

\begin{figure}
    \centering
    \input{figures/domain_for_GPR.tex}
    \caption{
    GPR setting for approximating the velocity field of an incompressible fluid flow around a profile in a rectangular computational domain with discrete wall boundary observations (blue dots \textcolor{blue}{$\bullet$}), flow inlet and outlet boundary data (orange dots \textcolor{orange}{$\bullet$}), and bulk flow data (red dots \textcolor{red}{$\bullet$}). No design observation points are used on the profile boundary.}
    \label{fig_4d}
\end{figure}

Let us consider each individual snapshot at a fixed time $t\in[0,T]$. Let $\Vboundaryset(t)=\smash{\{\Vboundary{i}(t)=\velocity(\Xboundary{i},t)\}_{i=1}^{\Nboundary}}$ be the discrete values of the flow velocity observed at positions $\Xboundaryset = \smash{\{\Xboundary{i}\}_{i=1}^{\Nboundary}}$ on the external boundary $\partial\DomG$ of $\DomG$. The sets $\Vboundaryset(t)$ and $\Xboundaryset$ correspond thus to a collection of discrete observation of the conditions \eqref{eq_3a} and \eqref{eq_3d}.
Additionally, we consider discrete values $\Vdomset(t)=\smash{\{\Vdom{i}(t)=\velocity(\Xdom{i},t)\}_{i=1}^{\Ndom}}$ for snapshots of the flow velocity at positions $\Xdomset = \smash{\{\Xdom{i}\}_{i=1}^{\Ndom}}$ in the interior of $\DomG$.

Then we define $\uapprox(\cdot,t)$ as the approximation of the velocity field by the mean of the GP posterior under the definition of the physics-informed GP prior $\velocity = \velocityGPobstacle$ from the previous \sref{sec:PIGP}. The given discrete observations are $\velocity(\Xboundaryset,t)= \Vboundaryset(t)$ and $\velocity(\Xdomset,t)= \Vdomset(t)$, for external boundary and bulk flow information. The velocity field can then be estimated as:
\begin{equation} \label{eq_3e}
    \uapprox(\x,t) = \esp\braces{\velocityGPobstacle(\x)\ |\  \velocityGPobstacle(\Yset) = \Uset(t) } \,,
\end{equation}
where $\Yset = \Xboundaryset \cup \Xdomset$ is the expanded set of total observation positions and $\Uset(t) = \smash{(\Vboundaryset(t)^\itr,\Vdomset(t)^\itr)}^\itr$ is the $\R^{2N}$ column vector of the corresponding velocity information, with $N=\Nboundary+\Ndom$. This reconstruction formula allows us to approximate the fluid flow velocity at any unobserved point $\x$ in the domain $\DomG$ while inherently satisfying the divergence-free and profile boundary conditions continuously at all locations, respectively. A representation formula for $\smash{\uapprox(\cdot,t)}$ can be established along the same lines as in \sref{sec:math} so that:
\begin{equation}\label{eq_3e_reg}
\uapprox(\x,t) = \Kobstacle(\x,\Yset)\,\BetaBCGP(t)\,,
\end{equation}
with regression coefficients:
\begin{equation}
\BetaBCGP(t) = \left(\Kobstacle(\Yset,\Yset) + \nugget\II\right)^{-1} \Uset(t)\,,
\end{equation}
and covariance matrix $\smash{\Kobstacle}$ given by \eqref{eq_2d}. This representation introduces the regularization parameter $\nugget>0$ (the so-called nugget) such that the formula \eqref{eq_3e_reg} is actually no longer strictly interpolatory. For each numerical experiment, we chose the order of the regularization parameter as the minimal order for which the Gram matrices are invertible.

The covariance counterpart for computing the posterior correlation between estimates from two locations $\x,\xp$ inside $\DomG$ can also be established as:
\begin{equation*}
\upostcov(\x,\xp) = \Cov\braces{\velocityGPobstacle(\x),\,\velocityGPobstacle(\xp)\ |\ \velocityGPobstacle(\Yset) = \Uset(t) }\,.
\end{equation*}
By using a matrix-valued formula similar to \eref{eq_2o}, this last expression becomes:
\begin{equation}\label{eq_posterior_cov_velocity}
\upostcov(\x,\xp) = \Kobstacle(\x,\xp) - \Kobstacle(\x,\Yset)\left(\Kobstacle(\Yset,\Yset) + \nugget\II\right)^{-1}\Kobstacle(\Yset,\xp)\,.
\end{equation}

Besides, other flow fields can be estimated from this GPR representation based on velocity information. Indeed, the stream velocity function $\vstream$ can be approximated by:
\begin{equation}\label{eq_stream_approx}
    \vstreamapprox(\x,t) = \curl_{\xp}\,\Gobstacle(\x,\Yset)\,\BetaBCGP(t)\,,\quad\x\in\DomG\,,
\end{equation}
whereas the vorticity field $ \vorticity$, defined by:
\begin{equation}
    \vorticity(\x,t) = \curl_\x\cdot\velocity(\x,t) = \curl_\x^\itr\velocity(\x,t)\,,
\end{equation}
can be directly reconstructed from:
\begin{equation} \label{eq_3i}
\begin{split}
    \vortapprox(\x,t) &= \Delta_\x(\curl_{\xp}\,\Gobstacle(\x,\Yset)\cdot\BetaBCGP(t)) \\
    &= \curl_{\xp}\Delta_\x\Gobstacle(\x,\Yset)\cdot\BetaBCGP(t)\,,\quad\x\in\DomG\,,
    \end{split}
\end{equation}
since $\curl_\x\cdot\curl_\x=\norm{\curl_\x}^2=\Delta_\x$.

We finally define two indicator functions to quantify the approximation of the boundary condition on the profile boundary using the BCGP scalar kernel $\Gobstacle$. The first corresponds to the $L^1$ norm of the stream function reconstruction over $\curveD$, that is:
\begin{equation}\label{eq_stream_indicator}
    \norm{\vstreamapprox(\cdot,t) }_{L^1(\curveD)} = \int_\gammadom \abs{\vstreamapprox(\gamma(s),t)}\, \dd s \,.
\end{equation}
The second one is a relative aggregated indicator of the normal components of the velocity field on the profile boundary $\curveD$:
\begin{equation}\label{eq_normal_indicator}
    \epsNormal = \frac{\norm{\uapprox(\cdot,t)\cdot\normalD}_{L^1(\curveD)} } { \norm{\uapprox(\cdot,t)}_{L^1(\curveD)} } \,.
\end{equation}
These indicators depend on the numerical spectral accuracy bound $\deltaSpec$ specified for the computation of BCGP kernels when evaluating the corresponding approximations $\vstreamapprox$ and $\uapprox$ in the BCGP-based approach.

\subsection{Approximation of constrained kernel derivatives}\label{sec:dBCGP}

We see from Eqs. \eqref{eq_3e_reg}, \eqref{eq_stream_approx} and \eqref{eq_3i} that derivatives of the BCGP kernels are needed in the proposed approach. Here we describe how they may be computed. An approximation of the kernel derivatives in \eref{eq_2v} is given in a similar manner as for \eref{eq_discrete_kernel_prev}:

\begin{equation}\label{eq_BCGP_derivatives_discrete}
    \partial^\alpha_\x\otimes\partial^\beta_{\xp}\,\Gobsdist{\Nint,\Neigen}(\x,\xp) = \partial^\alpha_\x\otimes\partial^\beta_{\xp}\,\G(\x,\xp)  - \sum_{n=0}^{\Neigen} \parenth{ \frac{1}{\sqrt{\lameigenD^{(\Nint)}_n} } \partial^{\alpha}_\x\eigenphiExt{n}^{(\Nint)}(\x) } \parenth{ \frac{1}{\sqrt{\lameigenD^{(\Nint)}_n} } \partial^{\beta}_{\xp}\eigenphiExt{n}^{(\Nint)}(\xp) },
\end{equation}
where $I$ and $J + 1$ stand for the number of integration intervals and spectral modes as in \sref{sec:Num}, and for each multi-index $\alpha$ and $n = 0,\ldots,\Neigen$:
\begin{equation}\label{eq_lifting_derivative}
    \partial^\alpha_\x\eigenphiExt{n}^{(\Nint)} (\x) = \sum_{j=0}^{\Nint-1} \partial^\alpha_\x\G\parenth{\x,\boldgamma(s_j)}\eigenphiD_{n,j}^{(\Nint)} h_{j} = \partial^\alpha_\x \G(\x,\Xobstacleset) \weightMat \boldsymbol{\eigenphiD}_n^{(\Nint)}\,.
\end{equation}
Similarly to \eref{eq_discrete_kernel}, the approximation of the kernel derivative can then be computed as:
\begin{equation}\label{eq_discrete_kernel_deriv}
    \partial^\alpha_\x\otimes\partial^\beta_{\xp}\,\Gobsdist{\Nint,\Neigen}(\x,\xp) = \partial^\alpha_\x\otimes\partial^\beta_{\xp}\G(\x,\xp) - \parenth{ \partial^\alpha_\x \G(\x,\Xobstacleset) \SpecFactor } ( \partial^\beta_{\xp} \G(\xp,\Xobstacleset) \SpecFactor)^\itr \,,
\end{equation}
with $\SpecFactor$ given by \eref{eq_spectral_factor_mat}. In all subsequent applications the numerical approximations $\Gobsdist{\Nint,\Neigen}$ and $\partial^\alpha_\x\otimes\partial^\beta_{\xp}\,\Gobsdist{\Nint,\Neigen}$ are computed using the preceding definitions, as outlined in Algorithm \ref{algo_BCGP_derivatives}.

\begin{algorithm}

    \caption{Computation of BCGP kernels with derivatives}\label{algo_BCGP_derivatives}
    \KwIn{Base kernel $\G$ with fixed hyperparameters, evaluation points $\Xset = \smash{\{\x_i\}_{i=1}^{N_\Xset}}$ and $\Xpset = \smash{ \{\xp_j\}_{j=1}^{N_\Xpset}}$, spectral factor matrix $\SpecFactor$, profile boundary nodes $\Xobstacleset = \smash{\{\Xobstacle{k}\}_{k=0}^{\Nint-1}}$, derivative order $\Nderiv$}
    
    \KwOut{BCGP kernel evaluation $\Gobstacle(\Xset,\Xpset)$ and all derivatives $\partial^{\alpha}_{\x}\otimes\partial^\beta_{\xp}\,\Gobstacle (\Xset,\Xpset)$ of order $ \abs{\alpha} + \abs{\beta} = 1,\ldots,\Nderiv$} 
    
    \For{each multi-index $\alpha$ of order $\abs{\alpha}\leq \Nderiv$}
    {
        Compute the kernel marginal evaluation $\partial^{\alpha}_{\x}\G(\Xset,\Xobstacleset)$ \;

        Compute $\partial^{\alpha}_{\x}\G(\Xset,\Xobstacleset) \SpecFactor$ using the spectral factor matrix from Algorithm \ref{algo_spectral_factor} \;
        
        \If{$\Xpset\neq \Xset$}{
            Compute the kernel marginal evaluation $\partial^{\alpha}_{\x}\G(\Xpset,\Xobstacleset)$ \;
    
            Compute $\partial^{\alpha}_{\x}\G(\Xpset,\Xobstacleset) \SpecFactor$\;
        }

    }

    \For {each pair of multi-indices $(\alpha,\beta)$ of order $\abs{\alpha} + \abs{\beta}\leq \Nderiv$}
    {
        Compute the base kernel derivative $\partial^\alpha_{\x}\otimes\partial^\beta_{\xp}\,\G (\Xset,\Xpset)$ \;
        Assemble the BCGP kernel derivative $\partial^\alpha_{\x}\otimes\partial^\beta_{\xp}\,\Gobstacle$ as in \eref{eq_discrete_kernel_deriv}.
    }
    
\end{algorithm}

\begin{remark}
    Consider the case of a parameterized curve $\curveD$ of class $C^k$, denoting as $\boldgamma$ its regular diffeomorphism. Then, as a consequence of the Mercer decomposition of $\G$ along $\curveD$, we have that:
$$
\smash{\partial^\alpha_s \otimes \partial^\beta_{s'}\,\Gobstacle(\boldgamma(s),\boldgamma(s'))} = 0\, ,\quad  \text{ for all } s,s'\in\boldgamma^{-1}(\curveD)\, ,
$$
and for any derivatives in the curve direction with multi-indices $\alpha$ and $\beta$ of order $k$. Nevertheless, it is worth to notice that given multi-indices $\alpha,\beta$, the derivatives $\smash{\partial^\alpha_\x\otimes\partial^\beta_{\xp}\,\Gobstacle(\x,\xp)}$ are not necessarily null for all $\x,\xp\in\DomG$ since the spectral expansion of $\G$ through the pairs $\smash{(\lameigenD_n,\eigenphiD_n)_{n=0}^{+\infty}}$ is obtained specifically from its restriction to $\curveD$, and not over all $\DomG$.
\end{remark}

\section{Application to fluid flows around aerodynamic profiles}\label{sec:Aero}

In this section, we perform GPR reconstructions of the velocity field of incompressible flows around aerodynamic profiles using the physics-informed framework presented above. The numerical implementation for the presented results on the cylinder profile and NACA 0412 airfoil is available at \href{https://github.com/adrirps/Physics-Informed-BCGPs-for-Flow-Fields-Reconstruction}{https://github.com/adrirps/Physics-Informed-BCGPs-for-Flow-Fields-Reconstruction}. This Python implementation uses symbolic computation, through the package SymPy \cite{SymPy2017}, for getting high precision in kernel computations, including derivatives. Reported CPU times were obtained on a standard laptop with Intel Core i5-1345U processor and 16 GB RAM.

\subsection{Profiles}\label{sec:profiles}

We consider two cases for the definition of the profile boundary:

\begin{enumerate}[label=(\roman*)]
    \item Cylinder: the profile boundary is the circle $\Dcylinder$ of radius $r>0$ centered at ${\boldsymbol c}\in\DomG$, $\Dcylinder = \{\x\in\DomG,\,\norm{\x-{\boldsymbol c}}=r\}$, and with associated transformation $s\mapsto\boldgamma(s)$:
    \begin{equation*}
   \gamma_1(s) = c_1 + r\cos s\,,\quad\gamma_2(s) = c_2 + r\sin s\,,\quad s\in\gammadom\,.
    \end{equation*}
    \item NACA 4-digit airfoil series \cite{Ladson1996}: the definition of the transformation $s\mapsto\boldgamma(s)$ for a NACA profile in terms of its mean camber line $\ycamber$ and thickness distribution function $\ythickness$ is as follows. For a four-digit codification ``MPT" with parameters:
    \begin{itemize}
        \item $\text{M}\in [0,0.1]$, the scaled camber line amplitude;
        \item $\text{P} = \smash{\argmax_{x\in(0,1)}\ycamber(x)}$, the relative position of the maximum camber line;
        \item $\text{T}$, the two-digit scaled thickness amplitude;
        \item and $c$, the airfoil chord length;
    \end{itemize}
we have that $\DNACA{\text{MPT}} = \boldgamma(\gammadom)$ with the transformation $\boldgamma(s)=c\smash{(\gamma_1(s),\gamma_2(s))^\itr}$ where $x(s) = \smash{\demi(1+\cos s)}$ for $s \in\gammadom$ and:
    \begin{equation}\label{eq_NACA_definition}
    \gamma_1(s)=x(s) - \signo(\sin s)\frac{\ythickness(x(s))\ycamber^\prime(x(s))}{\sqrt{\brackets{\ycamber^\prime(x(s))}^2+1}}\,,\quad
    \gamma_2(s)=\ycamber(x(s)) + \signo(\sin s)\frac{\ythickness(x(s)) }{\sqrt{\brackets{\ycamber^\prime(x(s))}^2+1}} \,,
    \end{equation}
    such that
    \begin{align}
    \ycamber(x) & = \begin{cases} \displaystyle 
    \frac{\text{M}}{\text{P}^2} \parenth{ 2\text{P}x - x^2 }, & \text{for } 0\leq x \leq \text{P}\,, \\
    \displaystyle \frac{\text{M}}{(1-\text{P})^2} \parenth{ 1 - 2\text{P} + 2\text{P}x - x^2 }, & \text{for } \text{P} < x \leq 1\,,
    \end{cases}
   \end{align}
with:
\begin{equation}
\ythickness(x) = 5\text{T} \parenth{ a_0 \sqrt{x} + a_1 x + a_2 x^2 + a_3 x^3 + a_4 x^4 },\quad x\in[0,1]\,.
\end{equation}
The  coefficients $a_0,\dots,a_4$ are defined using the standard NACA conventions \cite{Ladson1996} for a closed airfoil.
\end{enumerate}

\subsection{GPR estimate with no information design inside the domain}

\begin{figure}[h]
    \centering
    \begin{subfigure}[b]{0.28\textwidth}
        \includegraphics[width=\textwidth]{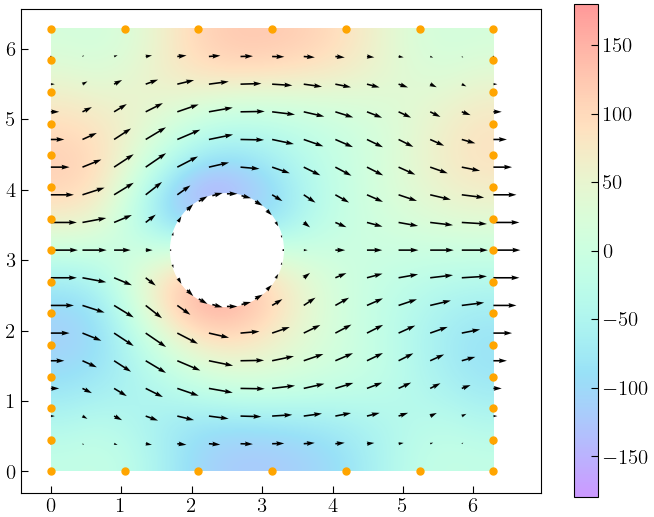} \caption{Cylinder ($r = 0.8$) with kernel parameters $\stdv = 1.5$, $\lcor = 1.5$, and $\alpha = 1.5$.}
    \end{subfigure}
    \hfill
    \begin{subfigure}[b]{0.28\textwidth}
        \includegraphics[width=\textwidth]{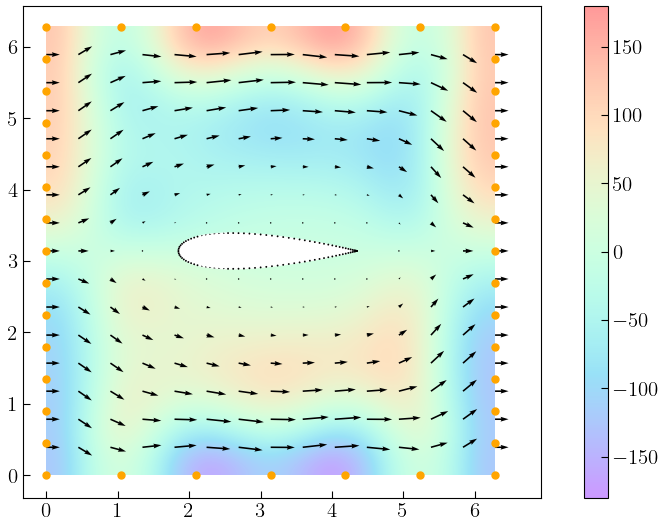}
        \caption{Symmetric NACA 0420 airfoil with kernel parameters $\stdv = 1.5$, $\lcor = 0.95$, and $\alpha = 0.95$.}
    \end{subfigure}
    \hfill
    \begin{subfigure}[b]{0.29\textwidth}
        \includegraphics[width=0.95\textwidth]{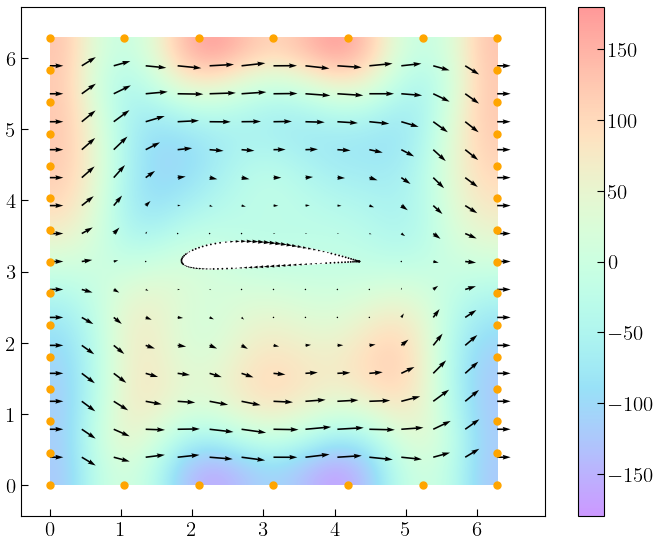}
        \caption{Non-symmetric NACA 4415 airfoil with kernel parameters $\stdv = 2.5$, $\lcor = 0.95$, and $\alpha = 0.95$.}
    \end{subfigure}
    \caption{Spatial GPR interpolations for velocity (arrows) and vorticity (colormap) fields around different profiles using BCGP kernels over the computational domain $\DomG = [0,2\pi]\times[0,2\pi]$ with $\Nboundary = 40$ boundary design points (orange dots $\textcolor{orange}{\bullet}$). The base kernel is given by \eref{eq_RBF}. No GPR design information is used in the interior of the domain, only the physics-informed kernel by means of BCGPs is used.}
    \label{fig_4a}
\end{figure}

As an illustration, we consider at first instance GPR reconstructions using design information only at the outer profile boundary $\curveD$. That is, the approximation formula of the velocity field is defined as in \eref{eq_3e} where $\X = \Xboundaryset$ and $\V = \Vboundaryset$ and similarly for the vorticity field estimated from \eqref{eq_3i}. The GPR reconstructions of the velocity and vorticity fields are presented in \fref{fig_4a} for different profile boundaries inside the computational domain. We observe that even if the GPR estimate is not informed with flow data inside the domain, the obtained fields satisfy the divergence-free condition and slip boundary condition on the profile. The scalar base kernel $\G :\DomG\times\DomG\to\R$ used for computing these estimates is the Gaussian anisotropic kernel defined in \eref{eq_RBF}, with only one scale for simplicity of computations. 
We recall that $\stdv$, $\lcor$ and $\alpha$ in \fref{fig_4a} are the kernel hyperparameters standing for the standard deviation, correlation length and anisotropy, respectively. In the next sections, we reconstruct the velocity field for different scenarios by using design observations issued from high-fidelity simulations of fluid flows.

\subsection{Flow around a cylinder profile} \label{sec:cylinder}

\begin{figure}
    \centering
    \includegraphics[width = 0.6\textwidth]{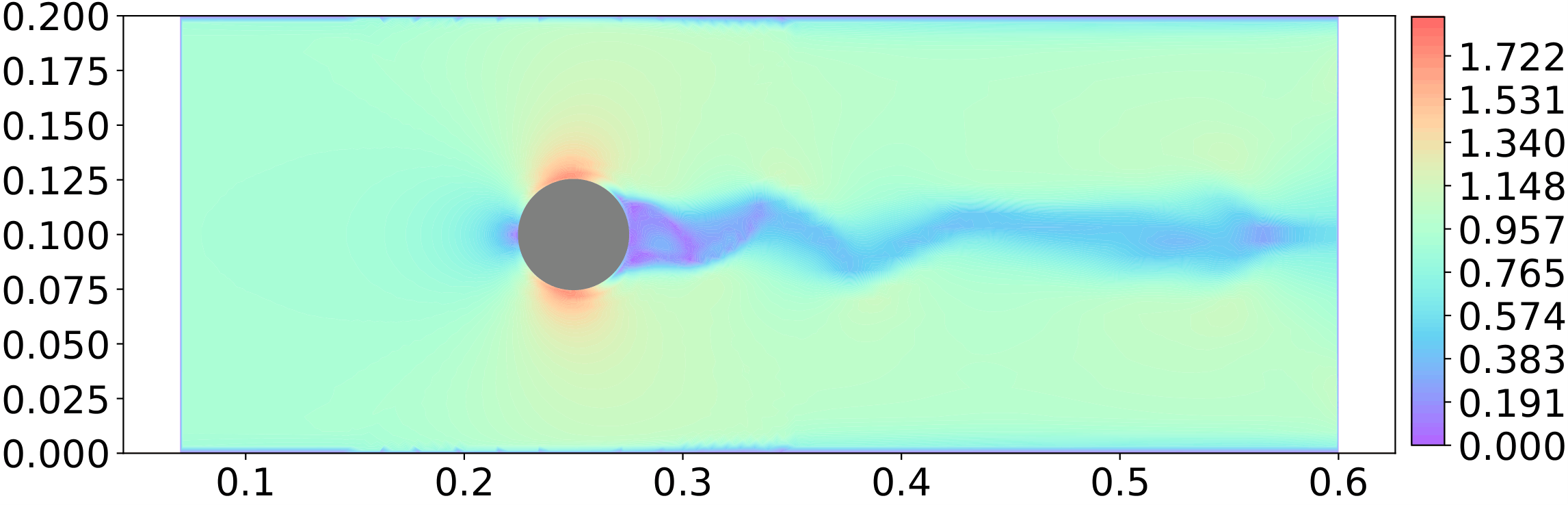}
    \caption{Ground truth velocity field of an incompressible fluid flow around a cylinder profile in a wind-tunnel setting. The colormap represents the norm of the velocity field in m/s.}
    \label{fig_4c}
\end{figure}

In this scenario, the finite volume solver OpenFOAM \cite{FOAMoriginal1998} is used for generating ground-truth data of an incompressible fluid flow around a cylinder profile at Reynolds number $\Reynolds=3\cdot 10^3$ in a wind-tunnel setting. The computational domain for interpolation is set as $\DomG = [0.07,0.6]\times [0,0.2]$, and the cylinder of radius $r=0.025$ is centered at ${\boldsymbol c}=(0.25,0.1)$. The boundary conditions are defined as:
\begin{itemize}
    \item no-slip conditions on the top and bottom walls: $\uwall(\x) = \vzero$ for $\x\in\partial\DomG_\text{w}$,
    \item slip conditions on the cylinder boundary: $\velocity (\x)\cdot\normalD(\x) = 0$ for $\x\in\Dcylinder$.
\end{itemize}
A snapshot of the ground-truth velocity field is shown in \fref{fig_4c}. The GPR reconstruction results are presented at this iteration.

Following \sref{sec:PIK}, the fluid flow in this configuration is modelled using GPR with a physics-informed kernel. We choose a base kernel with the same structure as in \eref{eq_RBF}, \emph{i.e.} an anisotropic Gaussian kernel, and set its hyperparameters as $\stdv = 0.04$, $\lcor= 0.045$ and $\alpha = 1.5$.
The nugget regularization parameter for ridge regression is set as $\nugget = 1\cdot 10^{-6}$ to ensure Gram matrix inversion. These values are fixed for the purpose of visualization. Since our main interest is to evaluate the normal fit indicator $\epsNormal$ from \eref{eq_normal_indicator}, we present a sensibility analysis of this indicator with respect to the hyperparameters $\stdv$ and $\lcor$ at the end of the section to support our findings.

\subsubsection{Continuous enforcement of the boundary conditions on the profile}

\begin{figure}[h]
    \centering
    \hspace{-1cm}
    \begin{subfigure}{0.6\linewidth}
        \centering
        \includegraphics[height=3.5cm]{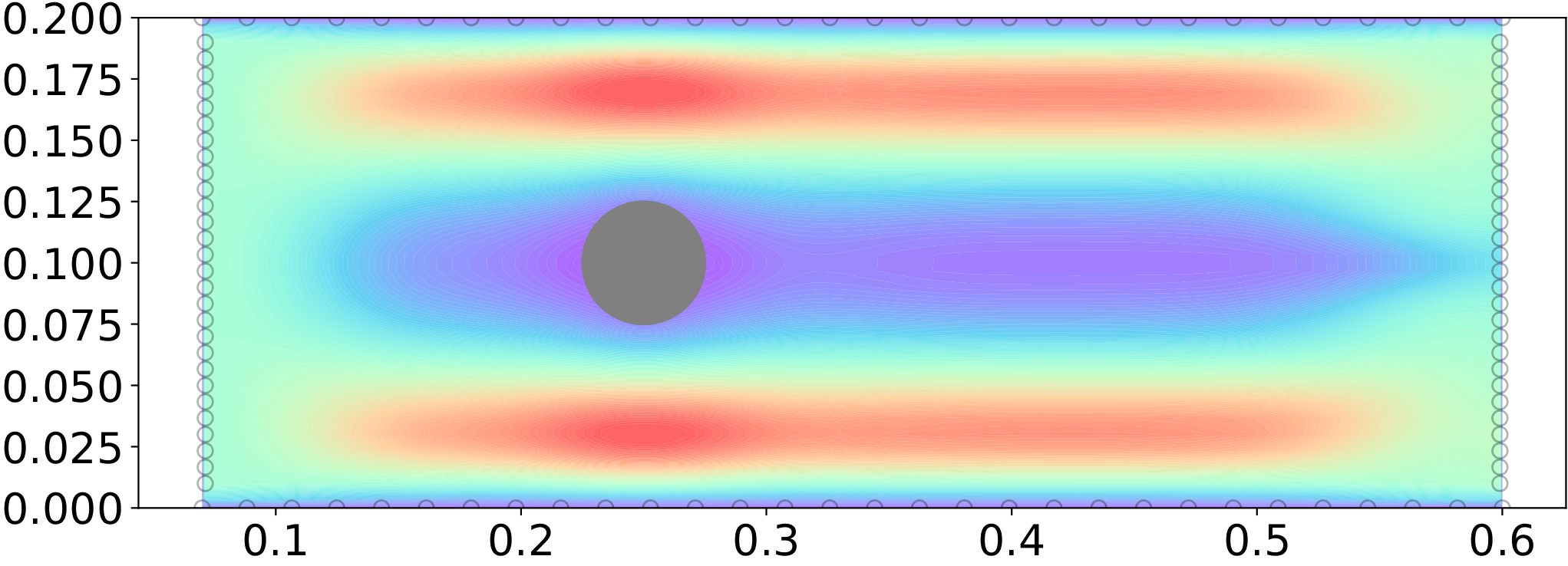}
        \vspace{0.3cm}
        \caption{Computational domain $\DomG$}
    \end{subfigure}\quad
    \begin{subfigure}{0.35\linewidth}
        \includegraphics[height=4cm]{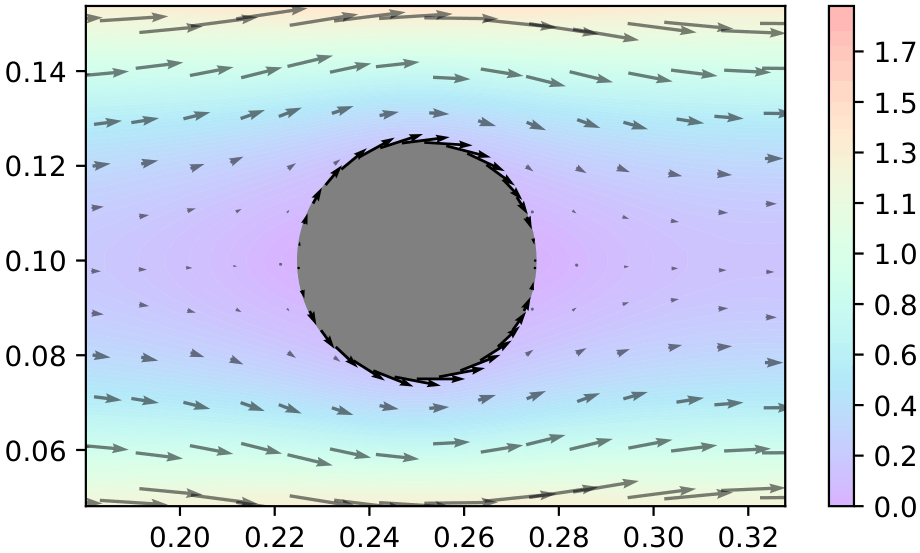}
        \caption{Region near the profile boundary. Arrows represent the velocity field.}
    \end{subfigure}
    \caption{Initial reconstruction by physics-informed GPR of the velocity field of an incompressible fluid flow ($\Reynolds = 3\cdot 10^3$) around a cylinder profile in a wind-tunnel setting using observations only at the outer boundary $\partial\DomG$ of the domain ($\Nboundary = 116$, black circles $\circ$). No design point is used on the profile boundary or in the interior of the computational domain $\DomG$. The colormap represents the norm of the velocity field in m/s.}
    \label{fig_4e}
\end{figure}

\begin{figure}[h]
    \centering
    \begin{subfigure}{0.49\textwidth}
        \centering
        \includegraphics[height=2.7cm]{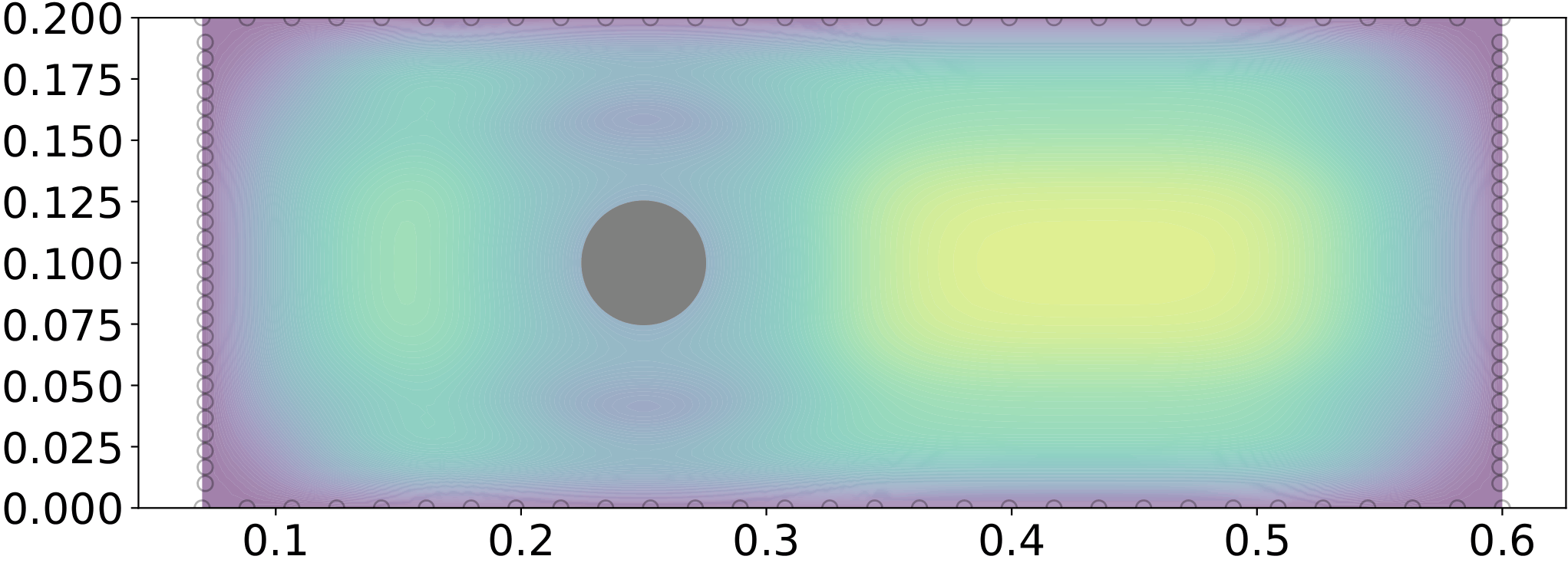}
        \subcaption{BCGP scalar kernel $\Gobstacle$}
    \end{subfigure}
    \begin{subfigure}{0.49\textwidth}
        \centering
        \includegraphics[height=2.7cm]{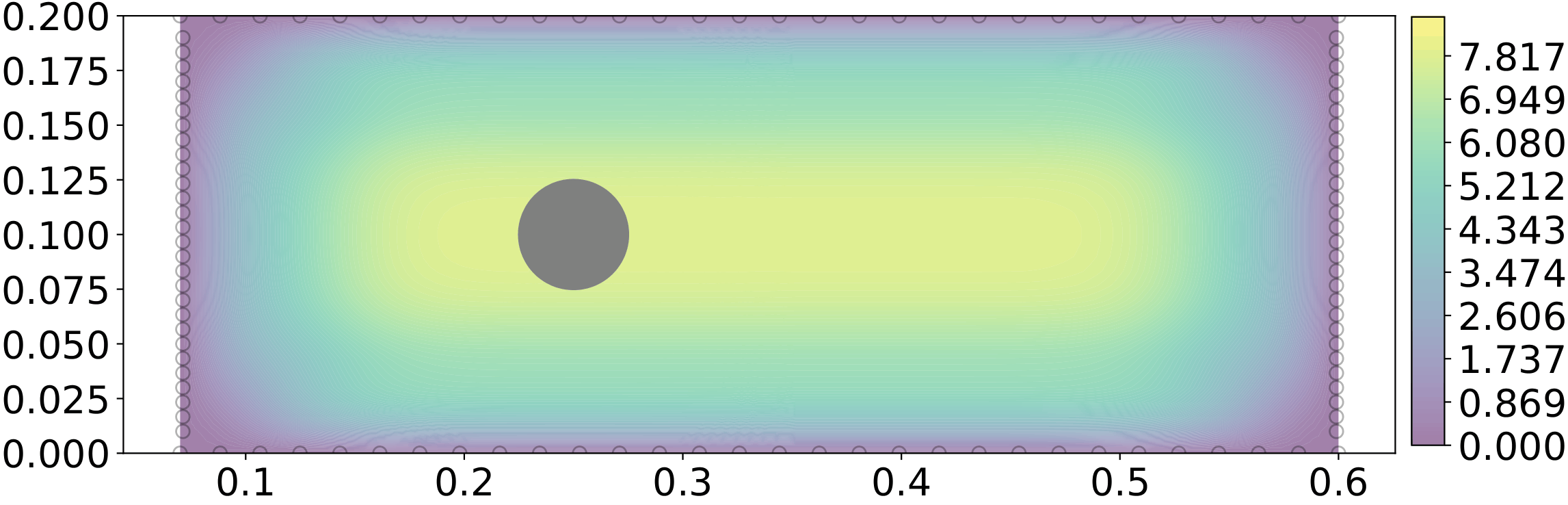}
        \subcaption{Base scalar kernel $\G$}
    \end{subfigure}
    \caption{Total standard deviation of the velocity field reconstructed by GPR with or without BCGP kernels. Design observations of the velocity are considered only at the outer boundary $\partial\DomG$ of the computational domain $\DomG$ ($\Nboundary = 116$, black circles $\circ$). (a) Estimate obtained by using the BCGP kernel $\Gobstacle$ from \eref{eq_scalar_BCGP_kernel} with the base kernel $\G$ from \eref{eq_RBF}; (b) estimate obtained by using the base kernel $\G$ from \eref{eq_RBF} solely, under the same observation design. The colormap represents the total standard deviation in m/s.}
    \label{fig_4f}
\end{figure}

We first present the reconstruction of $\velocity$ by means of the physics-informed scalar kernel $\Gobstacle$ using only design points for the inlet and outlet boundaries $\uinlet$ and $\uoutlet$, and for the boundary condition $\uwall=\vzero$ on the tunnel walls. That is, we do not consider observation points in the interior of the domain in this reconstruction. The interpolated velocity field is presented in \fref{fig_4e}. It can be observed that even though the information for flow reconstruction is insufficient at the interior of the computational domain compared to the ground truth, the slip boundary condition is already satisfied continuously along the profile boundary $\Dcylinder$.
Moreover, in \fref{fig_4f} the computation of the total standard deviation of the velocity field (defined as the square root of the trace of the matrix-valued posterior covariance of the form \eqref{eq_2p}) for each approach is presented. We observe that constraining the base GP with the proposed BCGP-based approach around the profile boundary reduces uncertainty over this area, whereas the uncertainty is higher and uniform in the interior of $\DomG$ if only the unconstrained base kernel $\G$ is used, since no physical information about the boundary condition is given.

\begin{figure}[h]
    \centering
    \includegraphics[width = 0.6\textwidth]{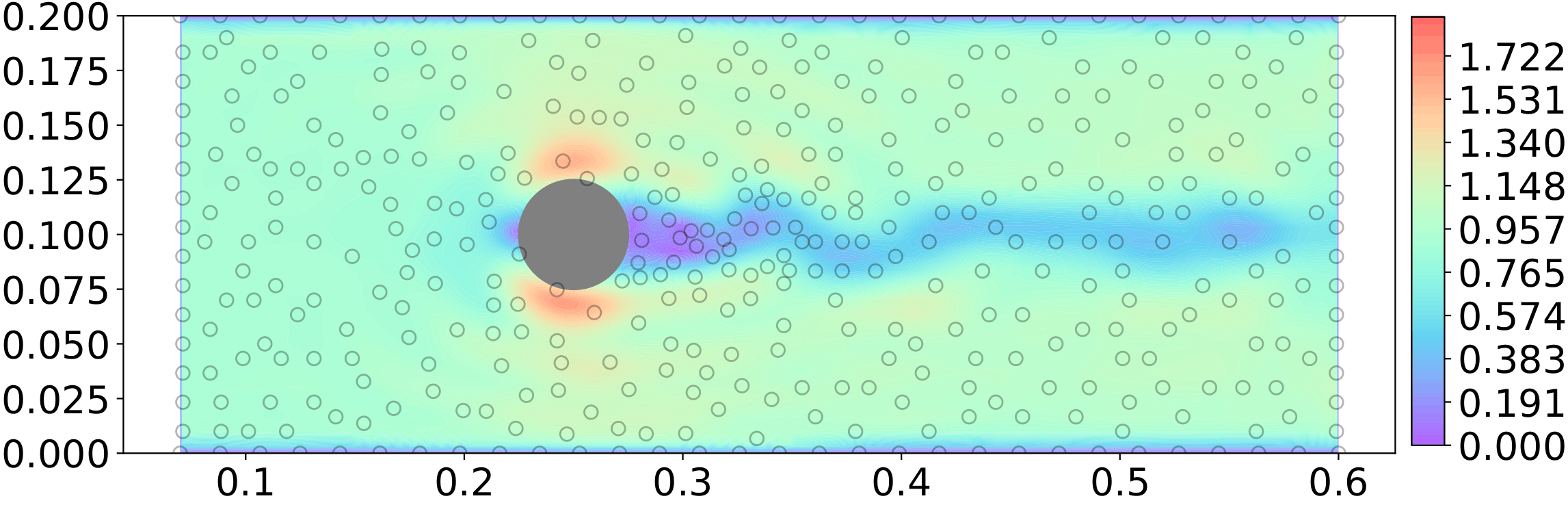}
    \caption{Reconstruction by physics-informed GPR of the velocity field of an incompressible fluid flow ($\Reynolds = 3\cdot 10^3$) around a cylinder profile in a wind-tunnel setting using $N = 415$ observations in the domain $\DomG$ (black circles $\circ$). No design point is used on the profile boundary. The colormap represents the norm of the velocity field in m/s.
    }
    \label{fig_4g}
\end{figure}

Next, we present the GPR reconstruction of the velocity field using our physics-informed framework by adding observation points inside the computational domain. That is, we consider the reconstruction formula $\eqref{eq_3e_reg}$ with $\Yset = \Xboundaryset \cup \Xdomset $ and 
$\Uset(t) = \smash{(\Vboundaryset(t)^\itr,\Vdomset(t)^\itr)}^\itr$ as in \sref{sec:GPR-Design}. A total of $N = 415$ observation points of the velocity was considered in the computational domain $\DomG$, of which $\Nboundary = 88$ and $\Ndom = 327$ observations belong to $\partial\DomG$ and $\DomG$, respectively. As before, no design observation on the profile boundary $\Dcylinder$ is included. In order to compute the spectral expansion for the definition of $\smash{\Gobsdist{\Nint,\Neigen}}$ from \eref{eq_discrete_kernel} and \eqref{eq_discrete_kernel_deriv}, a uniform regular discretization of $\gammadom = [0,2\pi]$ in $ \Nint = 400$ intervals and the largest $\Neigen + 1 = 19$ eigenvalues of the expansion were used (see below). The interpolation of the velocity field is presented in \fref{fig_4g}.

\begin{figure}[h]
    \centering
    \includegraphics[width=0.5\linewidth]{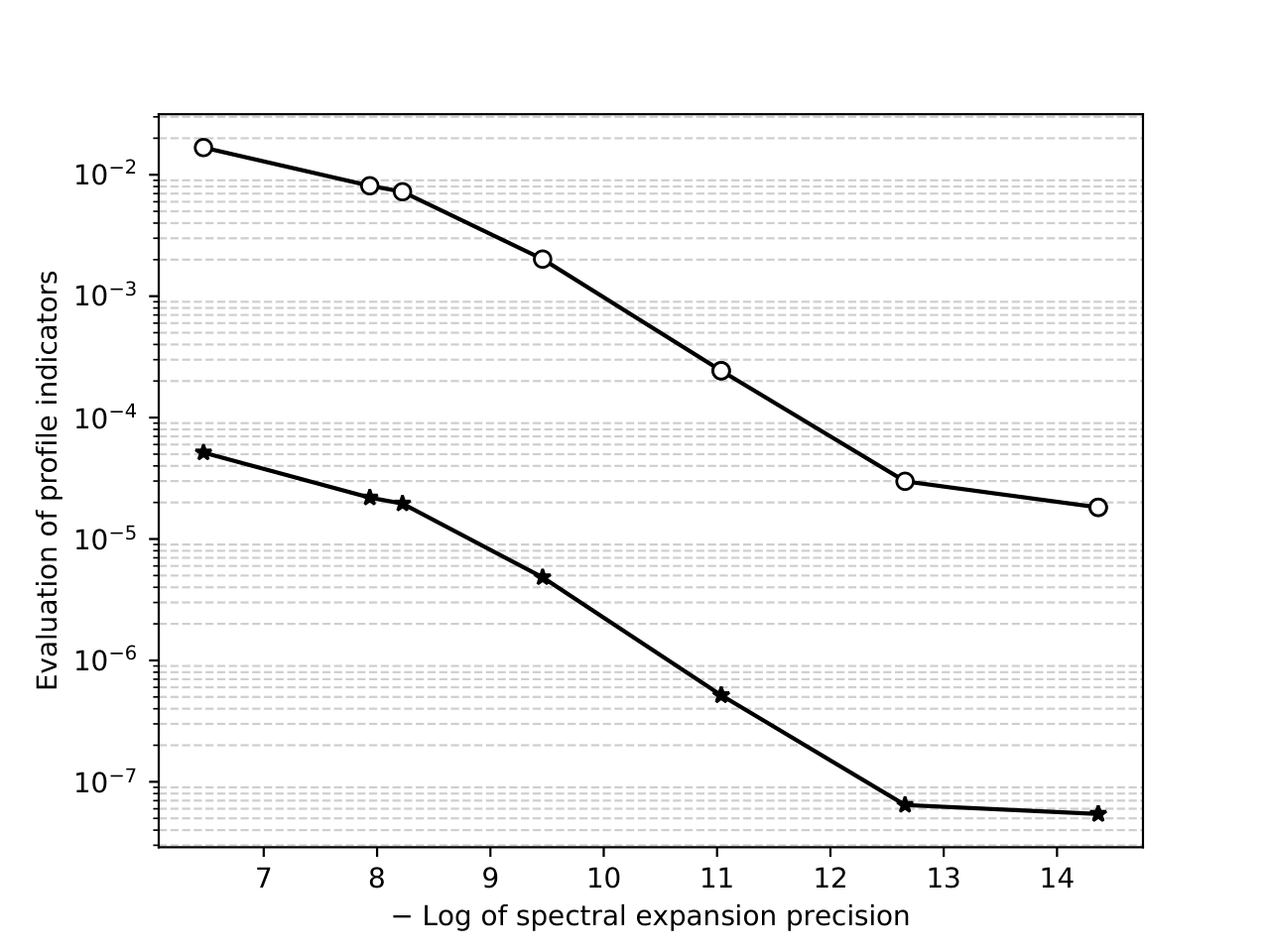}
    \caption{Evaluation of indicators of physical constraints around profile boundary with respect to spectral expansion convergence. The stream normal indicator of \eref{eq_stream_indicator} is shown with stars ($\star$) and the velocity normal indicator $\epsNormal$ of \eref{eq_normal_indicator} is shown with circles ($\circ$). Both of these functions were computed by fixing the GPR design framework with the same $N = 415$ observations of velocity inside $\DomG$ and kernel hyperparameters. The horizontal axis is the negative logarithm of $\epsSpec$ from \eref{eq_epsSpec} when varying the spectral bound $\deltaSpec$ in Algorithm \ref{algo_BCGP_derivatives}.}  \label{fig_spectral_convergence}
\end{figure}

As explained in \sref{sec:Num}, the computation of the BCGP kernel $\Gobstacle$ is established from the approximation of a spectral series. Thus, the satisfaction of the slip boundary condition \eqref{eq_3c} by the posterior estimate is achieved up to the numerical approximation of $\Gobstacle$ and its derivatives. Therefore, we present in \fref{fig_spectral_convergence} a convergence study of the profile boundary spectral indicators defined in \eref{eq_stream_indicator} and \eref{eq_normal_indicator}. The computation of the BCGP kernel $\Gobstacle$ is done with the choice of measure $\measureD$ as in \cref{Coro_1}. Notice that in this framework, the choice of the surface measure $\measureD$ from \cref{Coro_2} is equivalent to the uniform measure from \cref{Coro_1} up to a constant.
It is observed that a spectral accuracy bound $\deltaSpec = 10^{-12}$ (obtained for $\Nint = 400$ and at least $\Neigen = 18$) is needed to achieve $\epsNormal < 10^{-4}$ with the prescribed profile geometry and base kernel definition. In this case, the $L^1$ norm \eqref{eq_stream_indicator} along $\Dcylinder$ of the posterior estimate $\vstreamapprox$ of the stream function is of the order of $10^{-7}$.

\subsubsection{Discrete enforcement of the boundary conditions on the profile} \label{sec:discrete_kernel}

Alternatively to the reconstruction formulas in \sref{sec:GPR-Design} using a BCGP, a different approach for enforcing the slip boundary condition of \eref{eq_3c} can be considered in a discrete setting as follows. Let $\Vboundaryset$ and $\Vdomset$ be the observed velocity values at positions $\Xboundaryset$ and $\Xdomset$, corresponding to the outer domain boundary and inner domain observations, respectively. In addition, consider a discretization of the profile boundary $\Xobstacleset = \smash{\{\Xobstacle{i}\}_{i=1}^\Nobstacle}$. Then, by modelling the velocity stream function $\vstream$ directly from the base GP $\ZGP\sim\GP(0,\G)$, we can consider the following approach for the reconstruction of the velocity field:
\begin{equation}\label{eq_discrete_reconstruction}
    \uapprox(\x,t) = \esp \braces{\,\velocityGP(\x)\,|\, \velocityGP(\X) = \V(t),\ \velocityGP(\Xobstacleset)\cdot\normalD(\Xobstacleset) = 0\, }\,,
\end{equation}
where $\x$ is any unobserved point in $\DomG$. In this approach, the boundary constraint of \eref{eq_3c} is enforced in a discrete way in the posterior GP distribution.

\begin{figure}[h!]
    \begin{subfigure}{0.48\textwidth}
        \centering
        \includegraphics[height=2.7cm]{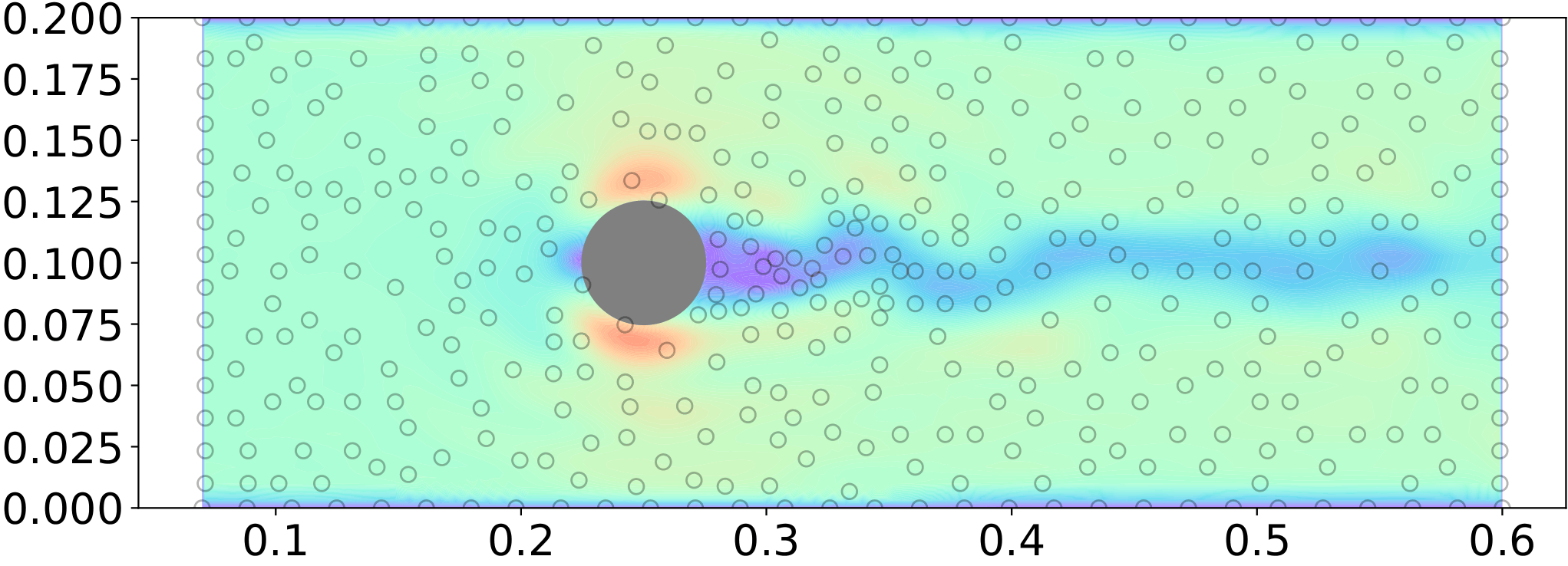}
        \subcaption{Velocity field estimate using BCGP kernel $\Gobstacle$.}
    \end{subfigure}
    \begin{subfigure}{0.48\textwidth}
        \centering
        \includegraphics[height=2.7cm]{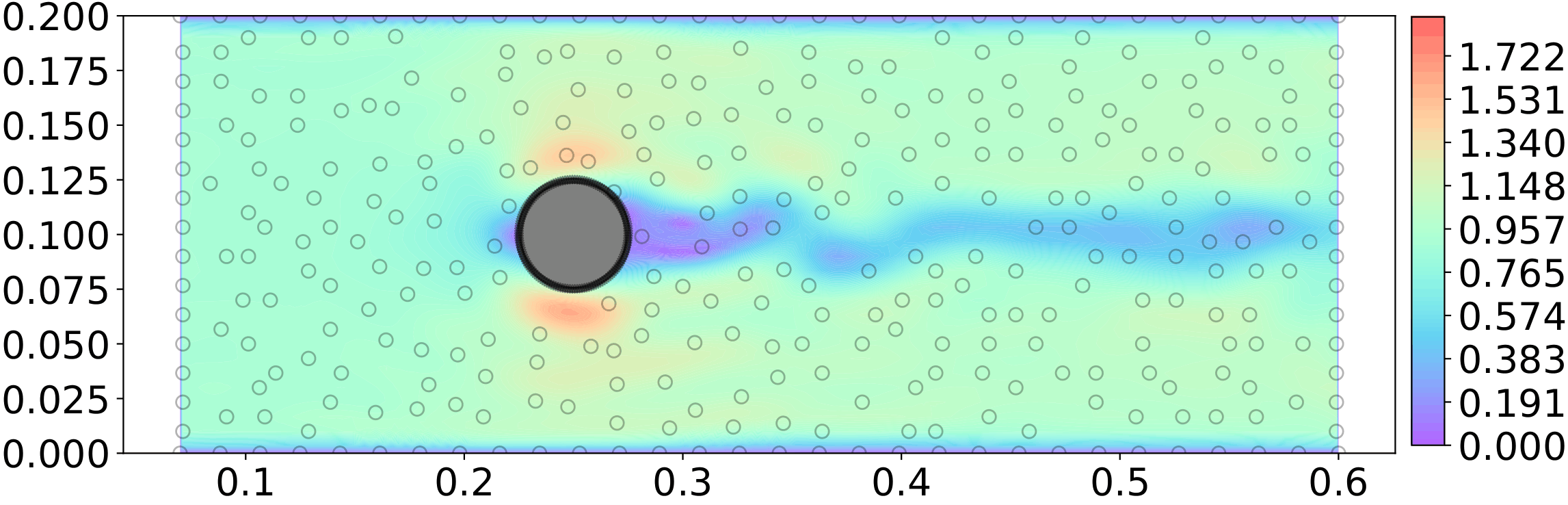}
        \subcaption{Velocity field estimate using base kernel $\G$.}
    \end{subfigure}\\

    \begin{subfigure}{0.48\textwidth}
        \centering
        \includegraphics[height=2.7cm]{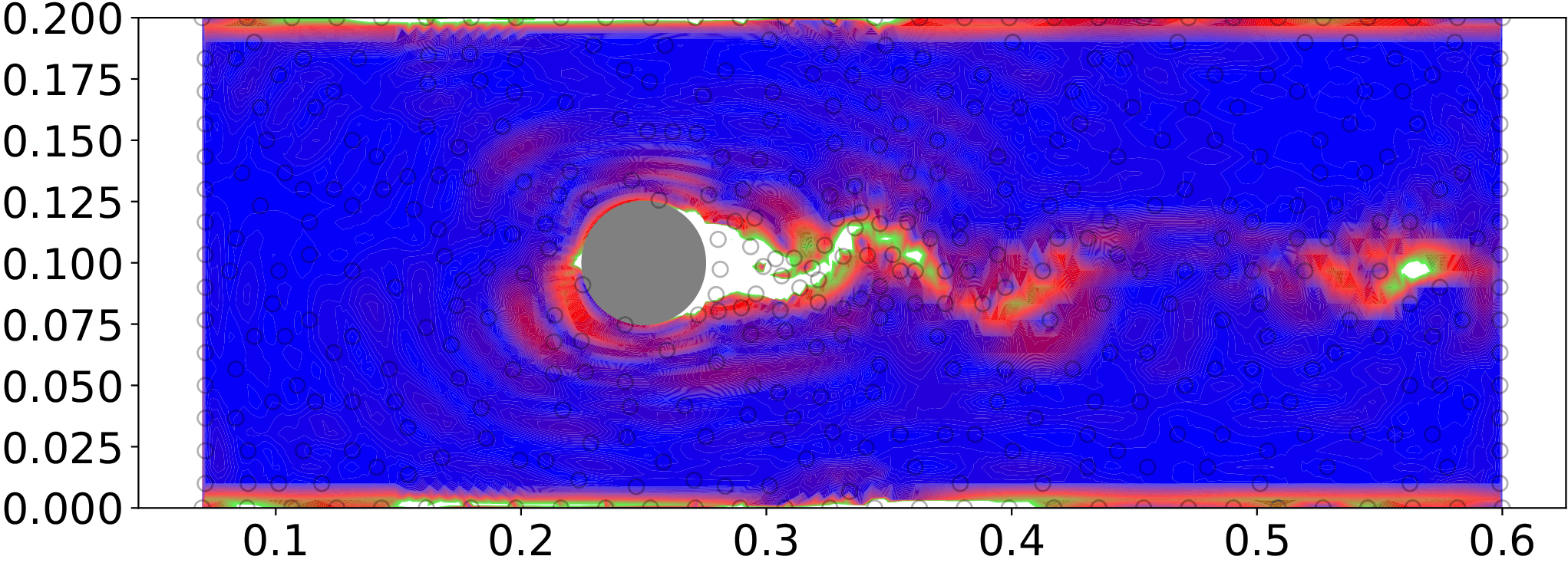}
        \subcaption{Relative error field using BCGP kernel $\Gobstacle$.}
    \end{subfigure}
    \begin{subfigure}{0.48\textwidth}
        \centering
        \includegraphics[height=2.7cm]{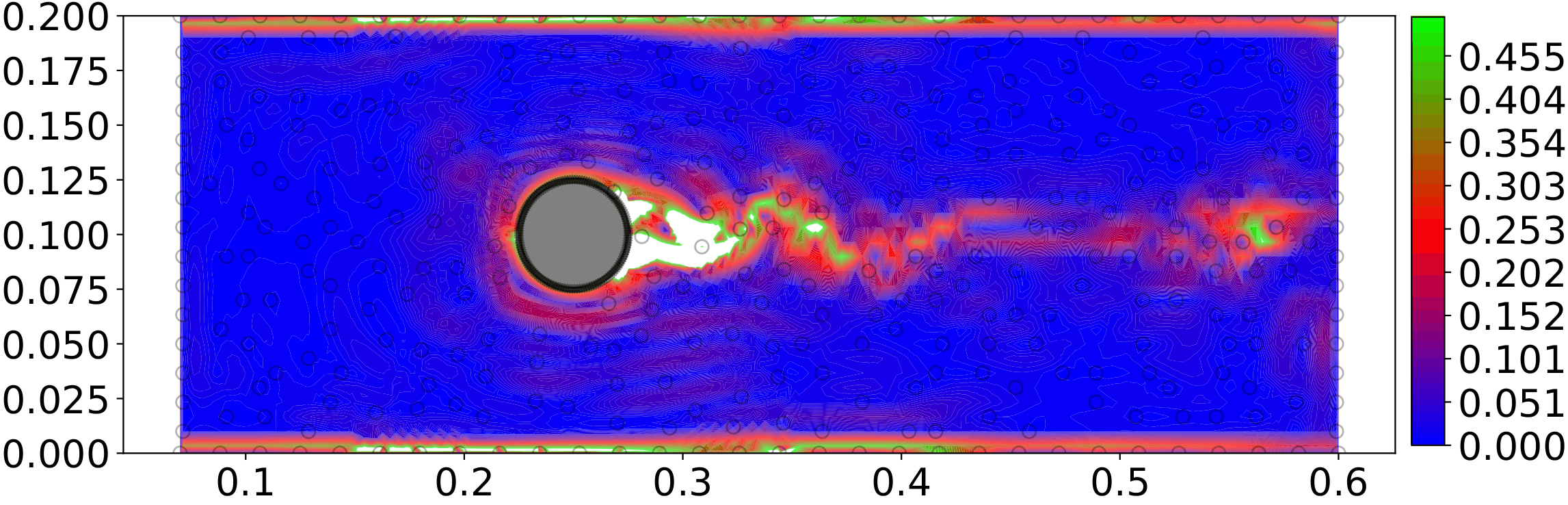}
        \subcaption{Relative error field using base kernel $\G$.}
    \end{subfigure}
    \caption{Velocity and relative error fields computed from physics-informed GPR estimates of an incompressible flow around a cylinder profile. The GPR estimates are obtained with a BCGP kernel $\Gobstacle$ from \eref{eq_scalar_BCGP_kernel} with the base kernel $\G$ from \eref{eq_RBF} in (a) and (c), and with the base kernel $\G$ from \eref{eq_RBF} solely in (b) and (d) for comparison. The circles ($\circ$) are the GPR observation points for the velocity inside the computational domain $\DomG$, whereas the stars ($\star$) are the observation points of the boundary condition on the profile. The colormap in (a) and (b) represents the norm of the velocity field in m/s, whereas the colormap in (c) and (d) represent the norm of the relative error field.} \label{fig_cylinder_comparison_interpolation}
\end{figure}

We now compare the BCGP-based approach and the discrete approach, while keeping a constant budget of observations for both cases. To do this, we fix this budget in terms of the size of the Gram matrix used for interpolation, say $\Ngram = 830$. Then, we give specific configurations of the design observations to each approach. For the BCGP-based method, we consider the same configuration of $N = 415$ observations of the velocity as in \fref{fig_4g}. In the case of the discrete approach, we consider $N = 344$ observations of the velocity inside the domain $\DomG$ and its outer boundary, and we add $\Nobstacle = 142$ observation points for the consideration of the slip boundary condition in \eref{eq_discrete_reconstruction}. The velocity field estimates from physics-informed GPR with $\Gobstacle$ and with $\G$ using discrete boundary constraints are presented in \fref{fig_cylinder_comparison_interpolation}. The relative error fields corresponding to each estimate, with respect to the ground truth from \fref{fig_4c}, are also depicted. Furthermore, the configuration of the GPR design for each approach is summed up in \tref{tab_cylinder_config_and_fit}, including the relative normal fit indicator $\epsNormal$ of \eref{eq_normal_indicator}. The reported CPU time for the offline computation of the BCGP spectral factor was 0.33 second, including precomputation of kernel derivatives. It is observed that the BCGP-based method allows to satisfy the slip boundary condition at a better order of accuracy than that of the discrete method for the same observation budget. Moreover, this budget allows to include observations of the velocity at other parts of the domain instead of adding discrete points on the profile boundary.

\begin{figure}[h!]
    \begin{subfigure}{0.33\textwidth}
        \centering
        \includegraphics[height=4.2cm]{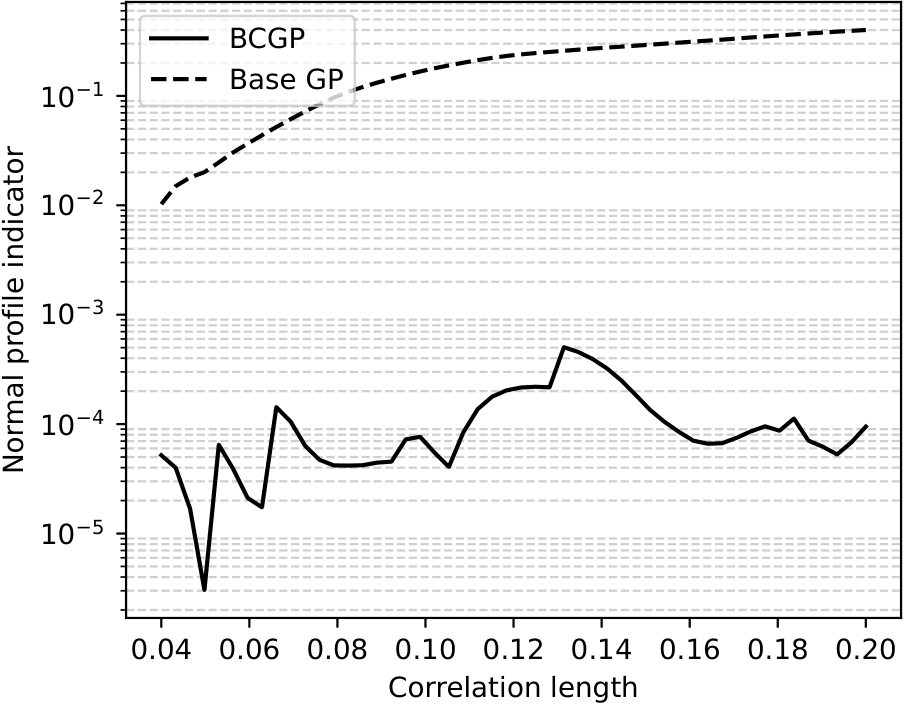}
        \subcaption{$\sigma = 0.01$}
    \end{subfigure}
    \begin{subfigure}{0.33\textwidth}
        \centering
        \includegraphics[height=4.2cm]{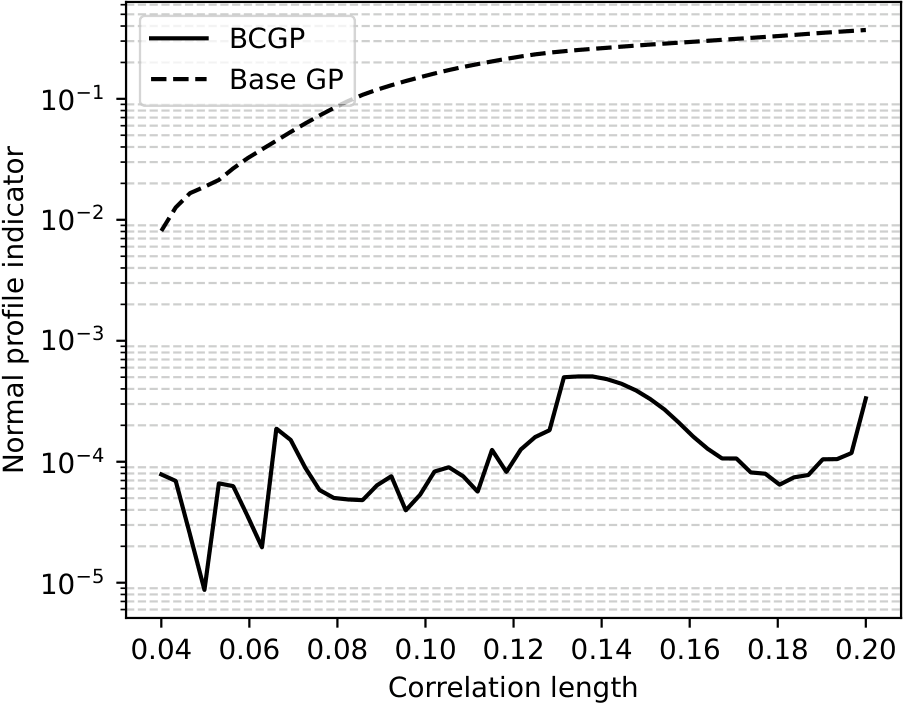}
        \subcaption{$\sigma = 0.05$}
    \end{subfigure}
    \begin{subfigure}{0.33\textwidth}
        \centering
        \includegraphics[height=4.2cm]{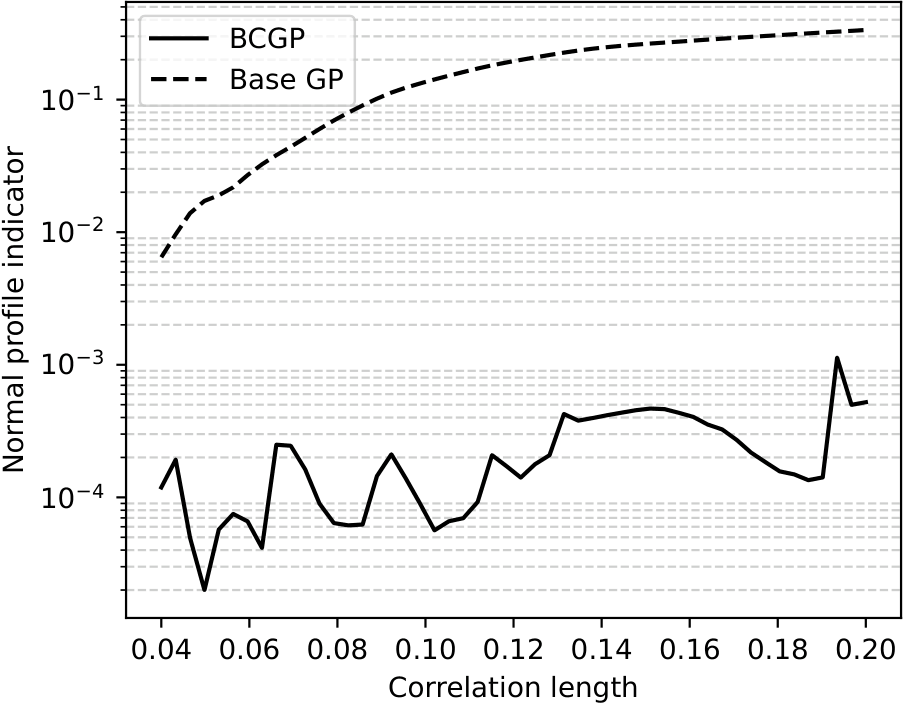}
        \subcaption{$\sigma = 0.5$}
    \end{subfigure}
    \caption{Sensitivity analysis of the normal fit indicator $\epsNormal$ from \eref{eq_normal_indicator} with respect to kernel hyperparameters of standard deviation $\sigma$ and correlation length $\lcor$. The anisotropic parameter $\alpha = 1.5$ is considered as a reference. For each value of $\stdv$, the evaluation of $\epsNormal$ vs. $\lcor$ is presented respectively from (a) to (c). We observe that the approach using the BCGP kernel $\Gobstacle$ from \eref{eq_scalar_BCGP_kernel} with the base kernel $\G$ from \eref{eq_RBF} (continuous line) is competitive with respect to discrete constraining of the slip boundary condition with the base kernel $\G$ from \eref{eq_RBF} solely (dashed line). } \label{fig_cylinder_comparison_params}
\end{figure}

Lastly, \fref{fig_cylinder_comparison_params} presents a sensitivity analysis with respect to the standard deviation $\sigma$ and correlation length $\lcor$ parameters. The lower bound $\lcor \geq 0.04$ was established to avoid instability of the velocity field reconstruction on the whole domain due to overfitting. The improvement in $\epsNormal$ using the BCGP kernel, in comparison to the discrete approach, remains valid for the tested configurations.

\begin{table}[h!]
    \centering
    \begin{tabular}{
        |>{\centering\arraybackslash}m{4cm}
        |>{\centering\arraybackslash}m{1.7cm}
        |>{\centering\arraybackslash}m{1.7cm}
        |>{\centering\arraybackslash}m{1.7cm}
        |>{\centering\arraybackslash}m{2cm}
        |>{\centering\arraybackslash}m{2.3cm}|}
    \hline
    GPR approach & $N$ & $\Nobstacle$ & $\Ngram$ & $\epsNormal$ & Posterior mean CPU time \\
    \hline
    RBF kernel $\G$ & 344 & 142 & 830 & $1.611\cdot 10^{-2}$ & 2.05 s \\
    \hline
    BCGP kernel $\Gobstacle$ & 415 & 0 & 830 & $1.823\cdot 10^{-5}$ & 2.86 s \\
    \hline
    \end{tabular}
    \caption{GPR design configurations and relative normal fit indicator $\epsNormal$ of \eref{eq_normal_indicator} for the GPR estimate of the velocity field around a cylinder profile using two approaches for the consideration of the slip boundary condition on the profile. The estimation using the scalar base kernel $\G$ is obtained with discrete boundary constraints.}
    \label{tab_cylinder_config_and_fit}
\end{table}

\subsection{Flow around a NACA 0412 airfoil} \label{sec:NACA}

\begin{figure}[h!]
    \begin{subfigure}{0.35\textwidth}
        \centering
        \includegraphics[height=4.1cm]{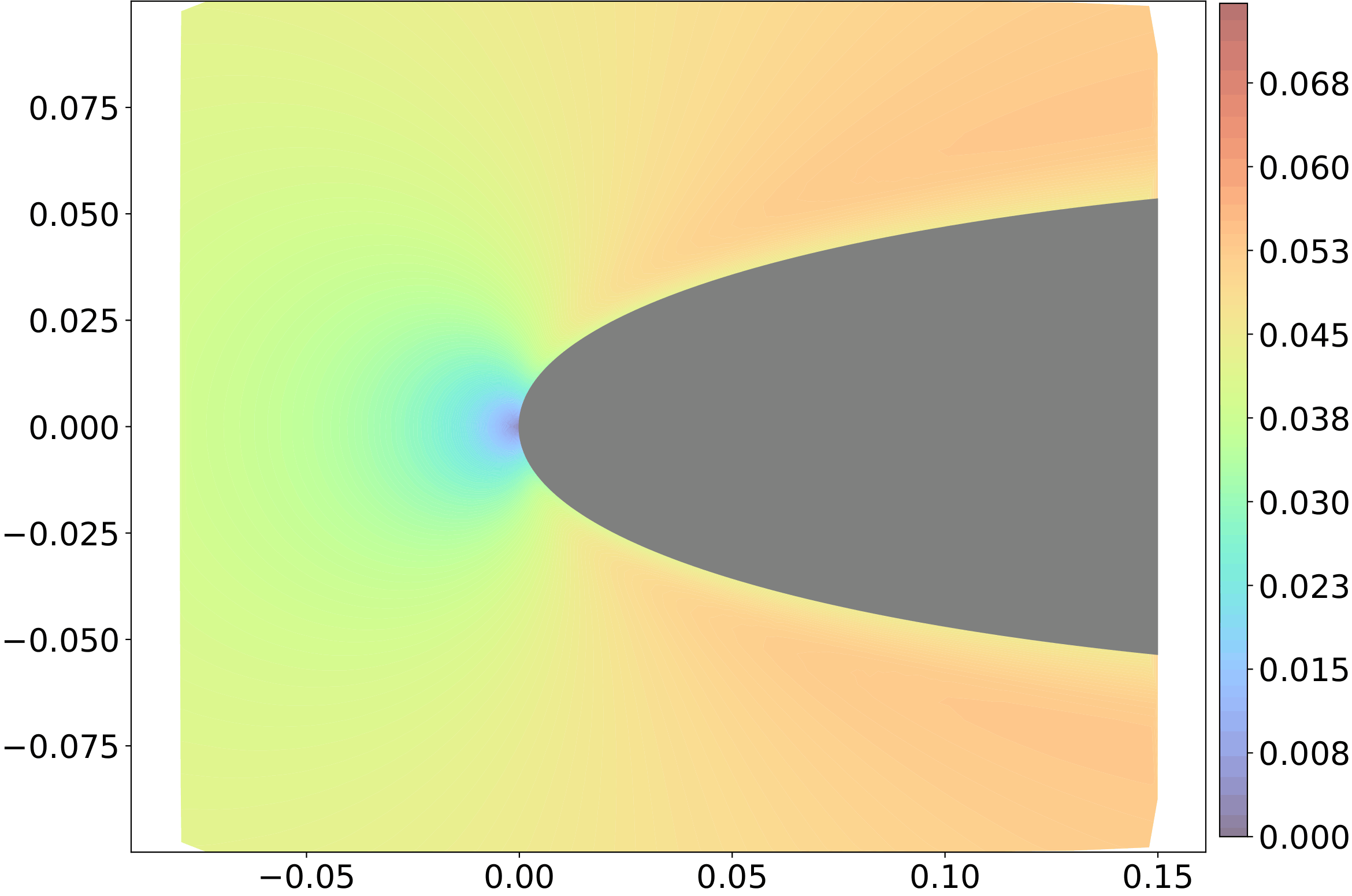}
        \subcaption{Leading edge}
    \end{subfigure}
    \begin{subfigure}{0.6\textwidth}
        \centering
        \includegraphics[height=4cm]{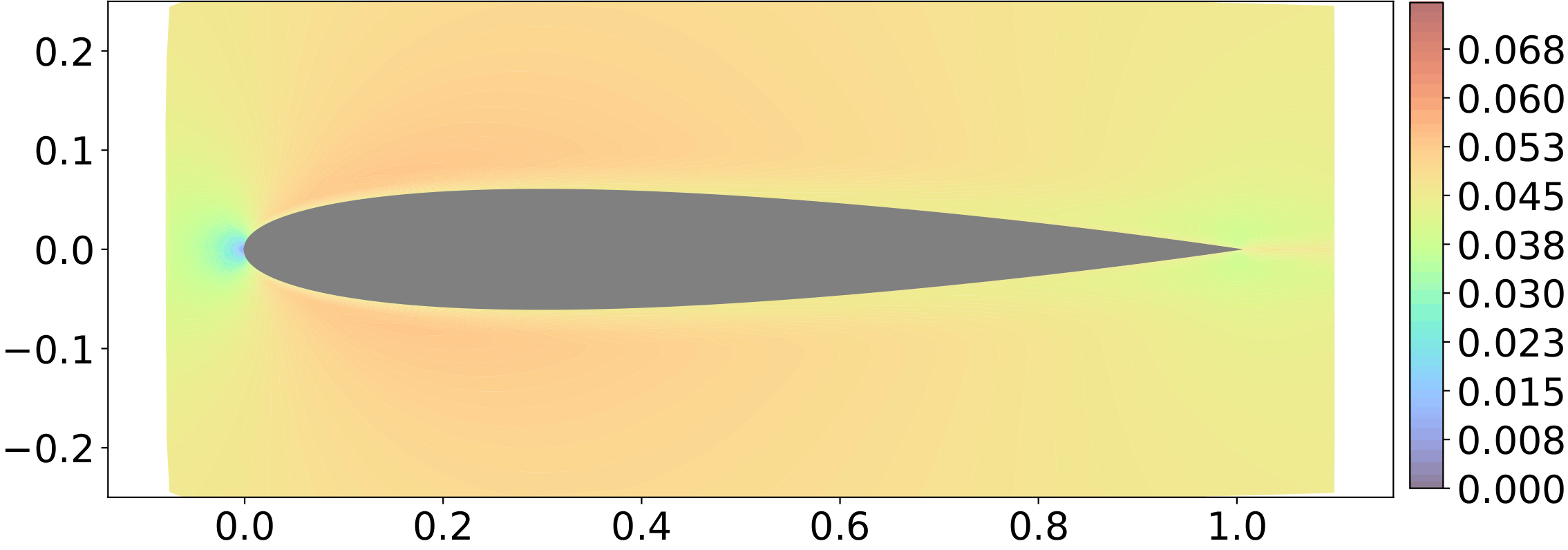}
        \subcaption{Complete airfoil}
    \end{subfigure}
    \caption{Stationary ground truth velocity field $\smash{\ugt}$ of an incompressible fluid flow around the NACA 0412 airfoil. The colormap represents the norm of the velocity field in m/s.}
    \label{fig_NACA_truth}
\end{figure}

We now turn to the GPR reconstruction of the velocity field of an incompressible fluid flow about a NACA 0412 airfoil at Reynolds number $\Reynolds=2.7\cdot 10^3$. The chord length of the profile $\DNACA{0412}=\boldgamma(\gammadom)$ is $c = 1$, where the definition of $\boldgamma$ is given by \eref{eq_NACA_definition}. Due to the small spatial scale of the high-fidelity case, we focus our study to the leading edge of the airfoil and set the parameterization domain to $\gammadom = [2.3,4]$. A slip condition is considered on the boundary of the airfoil, \textit{i.e.} $\velocity(\x) \cdot\normalD (\x) = 0$ for $\x\in\DNACA{0412}$. The ground-truth data were generated using the AirfRANS dataset library \cite{Bonnet2023} adapting for the slip boundary condition along the profile. A snapshot of the ground truth velocity field $\smash{\ugt}$ is depicted in \fref{fig_NACA_truth}. Since we consider a reconstruction problem with high resolution, the computational domain is set as the rectangular box $\DomG = [-0.08,0.15]\times [-0.1,0.1]$ around the leading edge of the airfoil. The presented results are computed using this specific stationary field.

Our objective is to compare different choices of GP priors used in the reconstruction formula \eqref{eq_3e_reg} by progressively informing their respective kernels about the physical constraints discussed in \sref{sec:PIK}. We use the baseline RBF kernel structure introduced in \eref{eq_kernel_additive}. The parameter $M$ stands for the number of fine scales used to capture the energy decay of the fluid dynamics. The following choices are investigated:

\begin{enumerate}
    \item RBF baseline: we set the velocity GP prior as a centered GP with covariance $\K$ defined by the following matrix-valued kernel using the scalar exponential kernel $\G$ from \eref{eq_kernel_additive} with only one coarse scale $M=0$:
    \begin{equation}\label{eq_separable_kernel}
        \K(\x,\xp) = \G(\x,\xp) {\begin{pmatrix} 1 & 1/2 \\ 1/2 & 1 \end{pmatrix}}\,, \quad \x,\xp\in\DomG\, ;
    \end{equation}
    \item Divergence-free RBF (DF-RBF): velocity is modelled as the curl of a stream function GP using the one-scale base kernel (\eref{eq_kernel_additive} with $M=0$) and no prior information is given about the slip boundary condition or energy decay;
    \item Multiple-scales RBF (M-RBF): the scalar base kernel is further informed with energy decay by incorporating multiple scales, as defined by \eref{eq_kernel_additive} with $M=3$;
    \item Physics-informed BCGP (PI-RBF): the multiple-scale kernel is further informed about the profile slip boundary condition from \eref{eq_BCGP}, through the BCGP procedure, where the base kernel is defined as in \eref{eq_kernel_additive} with $M=3$. The number of integration intervals is $\Nint = 300$ and the spectral accuracy bound $\deltaSpec$ is set to $\deltaSpec=10^{-12}$, which yields 112 spectral modes (see Algorithms \ref{algo_spectral_factor} and \ref{algo_BCGP_derivatives}). The spectral measure $\measureD$ is chosen as in \cref{Coro_2}. A comparison with respect to the measure from \cref{Coro_1} is presented later in \sref{sec:UQ}.
\end{enumerate}
The hyperparameters $\sigma_0$ and $\lcor_0$ for the standard deviation and correlation length, respectively, are estimated using a cross-validation methodology as depicted below in \sref{sec:CrossValidation}. The regularization parameter is set at $\nugget=\smash{10^{-8}}$ for the RBF and DF-RBF kernels and at $\nugget=\smash{10^{-10}}$ for the M-RBF and PI-RBF kernels. These choices are done in order to ensure that the Gram matrices are invertible.

The training set $\Yset$ gathers $N = 184$ velocity measurement positions which are selected randomly. No observation positions are considered near or at the boundary of the NACA airfoil, though. The test set $\smash{\Xtestset} = \smash{\{\Xtest{k}\}_{k=1}^\Ntest}$ gathers $\Ntest = 3439$ positions inside the computational domain. These positions have a higher concentration near the profile for a better resolution of the reconstructed fields. Note that the same training and test sets are used in order to compare all four GPR reconstruction approaches described above. The measure of goodness of fit over the test set is the root mean squared error $\smash{\errRMSE}$ defined as:
\begin{equation}\label{eq:eRMSE}
\errRMSE(\Xtestset;\uapprox) = \sqrt{\frac{1}{\Ntest}\sum_{k = 1}^{\Ntest} \norm{ \uapprox(\Xtest{k}) - \ugt(\Xtest{k}) }^2  }\,,
\end{equation}
where $\smash{\ugt}$ is the velocity field ground truth, and $\smash{\uapprox}$ is the velocity field from the reconstruction formula  \eqref{eq_3e_reg} associated to a specific kernel definition.

In order to quantify the uncertainty associated to these different reconstructions, we introduce the UQ coverage indicators $\smash{p_1}$ and $\smash{p_2}$ defined as the posterior coverage percentages for the horizontal and vertical velocity components, respectively. These quantities are computed as the ratios of test points lying inside their corresponding posterior $95\%$ confidence intervals. That is, given the set $\smash{\Xtestset}$ of $\Ntest$ locations within the computational domain and a reconstruction formula $\uapprox$, the indicators are defined by $\smash{p_i(\Xtestset;\uapprox)}=\smash{N_i/\Ntest}$ with:
\begin{equation}\label{eq_UQ_coverages}
N_i = \#\left\{\x\in\Xtestset\,:\,\abs{\smash{\vj^\star_i(\x) - \vj_i^\dagger(\x)}}\leq 1.96\times\upoststd{i}(\x)\right\}\,,
\end{equation}
for $i=1,2$. Here $\upoststd{i}(\x)=\smash{\sqrt{\upostStd{ii}(\x,\x)}}$ is the posterior standard deviation of the $i$-th velocity component at $\x\in\DomG$, for $\smash{\upostStd{ii}(\x,\x)}$ being the $i$-th element of the diagonal of the matrix-valued covariance $\upostcov(\x,\x)$ from \eref{eq_posterior_cov_velocity}.

Furthermore, to evaluate the fit of the incompressibility condition on the complete domain, we define the indicator $\epsDiver$ as the normalized $L^1$ norm of the divergence field:
\begin{equation}\label{eq_divergence_indicator}
    \epsDiver = \frac{1}{\abs{\DomG}}\norm{\diver\uapprox}_{L^1(\DomG)}\,.
\end{equation}
In practice, we evaluate numerically the velocity divergence field via finite differences using the corresponding reconstruction formula.

\begin{figure}[ht!]
    \centering

    \begin{subfigure}{0.45\textwidth}
        \centering
        \includegraphics[height=5.5cm]{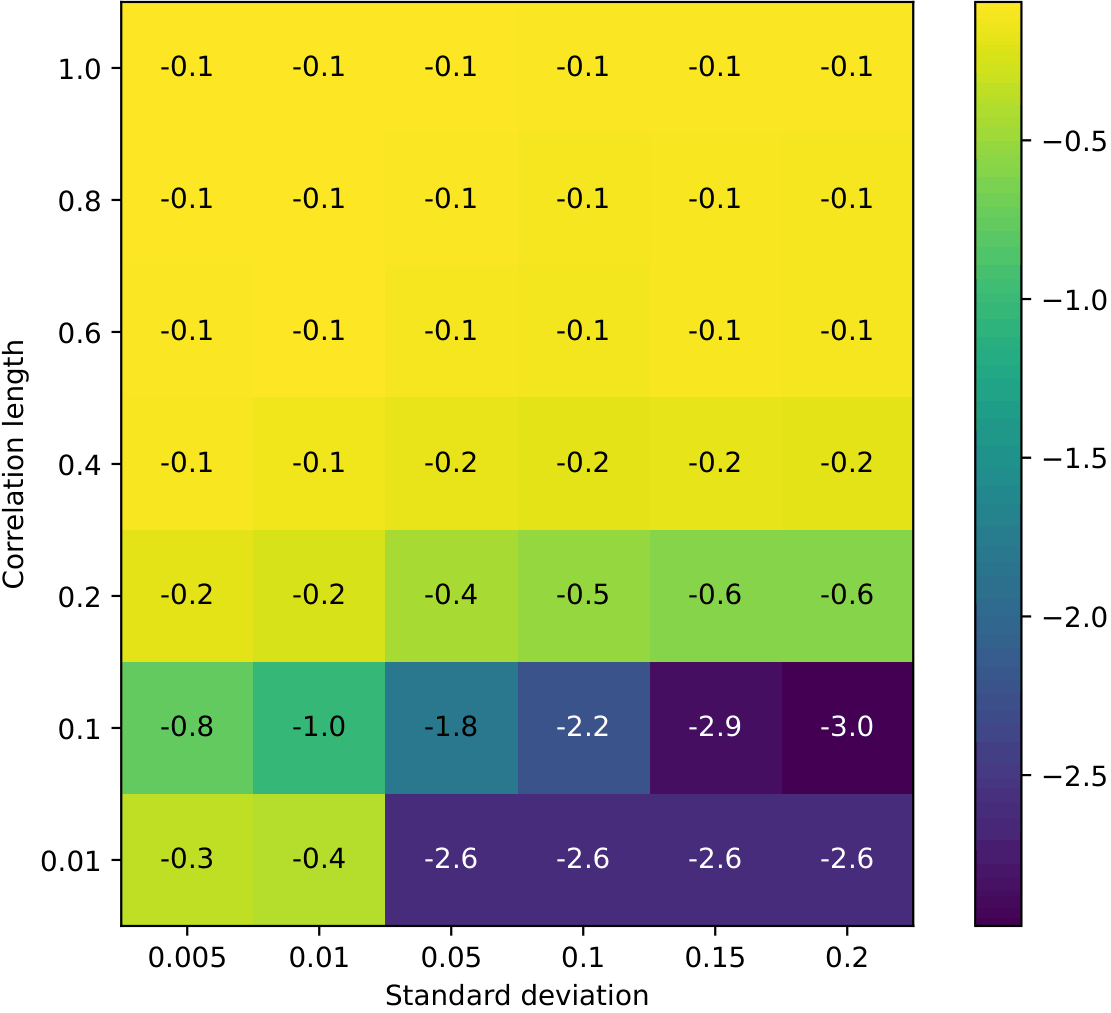}
        \subcaption{RBF kernel ($M=0$)}
    \end{subfigure}
    \begin{subfigure}{0.45\textwidth}
        \centering
        \includegraphics[height=5.5cm]{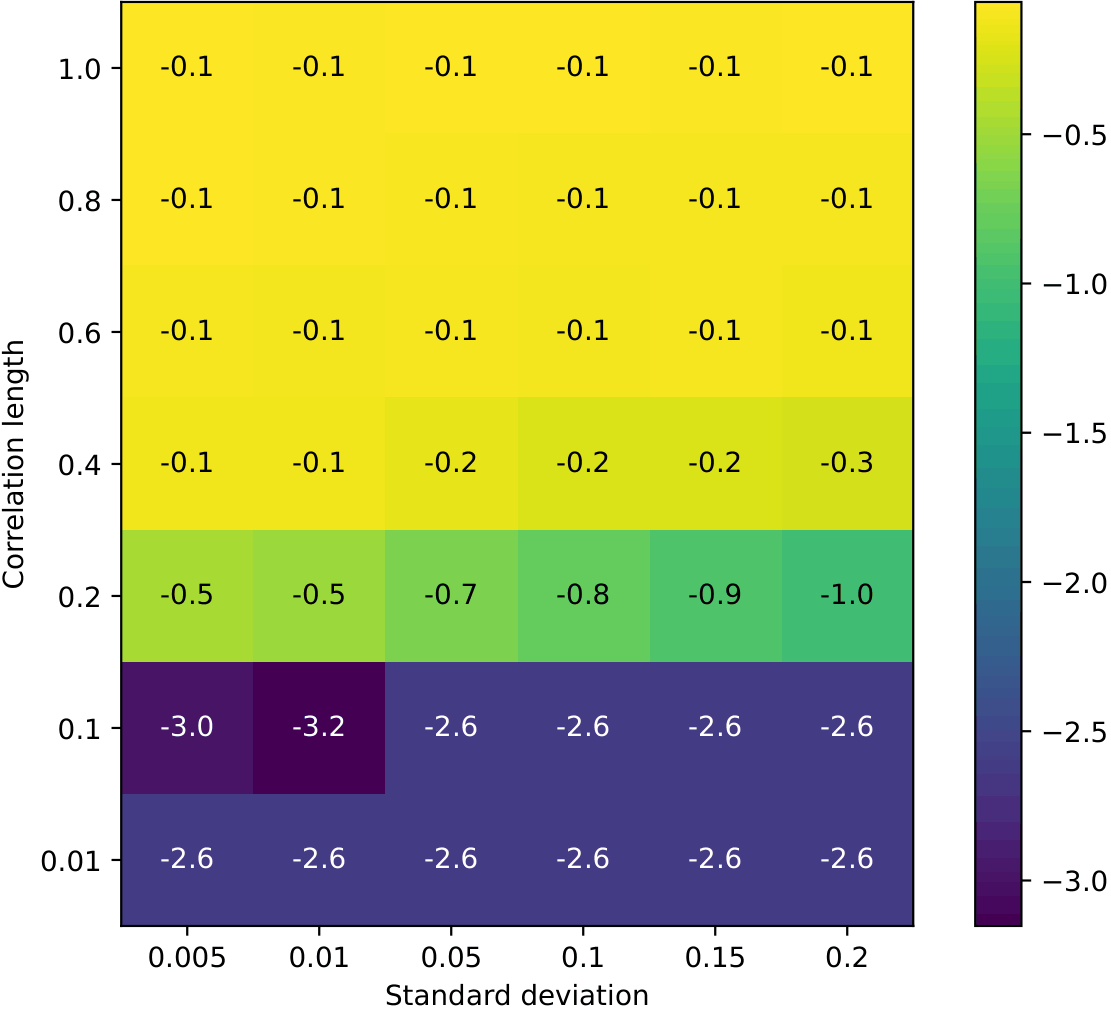}
        \subcaption{DF-RBF kernel ($M=0$)}
    \end{subfigure}\vspace{0.2cm}

    \begin{subfigure}{0.45\textwidth}
        \centering
        \includegraphics[height=5.5cm]{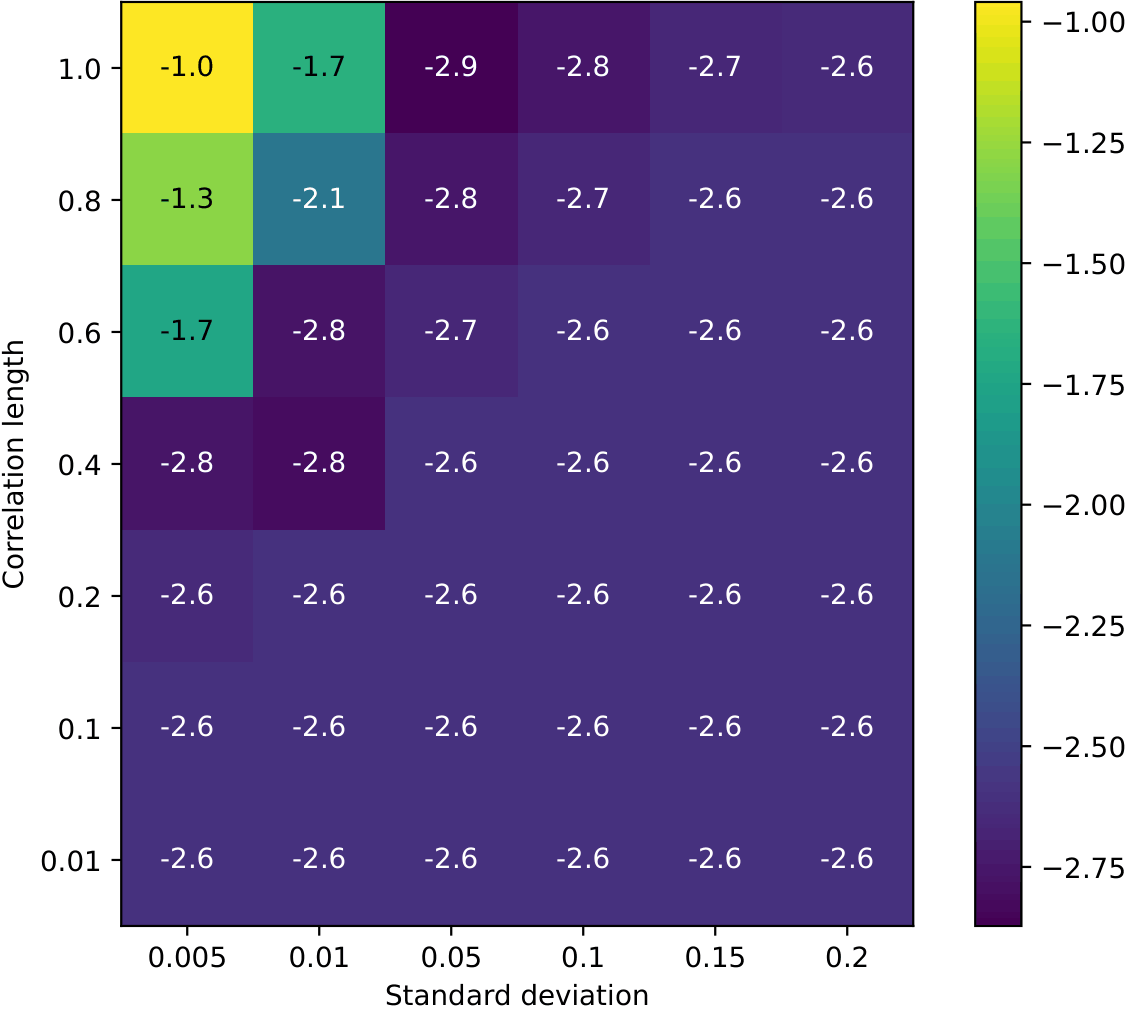}
        \subcaption{M-RBF kernel ($M=3$)}
    \end{subfigure}
    \begin{subfigure}{0.45\textwidth}
        \centering
        \includegraphics[height=5.5cm]{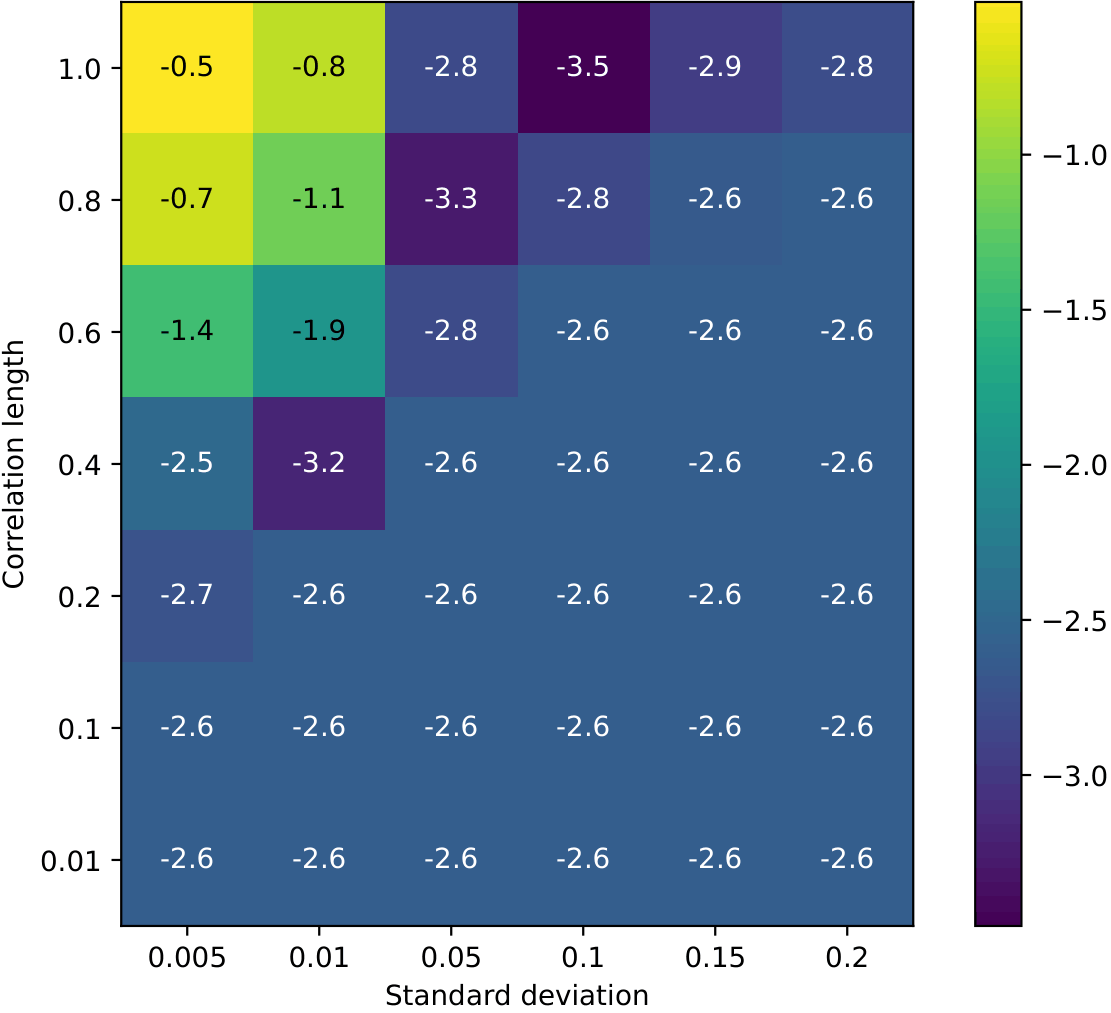}
        \subcaption{PI-RBF kernel ($M=3$)}
    \end{subfigure}

    \caption{
        Estimation of hyperparameters $\sigma_0$ and $\lcor_0$ through cross-validation using UQ coverage indicators. The base 10 logarithm of the mean loss function from \eref{eq_mean_loss_CV} obtained over the grid $\braces{0.005,\, 0.01.,\, 0.05,\, 0.1,\, 0.15,\, 0.2}\times \braces{ 0.01,\, 0.1,\, 0.2,\, 0.4 ,\,0.6 ,\,0.8,\, 1.0 }$ is depicted for each GP kernel definition (RBF, DF-RBF, M-RBF and PI-RBF). The minimal nodes are selected as the estimated hyperparameters for the reconstruction of the flow fields.
        }
    \label{fig_CV_gridsearch}
\end{figure}

\subsubsection{Kernel hyperparameter estimation with UQ-based cross-validation} \label{sec:CrossValidation}

In order to estimate the remaining hyperparameters, we use the training set to perform cross-validation with a loss function defined in terms of the UQ coverage indicators from \eref{eq_UQ_coverages}. To do so, we split the training set $\Yset$ into $K$ disjoint folds of equal size: $\Yset = \smash{\Yset^{(1)} \cup \cdots \cup \Yset^{(K)}}$. Then we define the individual loss function $\Loss$ given a validation fold $\smash{\Yset^{(k)}}$ and a selection of hyperparameters $\smash{\sigma_0,\lcor_0}$ (which are associated to a reconstruction formula $\uapprox$):
\begin{equation}
    \Loss(\Yset^{(k)};\sigma_0,\lcor_0) = \demi \brackets{ \left(p_1(\Yset^{(k)};\uapprox) - 0.95\right)^2 + \left(p_2(\Yset^{(k)};\uapprox) - 0.95\right)^2  }\,.
\end{equation}
We use \eqref{eq_UQ_coverages} for the computation of $\smash{p_1}$ and $\smash{p_2}$, replacing the test set $\smash{\Xtestset}$ by each validation fold $\Yset^{(k)}$. Finally, we perform grid search to minimize the mean loss function:
\begin{equation}\label{eq_mean_loss_CV}
    \Loss_K(\sigma_0,\lcor_0) = \frac{1}{K} \sum_{k=1}^K\Loss(\Yset^{(k)};\sigma_0,\lcor_0)\,.
\end{equation}

In our experiment, we use $K=4$ folds to get estimates of the hyperparameters $\sigma_0$ and $\lcor_0$ for the four investigated kernels by minimizing over the grids displayed in \fref{fig_CV_gridsearch}. The logarithm of the mean loss function over these grids is depicted for each kernel. The minimal nodes are selected for computing the reconstructions in the next sections.

\subsubsection{Flow reconstruction with uncertainty quantification}\label{sec:UQ}

\begin{figure}[h!]
    \centering


    \begin{subfigure}{0.32\textwidth}
        \centering
        \includegraphics[height=3.75cm]{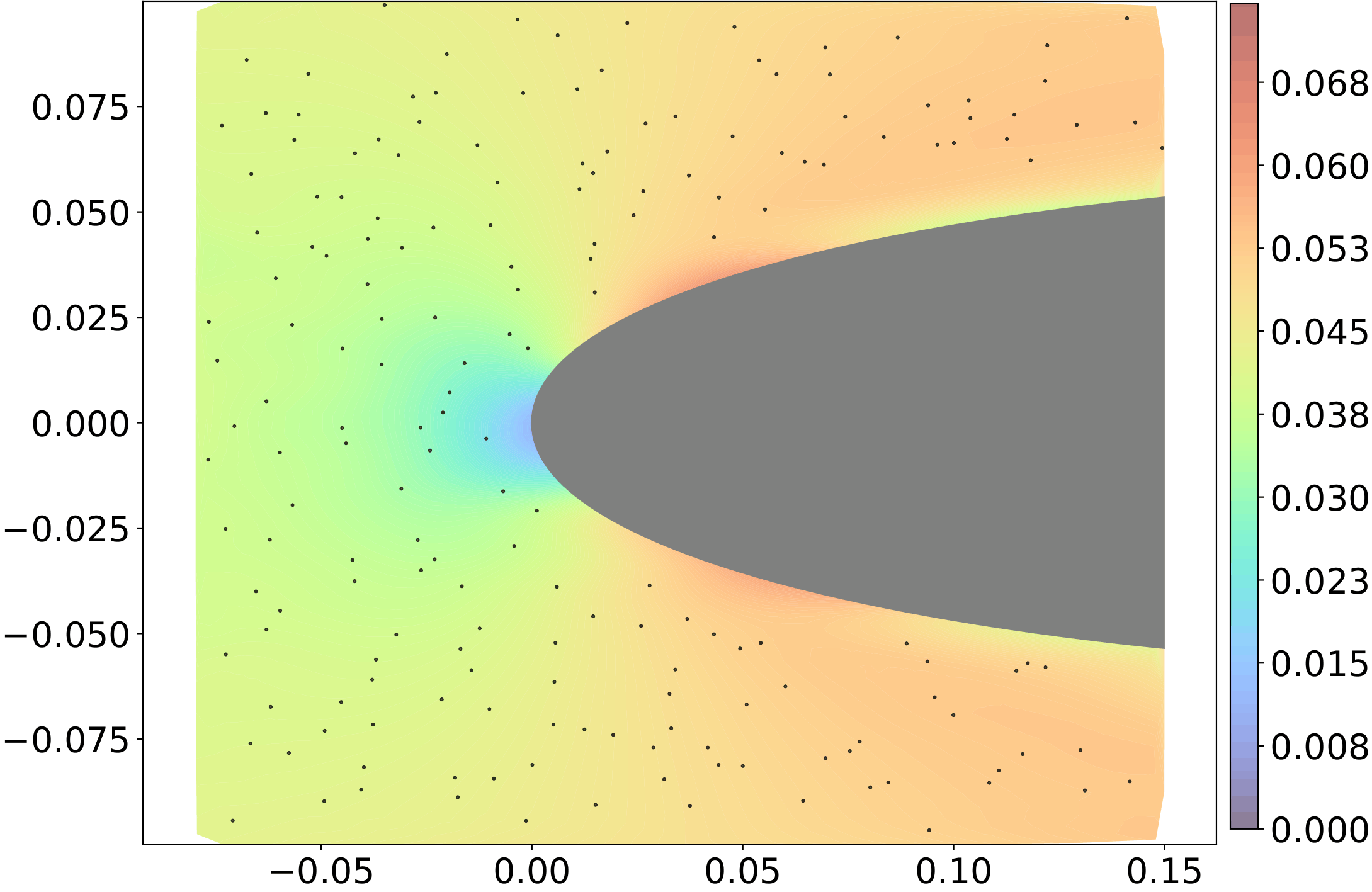}
        \subcaption{ RBF kernel ($M=0$)}
    \end{subfigure}
    \begin{subfigure}{0.32\textwidth}
        \centering
        \includegraphics[height=3.75cm]{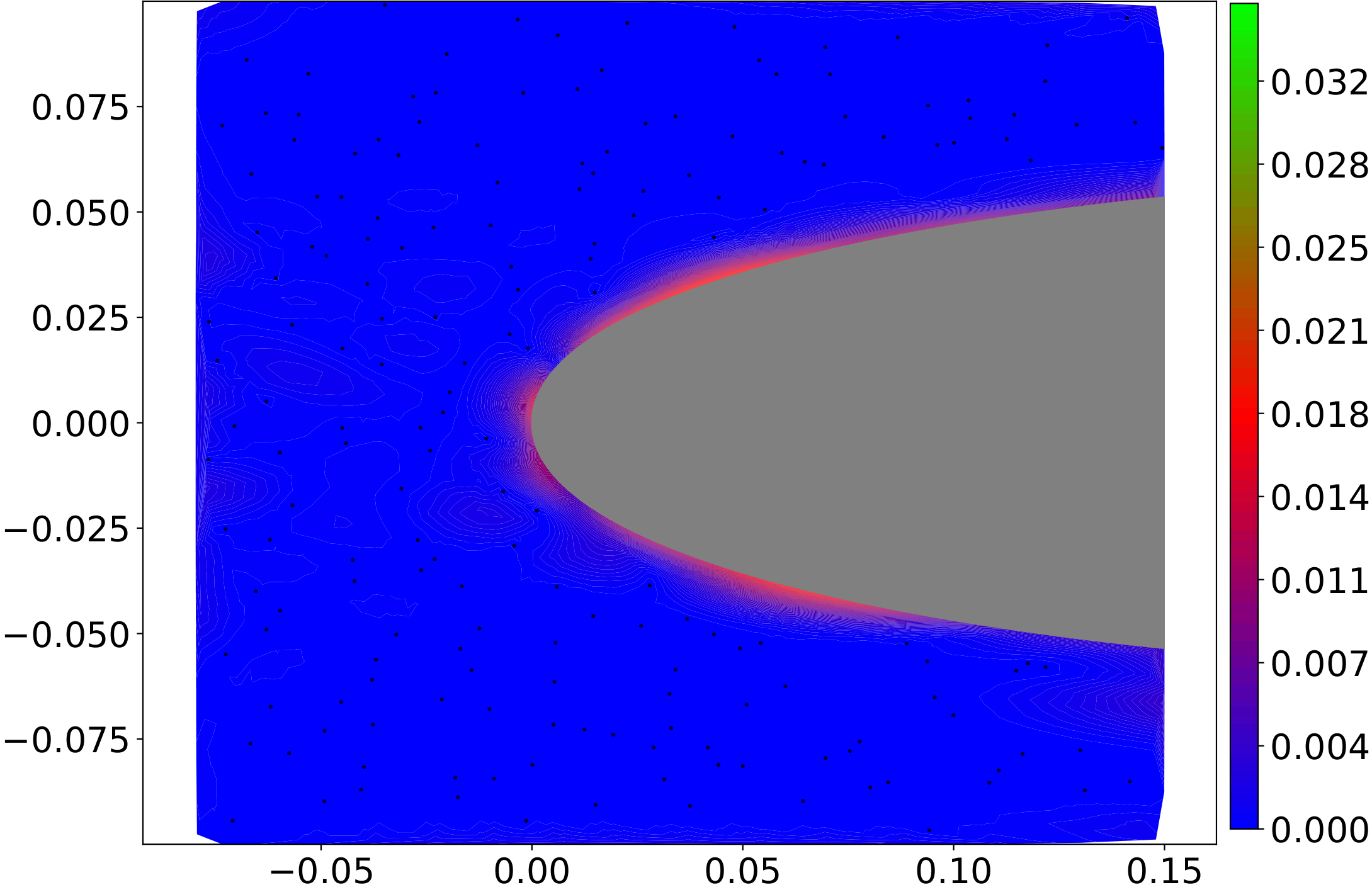}
        \subcaption{ RBF kernel ($M=0$)}
    \end{subfigure}
    \begin{subfigure}{0.32\textwidth}
        \centering
        \includegraphics[height=3.75cm]{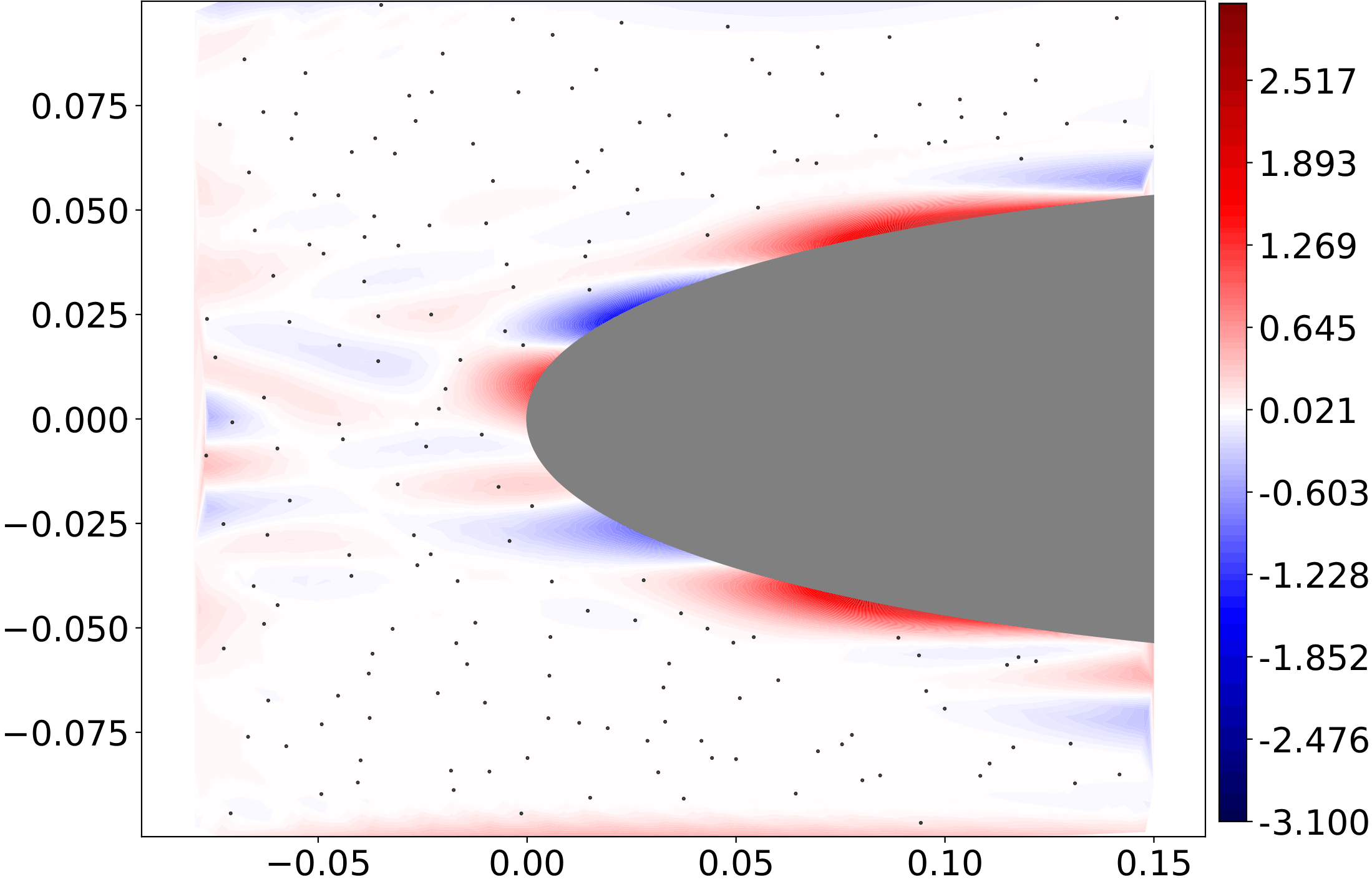}
        \subcaption{ RBF kernel ($M=0$)}
    \end{subfigure}

    \vspace{0.3cm}


    \begin{subfigure}{0.32\textwidth}
        \centering
        \includegraphics[height=3.75cm]{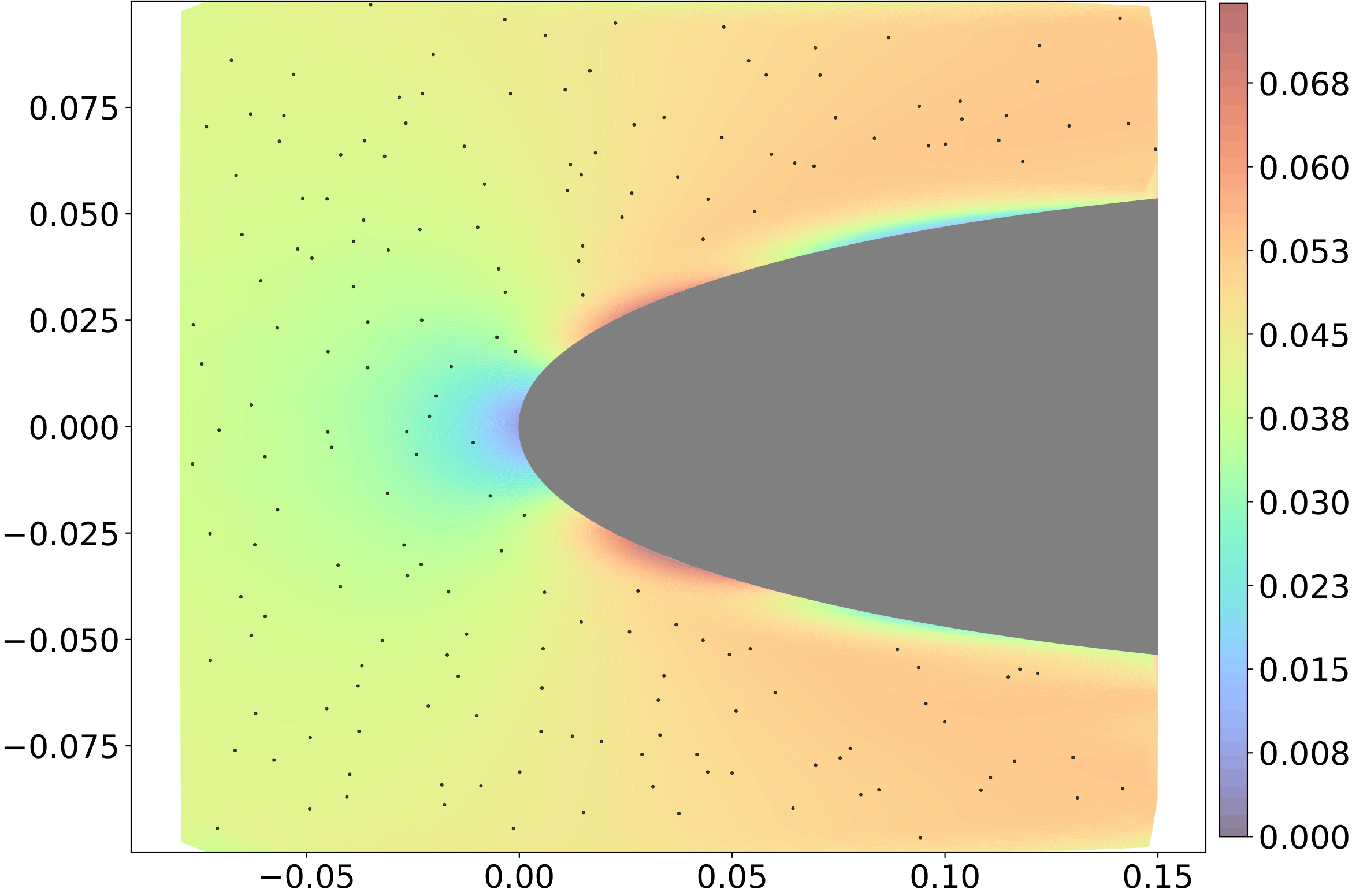}
        \subcaption{DF-RBF kernel ($M=0$)}
    \end{subfigure}
    \begin{subfigure}{0.32\textwidth}
        \centering
        \includegraphics[height=3.75cm]{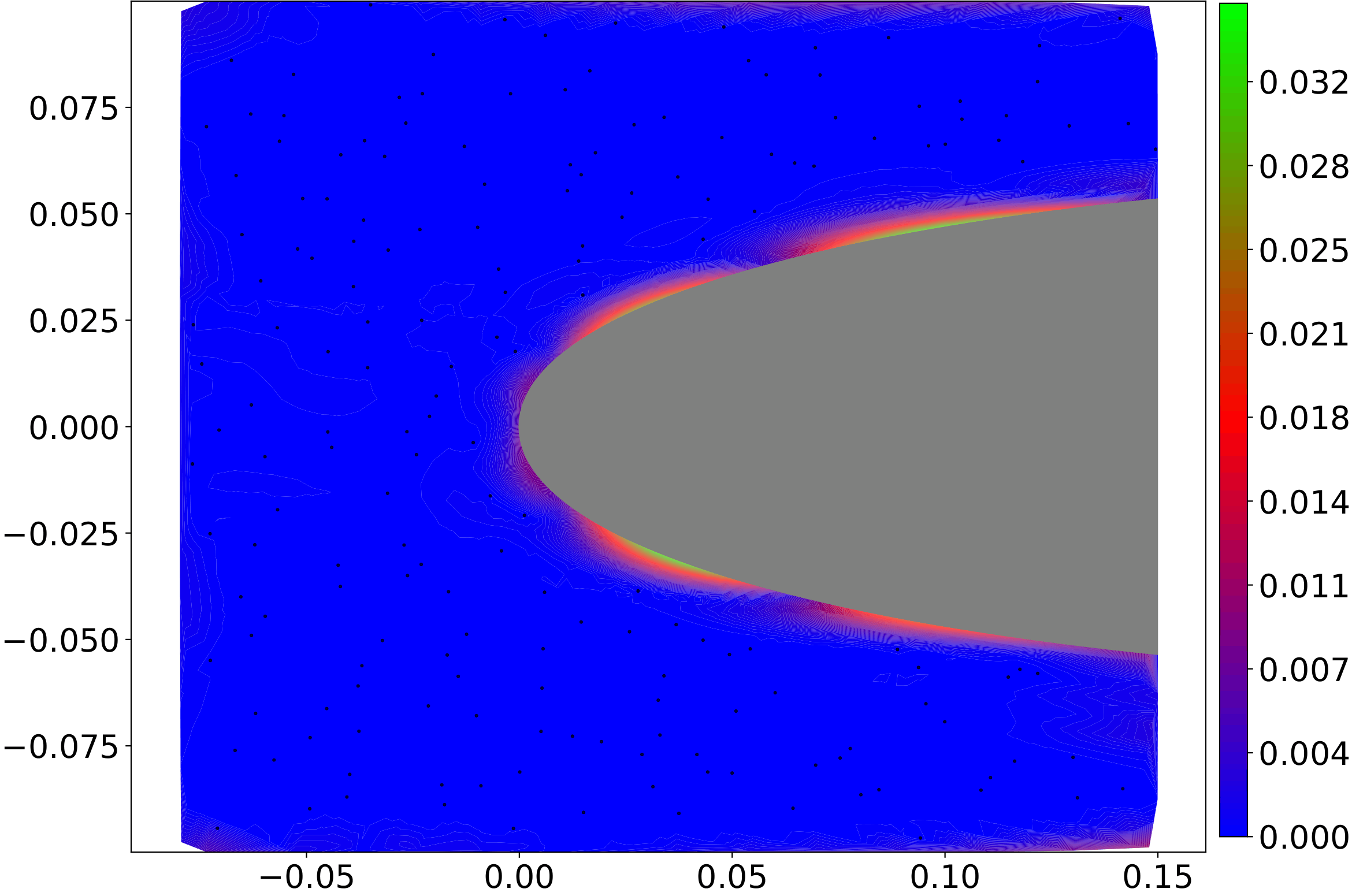}
        \subcaption{DF-RBF kernel ($M=0$)}
    \end{subfigure}
    \begin{subfigure}{0.32\textwidth}
        \centering
        \includegraphics[height=3.75cm]{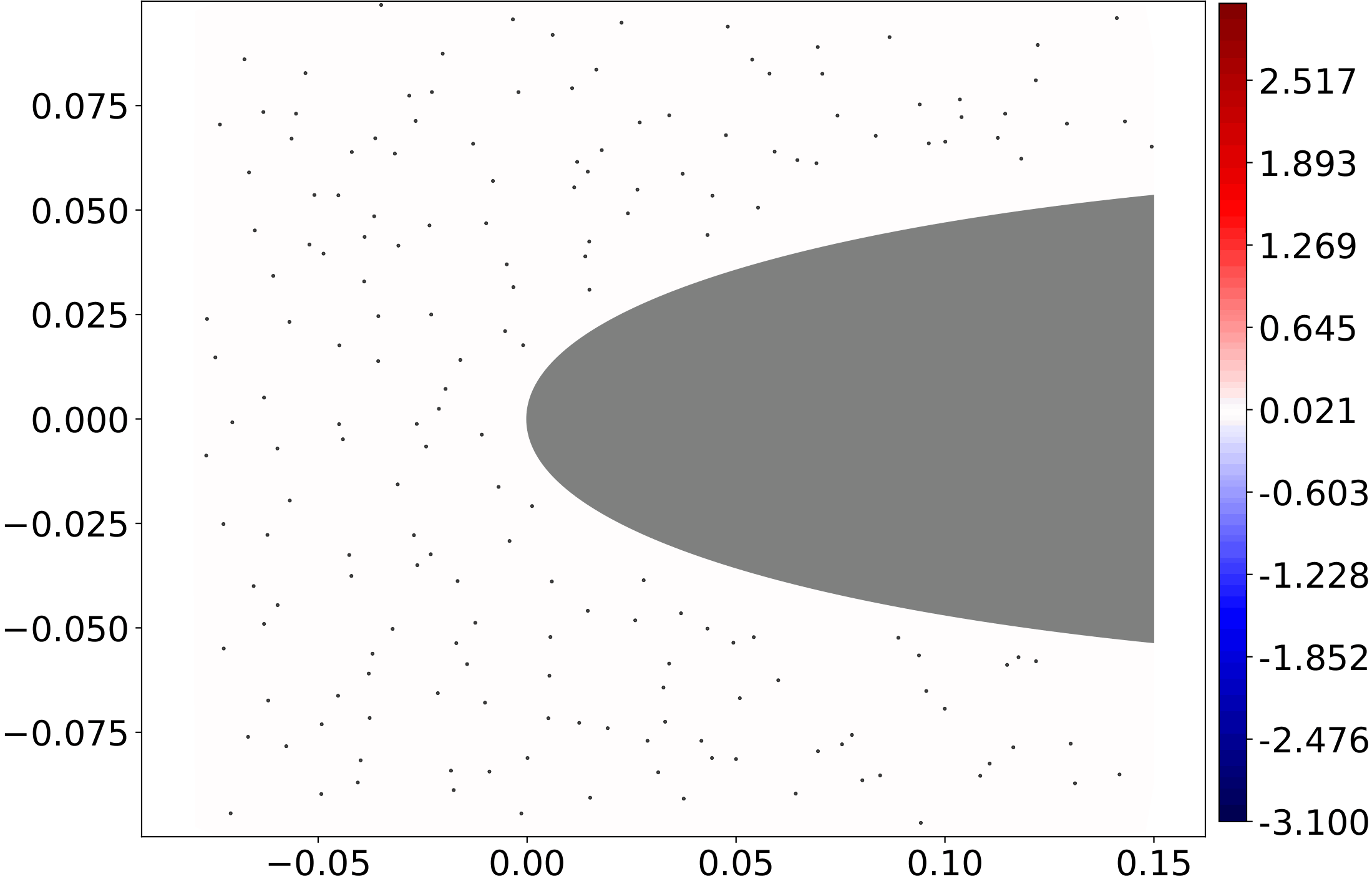}
        \subcaption{DF-RBF kernel ($M=0$)}
    \end{subfigure}
    
    \vspace{0.3cm}


    \begin{subfigure}{0.32\textwidth}
        \centering
        \includegraphics[height=3.75cm]{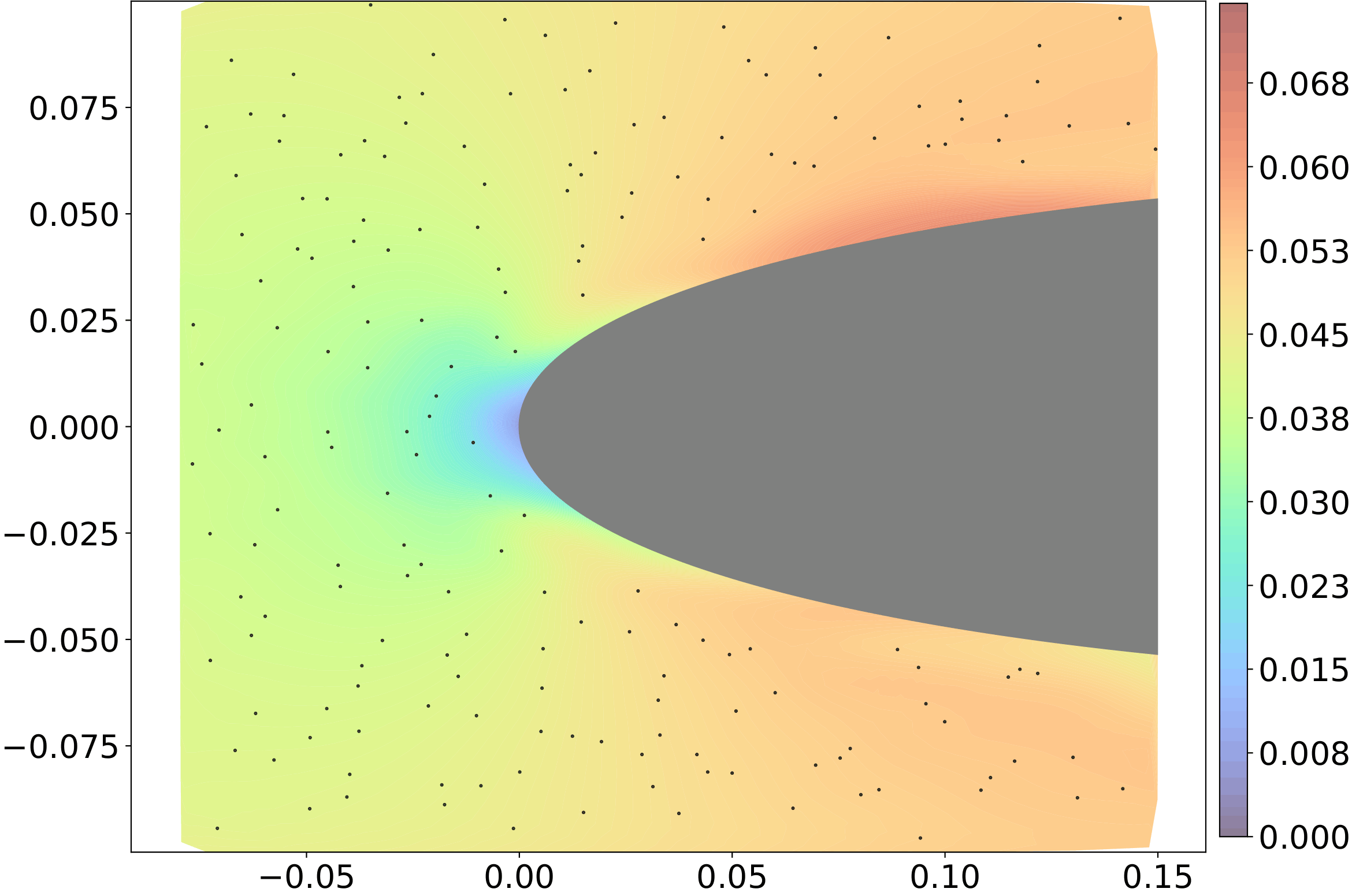}
        \subcaption{M-RBF kernel ($M=3$)}
    \end{subfigure}
    \begin{subfigure}{0.32\textwidth}
        \centering
        \includegraphics[height=3.75cm]{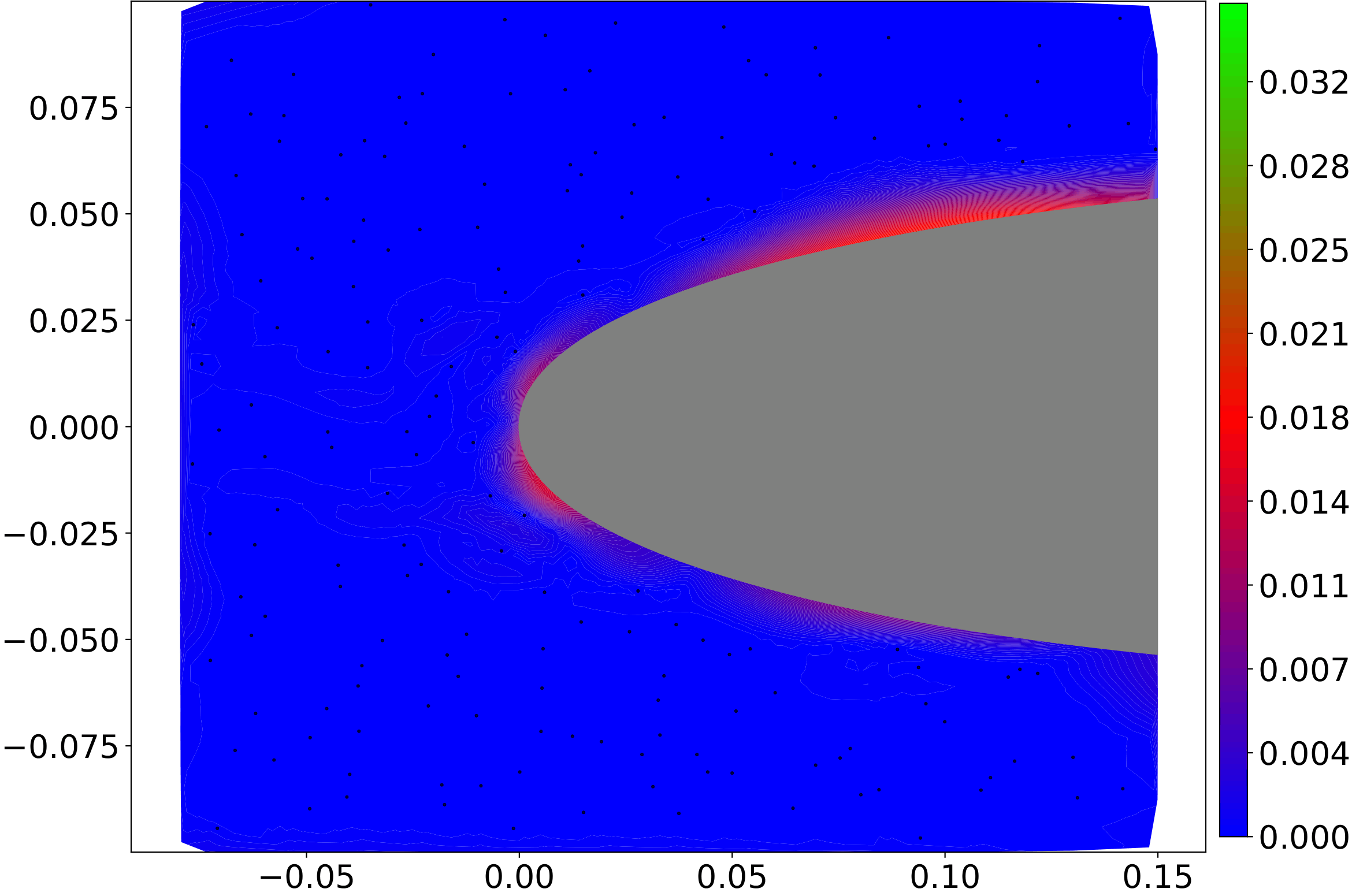}
        \subcaption{M-RBF kernel ($M=3$)}
    \end{subfigure}
    \begin{subfigure}{0.32\textwidth}
        \centering
        \includegraphics[height=3.75cm]{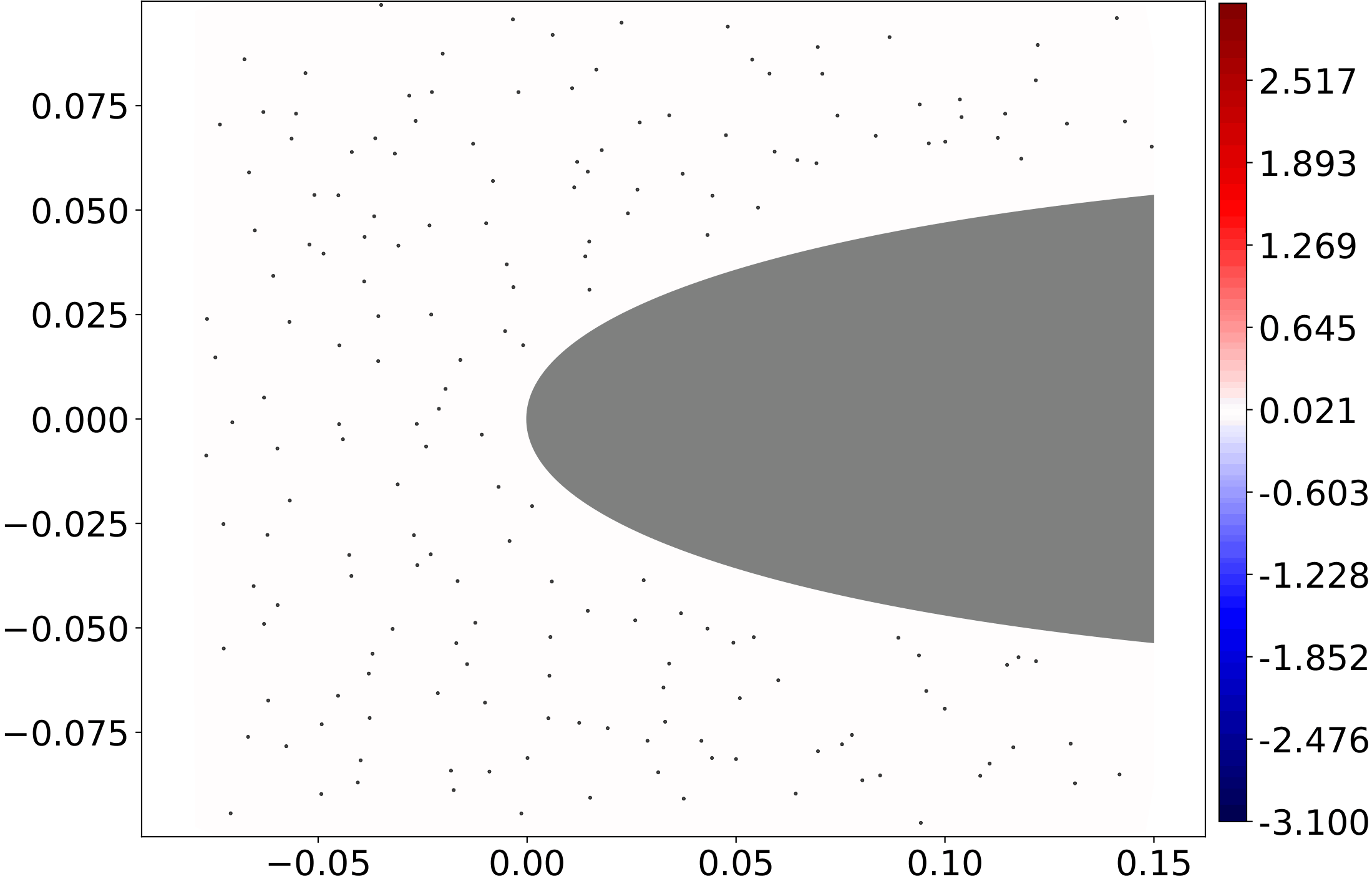}
        \subcaption{M-RBF kernel ($M=3$)}
    \end{subfigure}

    \vspace{0.3cm}


    \begin{subfigure}{0.32\textwidth}
        \centering
        \includegraphics[height=3.75cm]{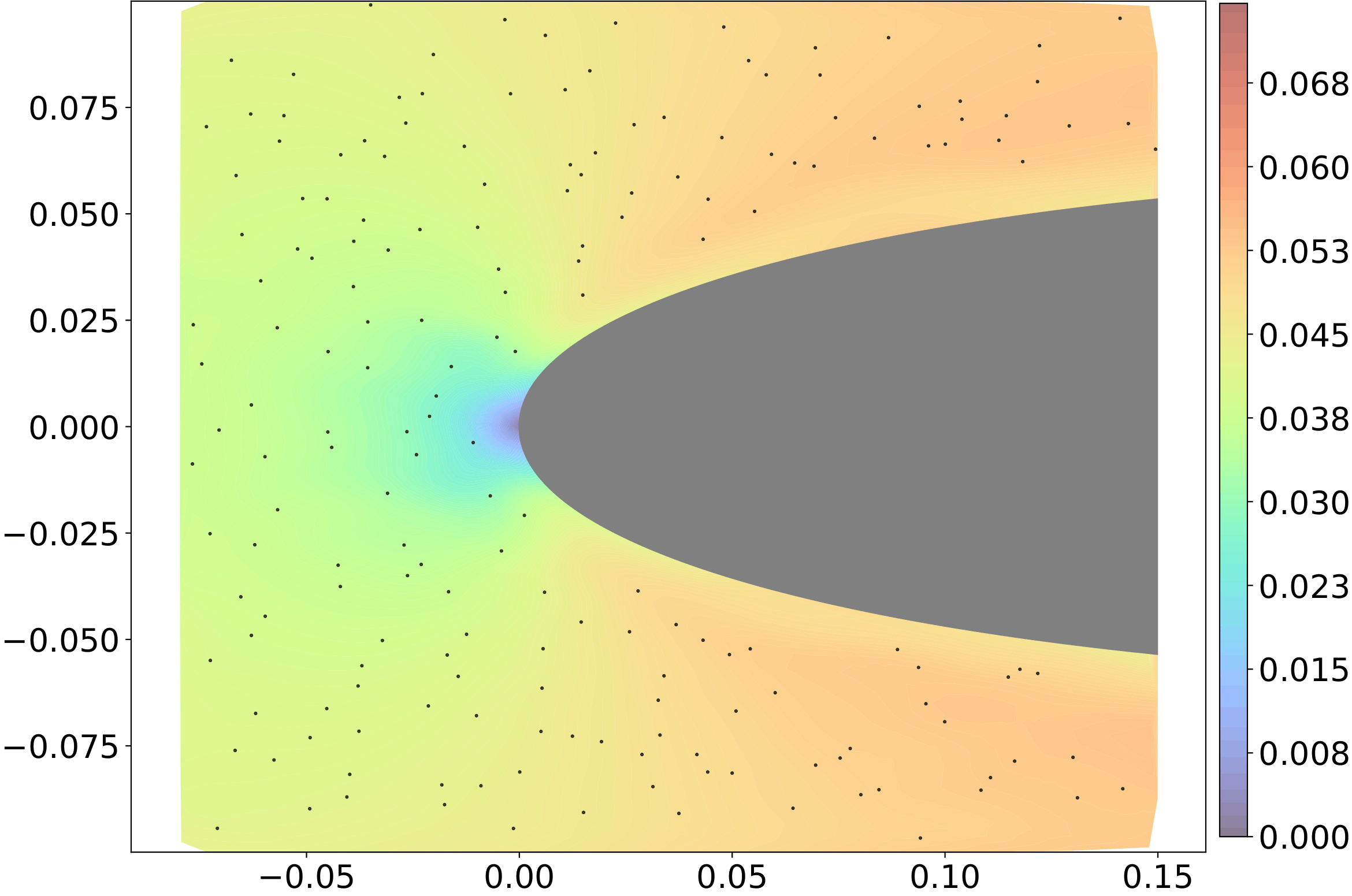}
        \subcaption{PI-RBF kernel ($M=3$)}
    \end{subfigure}
    \begin{subfigure}{0.32\textwidth}
        \centering
        \includegraphics[height=3.75cm]{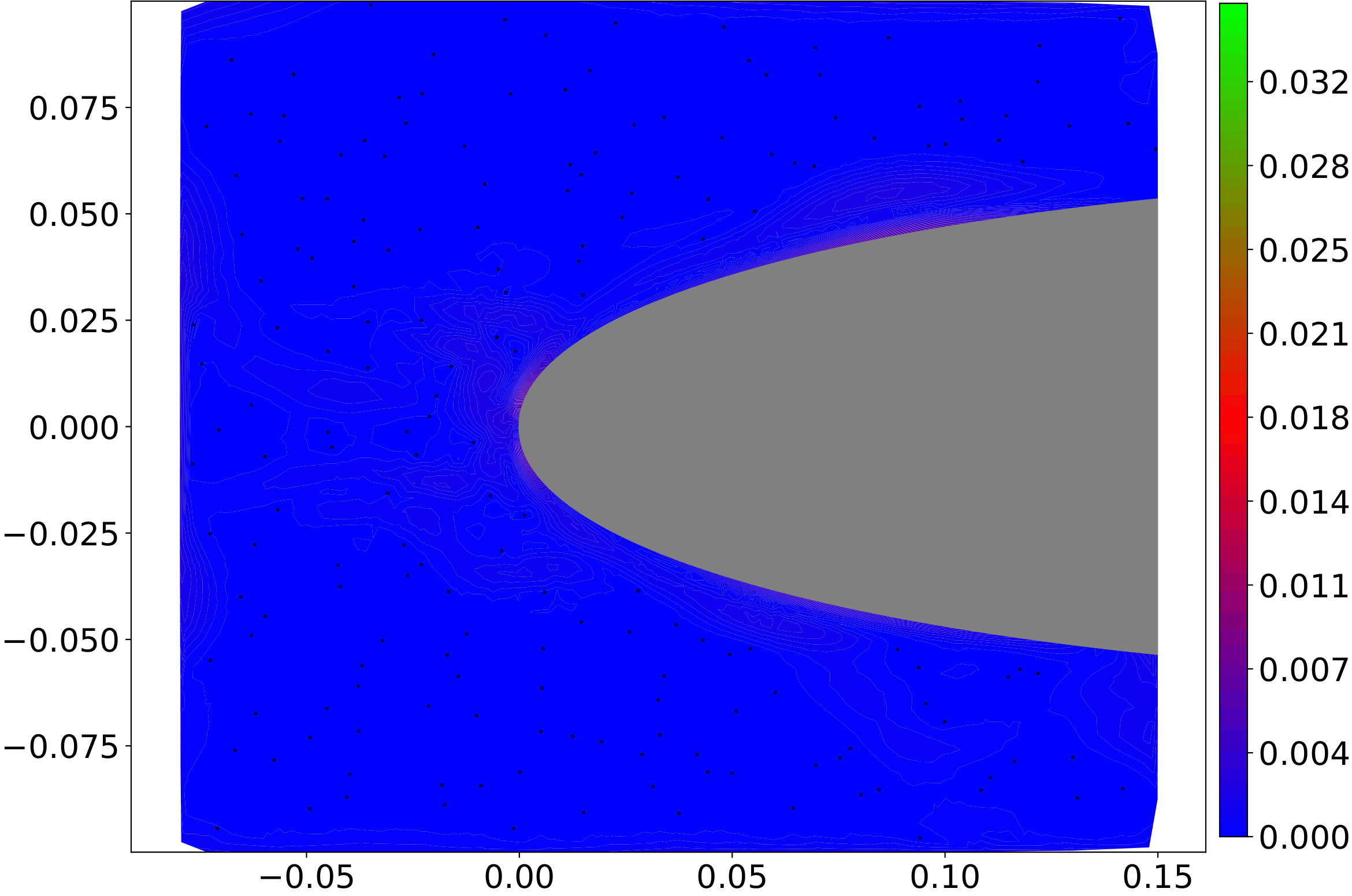}
        \subcaption{PI-RBF kernel ($M=3$)}
    \end{subfigure}
    \begin{subfigure}{0.32\textwidth}
        \centering
        \includegraphics[height=3.75cm]{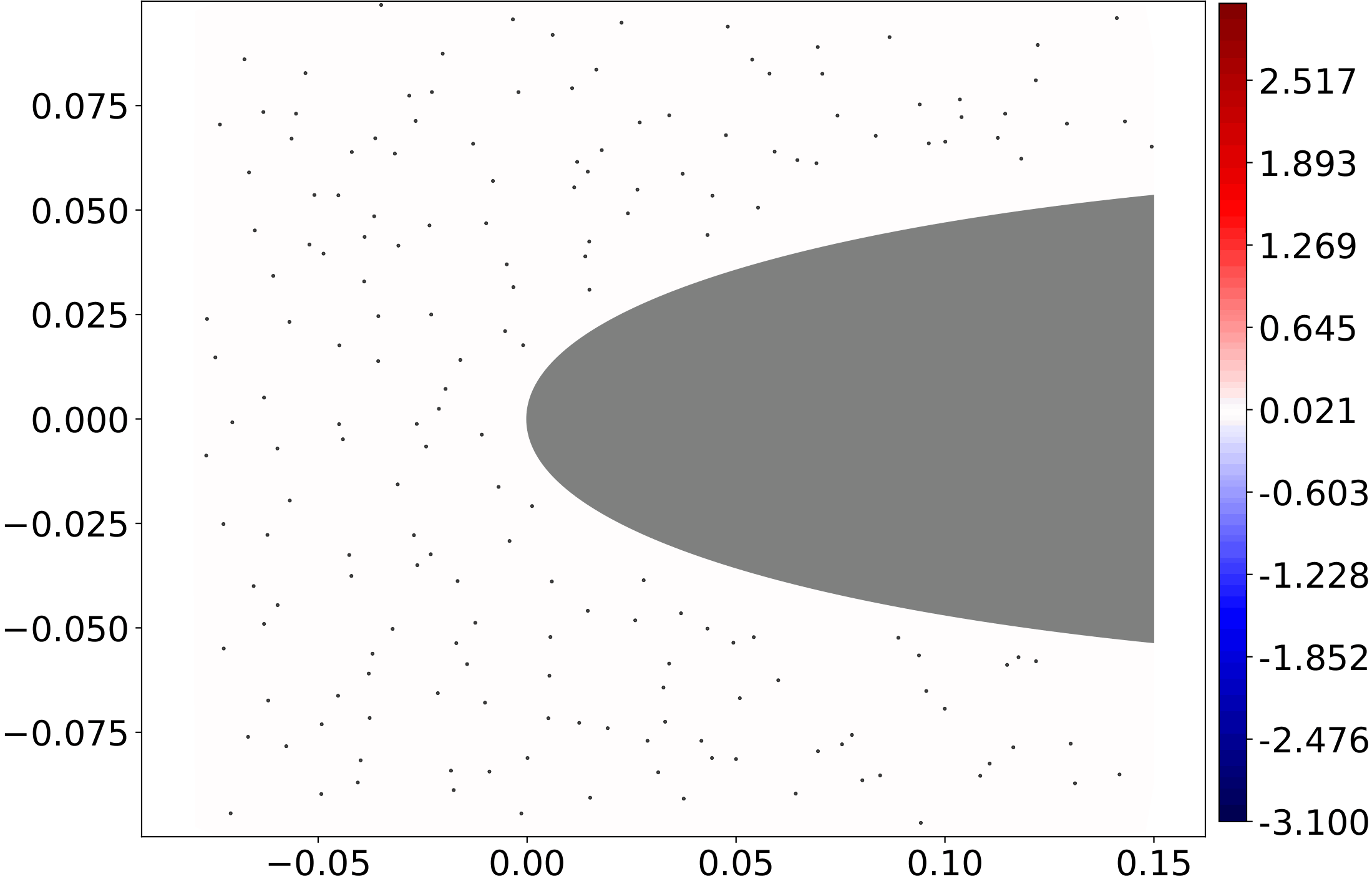}
        \subcaption{PI-RBF kernel ($M=3$)}
    \end{subfigure}
    
    \vspace{0.2cm}


    \caption{
        Reconstruction of the velocity field of an incompressible flow at the leading edge of a NACA 0412 airfoil using GPs with different kernels. In the first column (a, d, g and j), the norm of the reconstructed velocity field $\uapprox$ is presented in m/s. In the second column (b, e, h and k), the norm of the residual field $\smash{\uapprox-\ugt}$ is presented in m/s. The third column (c, f, i and l) displays the divergence field $\smash{\diver\uapprox}$ in 1/s units. The choices of GP kernels are: one-scale RBF (a)-(c); divergence-free DF-RBF (d)-(f); multiple-scale M-RBF accounting for energy decay (g)-(i); and physics-informed multiple-scale PI-RBF (informed about the profile boundary condition through the BCGP procedure) (j)-(l). The black dots ($\bullet$) are the velocity observation positions corresponding to the training set within the computational domain $\DomG$.
        }
    \label{fig_NACA_0412_interpolation}
\end{figure}

The first column in \fref{fig_NACA_0412_interpolation} displays the reconstructed velocity fields $\uapprox$ within the computational domain $\DomG$ using each one of the four kernels (RBF, DF-RBF, M-RBF, and PI-RBF). The corresponding residual fields $\smash{\uapprox-\ugt}$ are depicted in the second column, and the velocity divergence fields are displayed in the third column of \fref{fig_NACA_0412_interpolation}. These reconstructions are performed using the same training set $\Yset$, stressing again that no velocity measurements are set at the boundary of the profile at all. A summary of the regression configurations as well as the fit evaluation indicators are gathered in \tref{tab_NACA_0412}. The UQ coverage indicators are computed on the test set $\Xtestset$. The reported CPU time for the offline computation of the BCGP spectral factor was 0.94 second, including precomputation of kernel derivatives. We observe that the baseline approximation (RBF) is relatively accurate but does not satisfy physical conditions. The reconstruction using the DF-RBF kernel satisfies incompressibility but is not accurate enough around the airfoil boundary.  We notice that the reconstruction improves from DF-RBF to M-RBF by the inclusion of larger scales and a proper choice of hyperparameters. However, there is still no means of getting information about the profile boundary with either of these two approaches. The boundary condition fit improves significantly with the inclusion of the BCGP procedure. In this way, even if there is no discrete observation on the profile boundary for the computation of the posterior estimates, the PI-RBF kernel is able to recover this information. The numerical indicator $\epsNormal$ of \eref{eq_normal_indicator} quantifying the goodness of fit of the boundary condition is significantly decreased (by three orders of magnitude) compared to the other kernels. The choice of a PI-RBF kernel improves the reconstructed velocity fields at the leading edge of the profile as well.

\begin{table}[h!]
    \centering
    \begin{tabular}{
        |>{\centering\arraybackslash}m{4cm}
        |>{\centering\arraybackslash}m{2.7cm}
        |>{\centering\arraybackslash}m{2.7cm}
        |>{\centering\arraybackslash}m{2.7cm}
        |>{\centering\arraybackslash}m{3.2cm}|}
    \hline
    GP kernel approach & RBF & DF-RBF & M-RBF & PI-RBF \\
    \hline
    Prior information from physics & Default anisotropy & Incompressibility &  \makecell{ Incompressibility \\ Energy decay} & \makecell{ Incompressibility \\ Energy decay\\ Slip boundary\\ condition} \\
    \hline
    Kernel hyperparameters & \makecell{$M = 0$ \\ $\sigma_0 = 0.20$ \\ $\lcor_0 = 0.1$} & \makecell{$M = 0$ \\ $\sigma_0 = 0.01$ \\ $\lcor_0 = 0.1$} & \makecell{$M = 3$ \\ $\sigma_0 = 0.05$ \\ $\lcor_0 = 1.0$} & \makecell{ $M = 3$ \\ $\sigma_0 = 0.10$ \\ $\lcor_0 = 1.0$} \\
    \hline
    Regularization nugget $\nugget$ & $10^{-8}$ & $10^{-8}$ & $10^{-10}$ & $10^{-10}$ \\
    \hline
    $\epsNormal$ from \eref{eq_normal_indicator} & $1.5\cdot 10^{-1}$  & $7.9\cdot 10^{-2}$ & $4.8\cdot 10^{-2}$ & $9.3\cdot 10^{-6}$ \\
    \hline
    $\errRMSE$ from \eref{eq:eRMSE} & $2.6\cdot 10^{-3}$ & $4.0\cdot 10^{-3}$ & $2.7\cdot 10^{-3}$ & $7.6\cdot 10^{-4}$ \\
    \hline
    $\epsDiver$ from \eref{eq_divergence_indicator} &  $8.6\cdot 10^{-2}$  &  $6.4\cdot 10^{-7}$  &  $3.9\cdot 10^{-7}$  & $7.8\cdot 10^{-6}$  \\
    \hline
    \makecell{UQ coverage indicators\\ on $\Xtestset$} & \makecell{ $p_1  = 78.9\%$ \\ $p_2  = 80.8\%$}  & \makecell{ $p_1  = 81.7\%$ \\ $p_2  = 78.2\%$ } & \makecell{ $p_1  = 78.2\%$ \\ $p_2  = 84.5\%$}  & \makecell{$p_1  = 96.7\%$ \\ $p_2  = 95.3\%$} \\
    \hline
    Posterior mean CPU time & 0.33 s & 0.61 s  & 1.25 s  &  2.77 s   \\
    \hline
    \makecell{Posterior covariance\\ CPU time} & 1.49 s & 4.59 s  &  17.06 s  &  21.89 s   \\
    \hline
    \end{tabular}
    \caption{
        Configuration and evaluation for the reconstruction of the velocity field of an incompressible flow at the leading edge of a NACA 0412 airfoil using GPs with different kernels. The choices of GP kernels are: one-scale RBF; divergence-free DF-RBF; multiple-scale M-RBF accounting for energy decay; and physics-informed multiple-scale PI-RBF (informed about the profile boundary condition through the BCGP procedure). The kernel hyperparameters are set through cross-validation using the training set; see \sref{sec:CrossValidation}. The UQ coverage indicators of each velocity component are computed using the test set and the corresponding velocity reconstruction formula of each configuration.
    }
    \label{tab_NACA_0412}
\end{table}

\begin{figure}[h!]
    \centering

    \begin{subfigure}{0.47\textwidth}
        \centering
        \includegraphics[height=4.5cm]{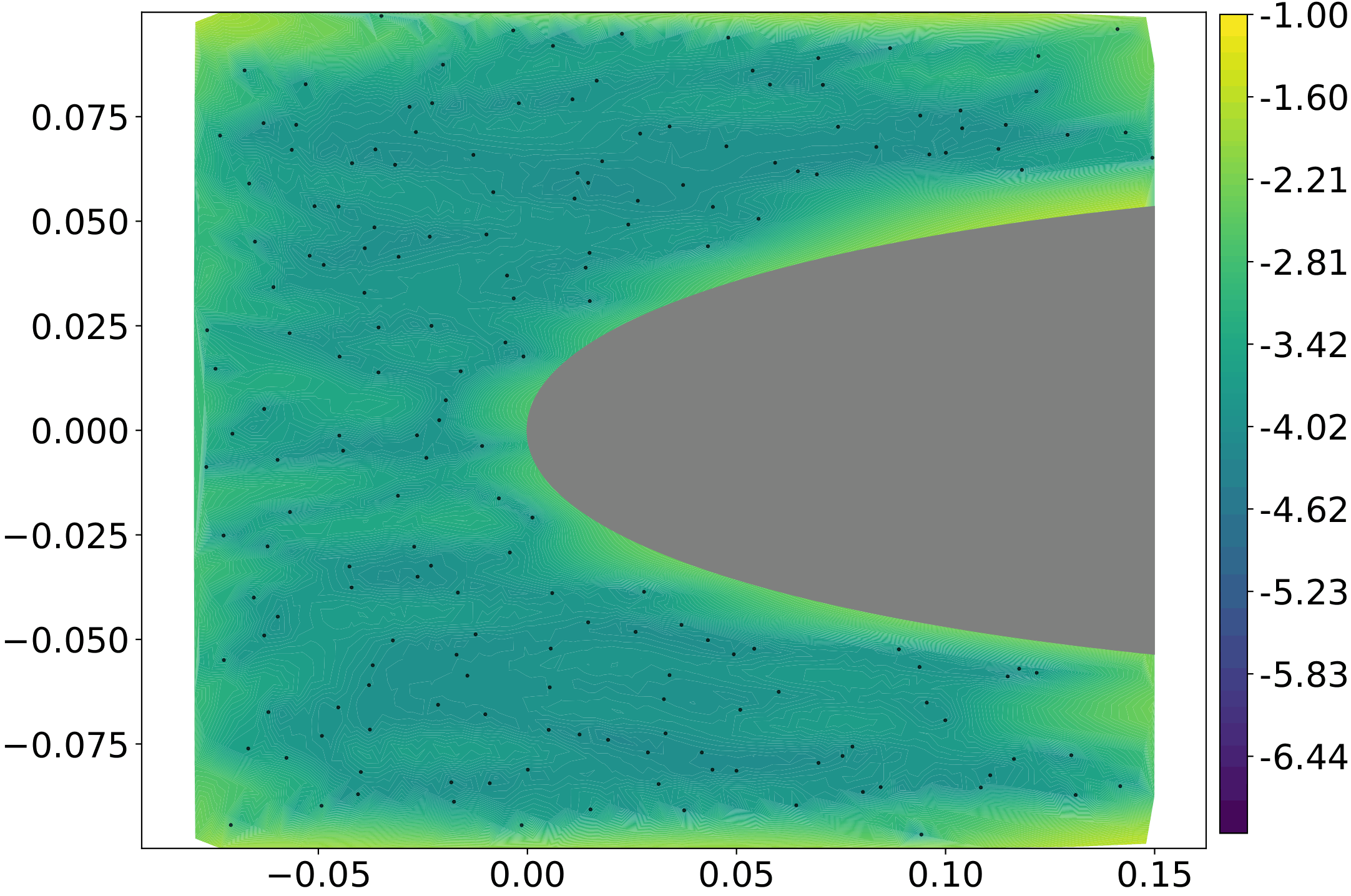}
        \subcaption{ RBF kernel ($M=0$)}
    \end{subfigure}
    \begin{subfigure}{0.47\textwidth}
        \centering
        \includegraphics[height=4.5cm]{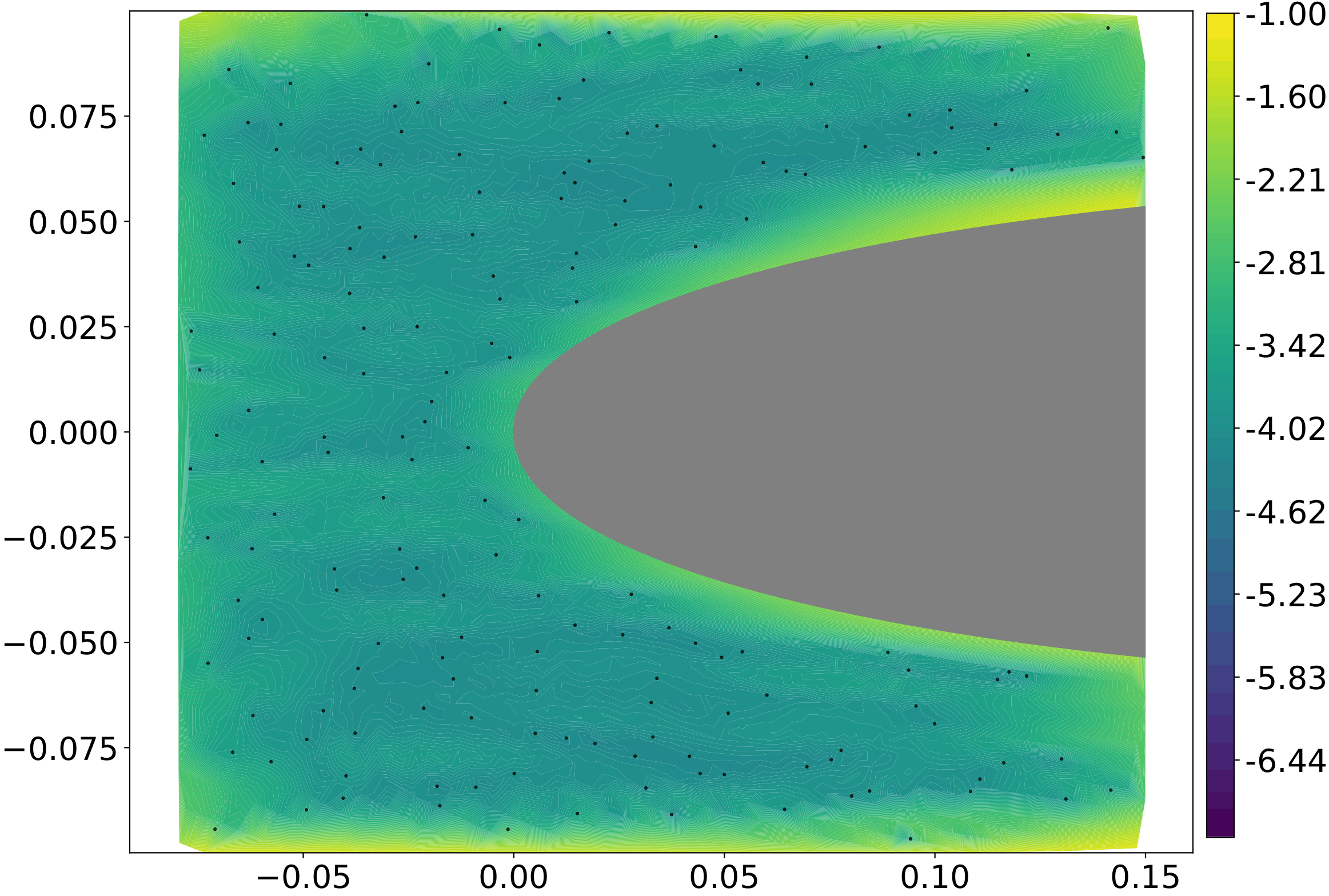}
        \subcaption{DF-RBF kernel ($M=0$)}
    \end{subfigure}

    \vspace{0.2cm}

    \begin{subfigure}{0.47\textwidth}
        \centering
        \includegraphics[height=4.5cm]{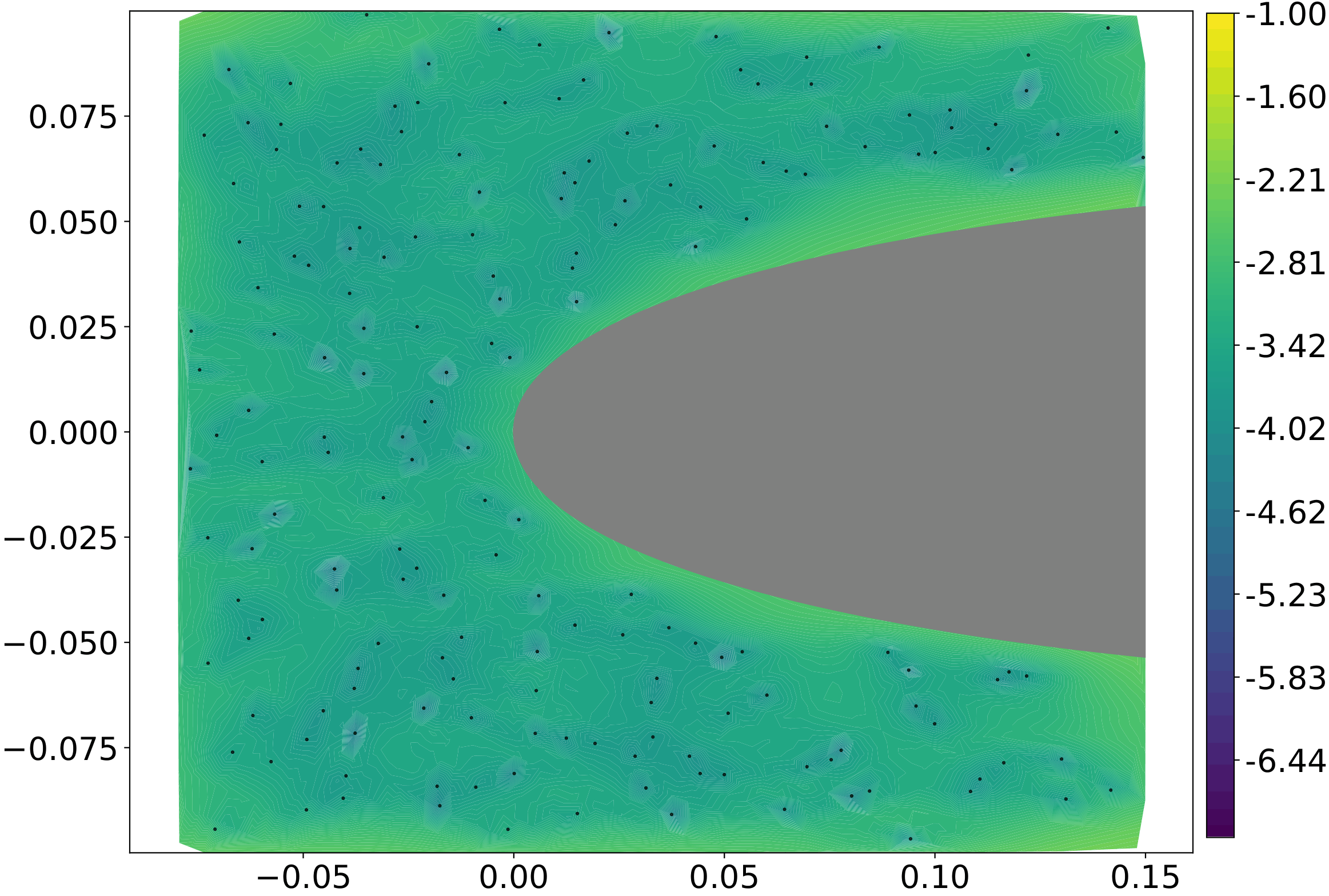}
        \subcaption{M-RBF kernel ($M=3$)}
    \end{subfigure}
    \begin{subfigure}{0.47\textwidth}
        \centering
        \includegraphics[height=4.5cm]{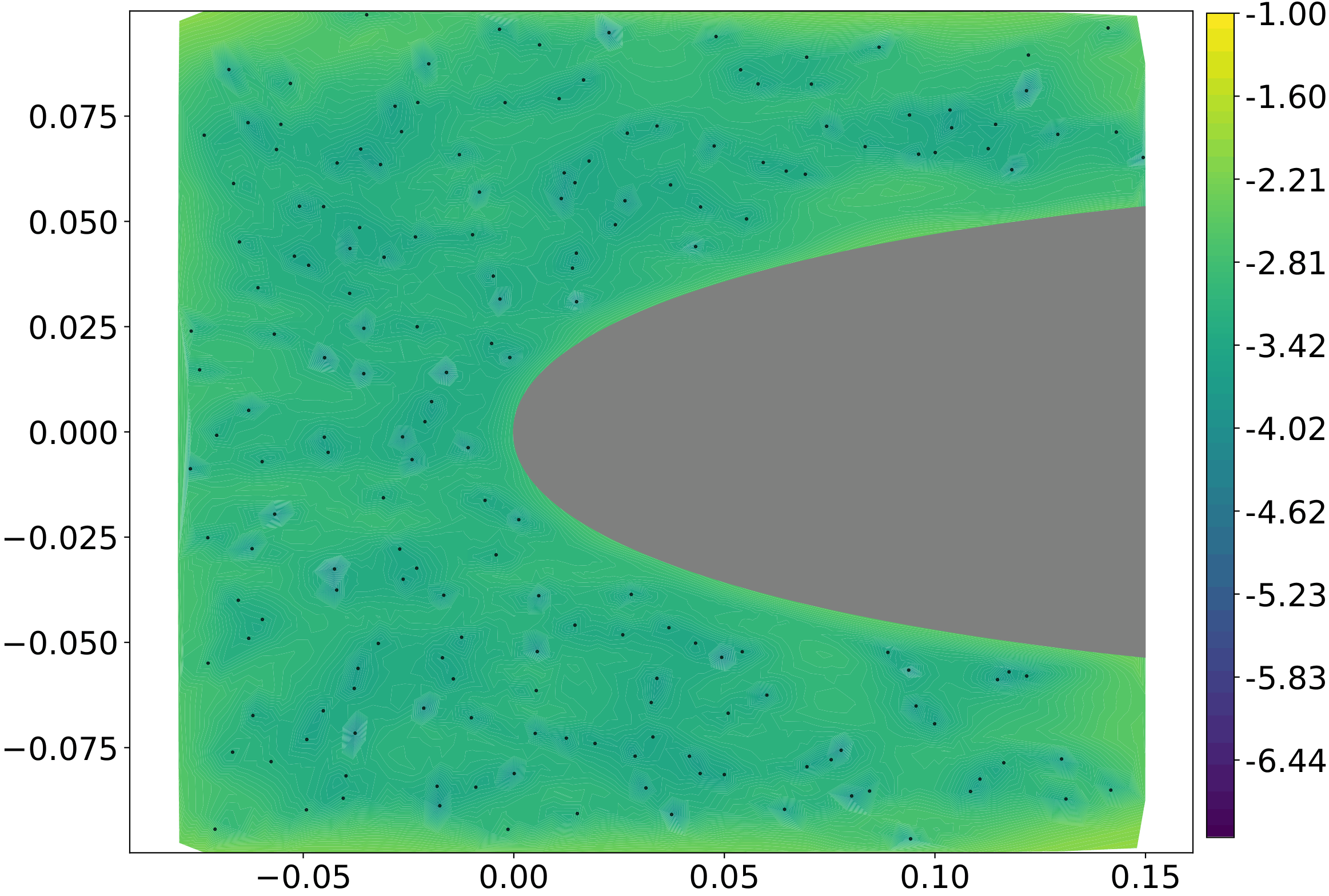}
        \subcaption{PI-RBF kernel ($M=3$)}
    \end{subfigure}

    \vspace{0.2cm}

    \caption{
        UQ of the reconstructed velocity field of an incompressible flow at the leading edge of a NACA 0412 airfoil using GPs with different kernels. The depicted fields are the base 10 logarithm of the posterior total standard deviation. The choices of GP kernels are: one-scale RBF; divergence-free DF-RBF; multiple-scale M-RBF accounting for energy decay; and physics-informed multiple-scale PI-RBF (informed about the profile boundary condition through the BCGP procedure). The black dots ($\bullet$) are the velocity observation positions corresponding to the training set within the computational domain $\DomG$.
        }
    \label{fig_UQ_velocity}
\end{figure}

\begin{figure}[h!]
    \centering

    \begin{subfigure}{0.24\textwidth}
        \centering
        \includegraphics[height=3.2cm]{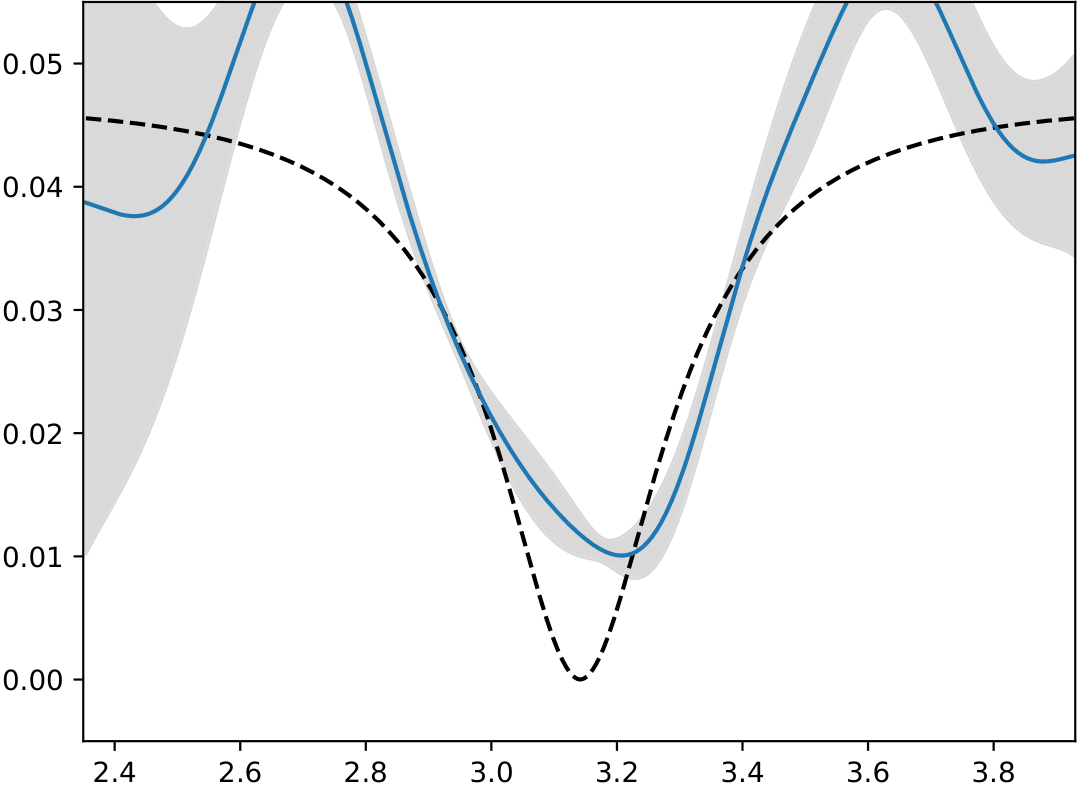}
        \subcaption{ RBF kernel ($M=0$)}
    \end{subfigure}
    \begin{subfigure}{0.24\textwidth}
        \centering
        \includegraphics[height=3.2cm]{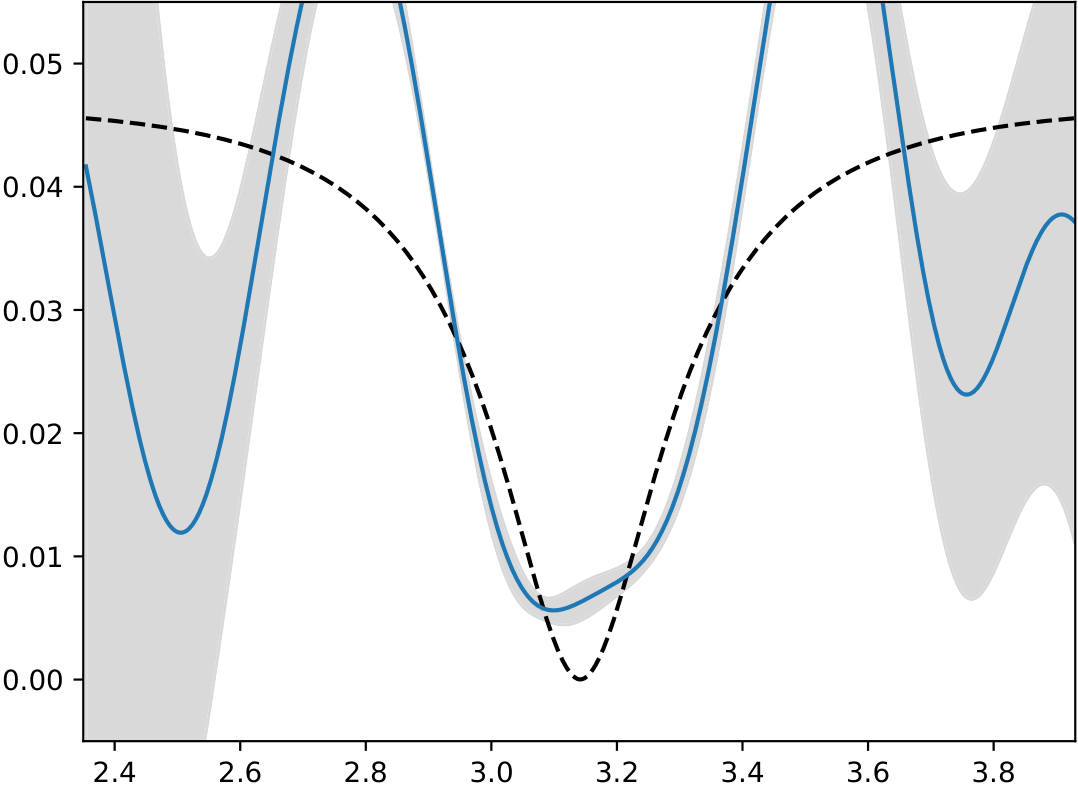}
        \subcaption{ DF-RBF kernel ($M=0$)}
    \end{subfigure}
    \begin{subfigure}{0.24\textwidth}
        \centering
        \includegraphics[height=3.2cm]{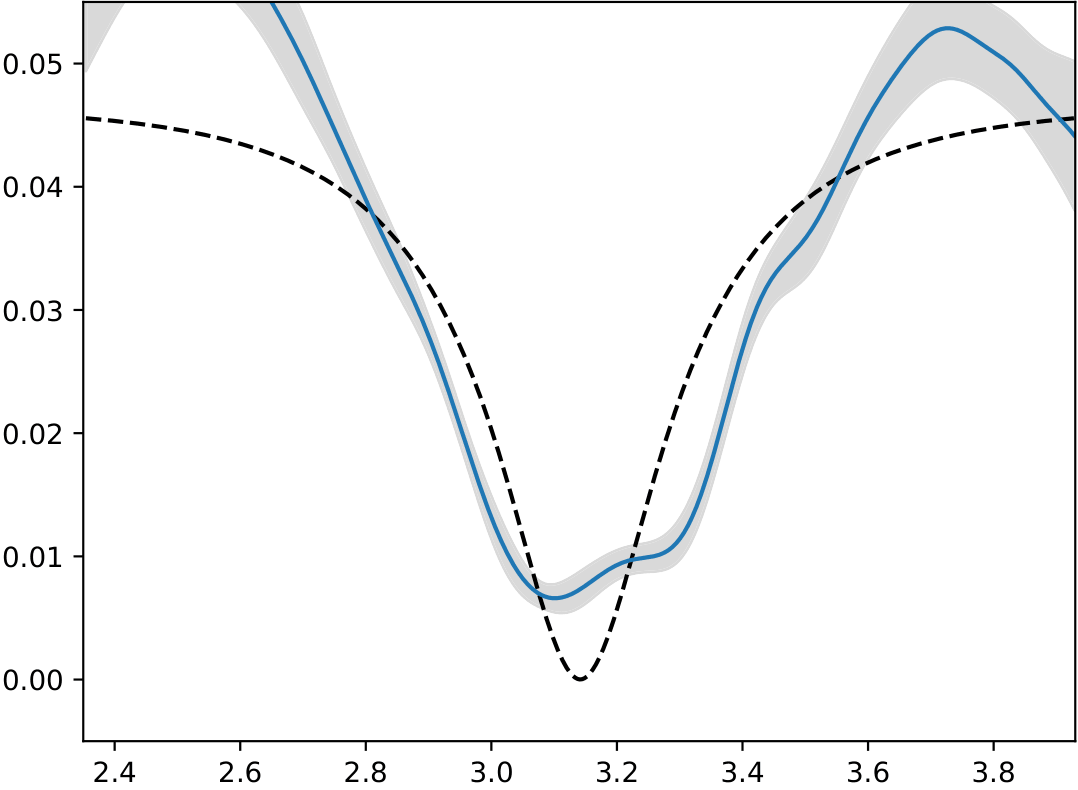}
        \subcaption{M-RBF kernel ($M=3$)}
    \end{subfigure}
    \begin{subfigure}{0.24\textwidth}
        \centering
        \includegraphics[height=3.2cm]{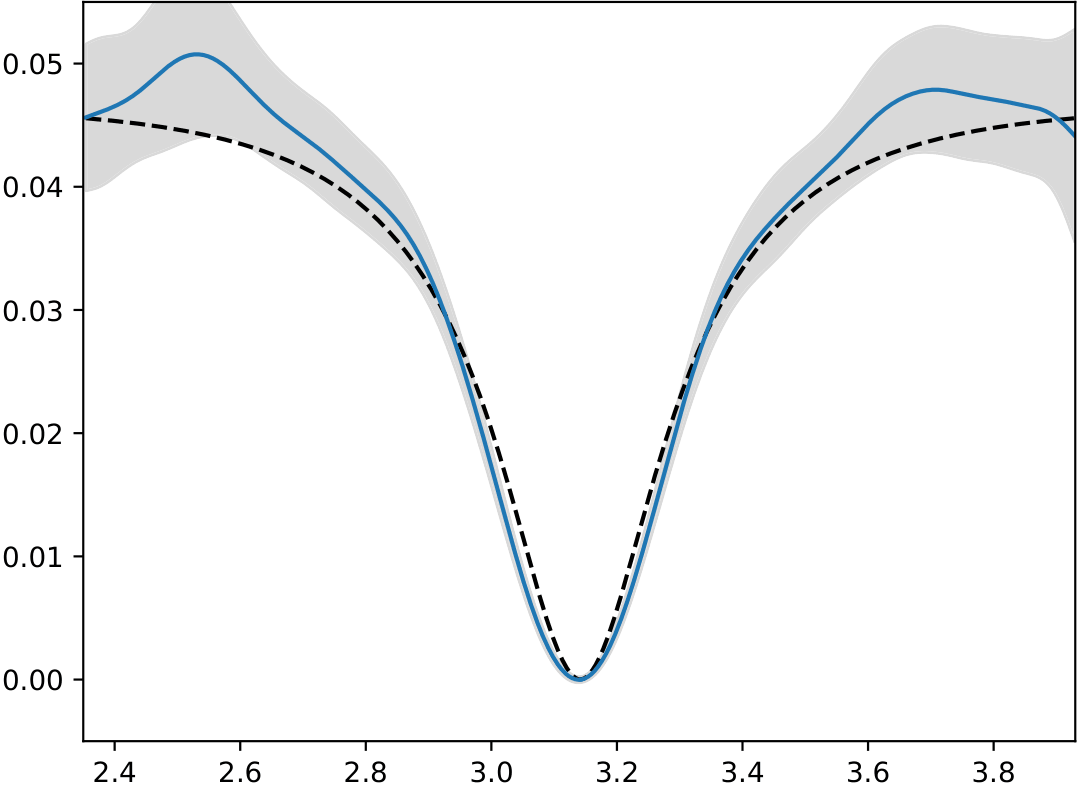}
        \subcaption{PI-RBF kernel ($M=3$)}
    \end{subfigure}

    \vspace{0.2cm}
    
    \begin{subfigure}{0.24\textwidth}
        \centering
        \includegraphics[height=3.2cm]{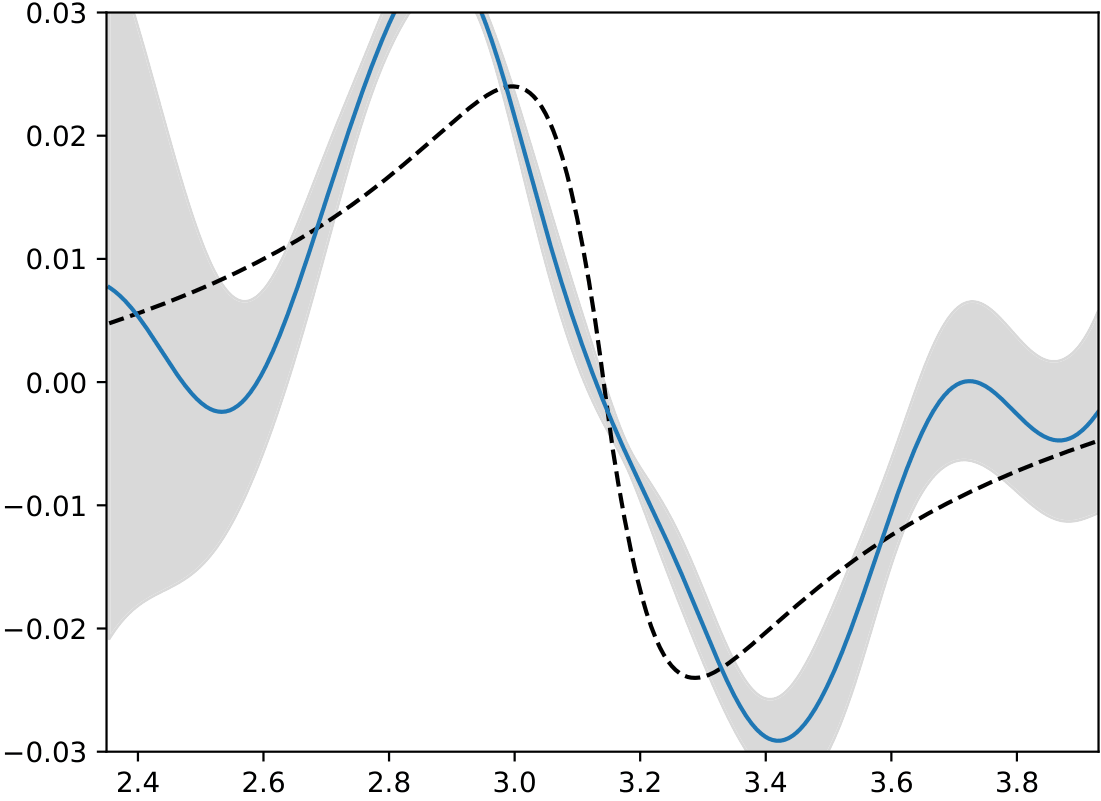}
        \subcaption{ RBF kernel ($M=0$)}
    \end{subfigure}
    \begin{subfigure}{0.24\textwidth}
        \centering
        \includegraphics[height=3.2cm]{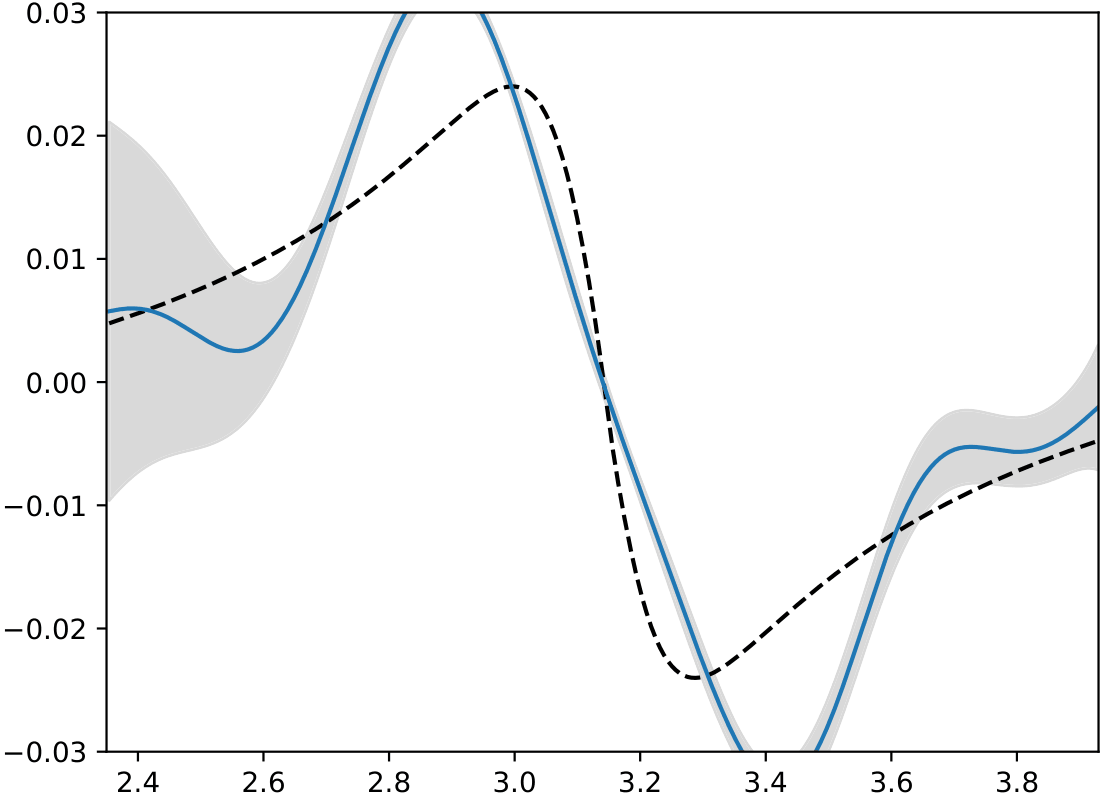}
        \subcaption{ DF-RBF kernel ($M=0$)}
    \end{subfigure}
    \begin{subfigure}{0.24\textwidth}
        \centering
        \includegraphics[height=3.2cm]{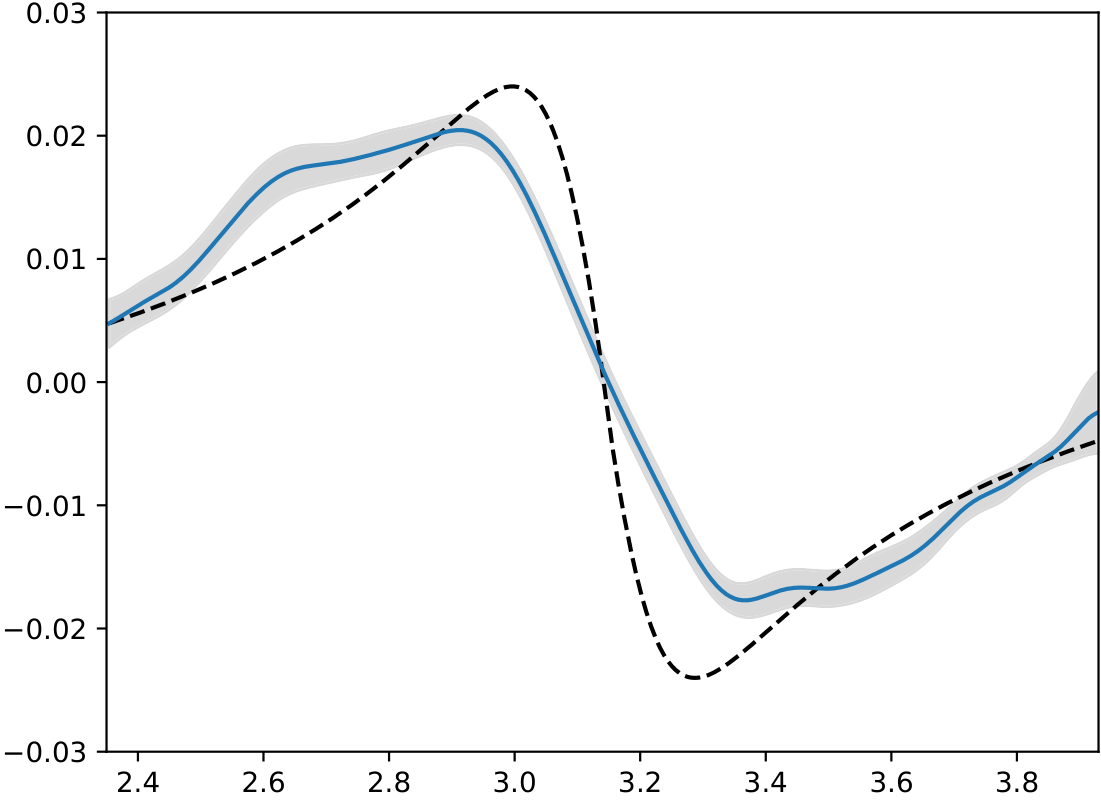}
        \subcaption{M-RBF kernel ($M=3$)}
    \end{subfigure}
    \begin{subfigure}{0.24\textwidth}
        \centering
        \includegraphics[height=3.2cm]{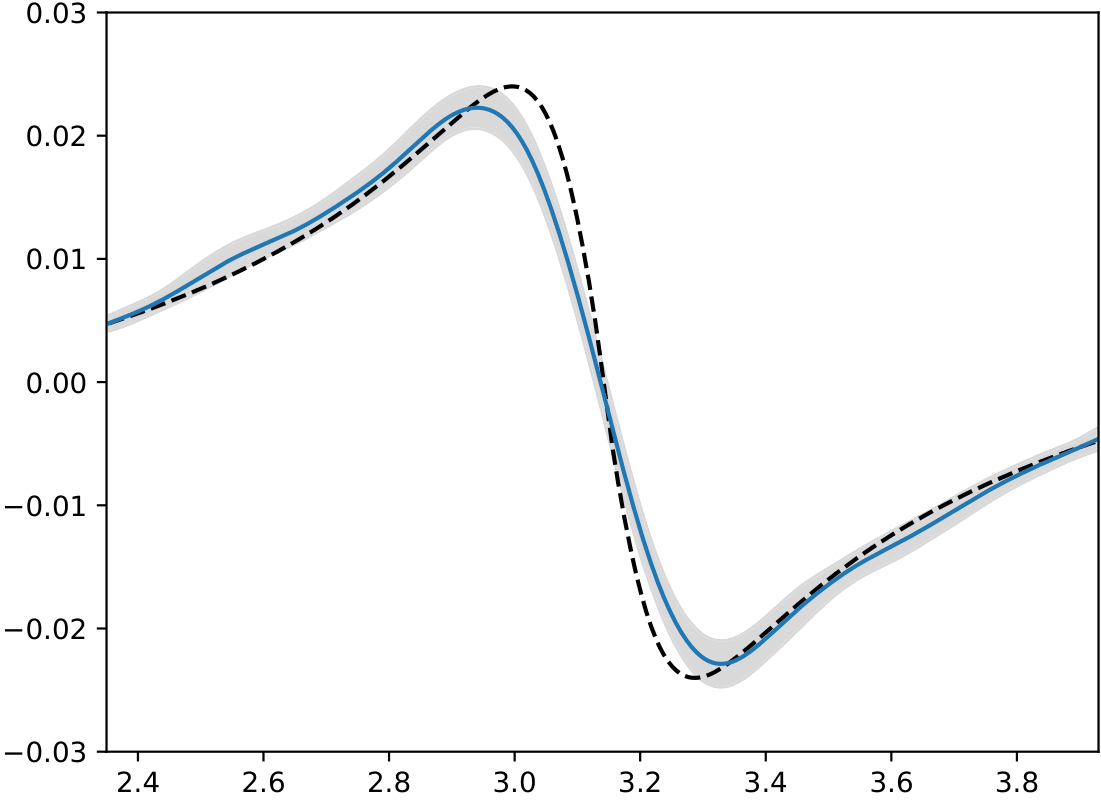}
        \subcaption{PI-RBF kernel ($M=3$)}
    \end{subfigure}

    \caption{
        Estimates and confidence intervals of the horizontal (from (a) to (d)) and vertical (from (e) to (h)) components of the velocity field at the leading edge of a NACA 0412 airfoil. The horizontal coordinates are the parameterization domain $\gammadom = [2.3,4]$. The choices of GP kernels are: one-scale RBF; divergence-free DF-RBF; multiple-scale M-RBF accounting for energy decay; and physics-informed multiple-scale PI-RBF (informed about the profile boundary condition through the BCGP procedure). The black dashed line represents the ground truth of the velocity field components. The continuous blue line represents the posterior mean components. The gray-shaded areas are the posterior $95\%$ confidence intervals.
    }
    \label{fig_profile_posterior_variances}
\end{figure}

In order to quantify the uncertainty associated to the velocity field estimates, the (logarithmic) root-mean-square standard deviations $\x\mapsto\smash{\log_{10}\sqrt{\Trace\upostcov(\x,\x)}}$ are displayed in \fref{fig_UQ_velocity}. It can be observed that the inclusion of boundary information reduces the uncertainty in the part of the computational domain surrounding the airfoil. Moreover, the posterior confidence intervals of each velocity component are displayed in \fref{fig_profile_posterior_variances}. The inclusion of multiple energy scales and the profile boundary condition significantly improve the fit on both components of velocity and yield a consistent calibration of the confidence intervals. As explained in the previous sections, the BCGP procedure enforces the information from \eref{eq_BCGP} on all points of the profile boundary. The tangential component of velocity are left to be learned from data in the posterior distribution only. The posterior variance on this boundary is therefore not null, but significantly consistent with the $95\%$ confidence precision.

\begin{figure}[h!]
    \centering
    \begin{subfigure}{0.49\textwidth}
        \centering
        \includegraphics[height=6cm]{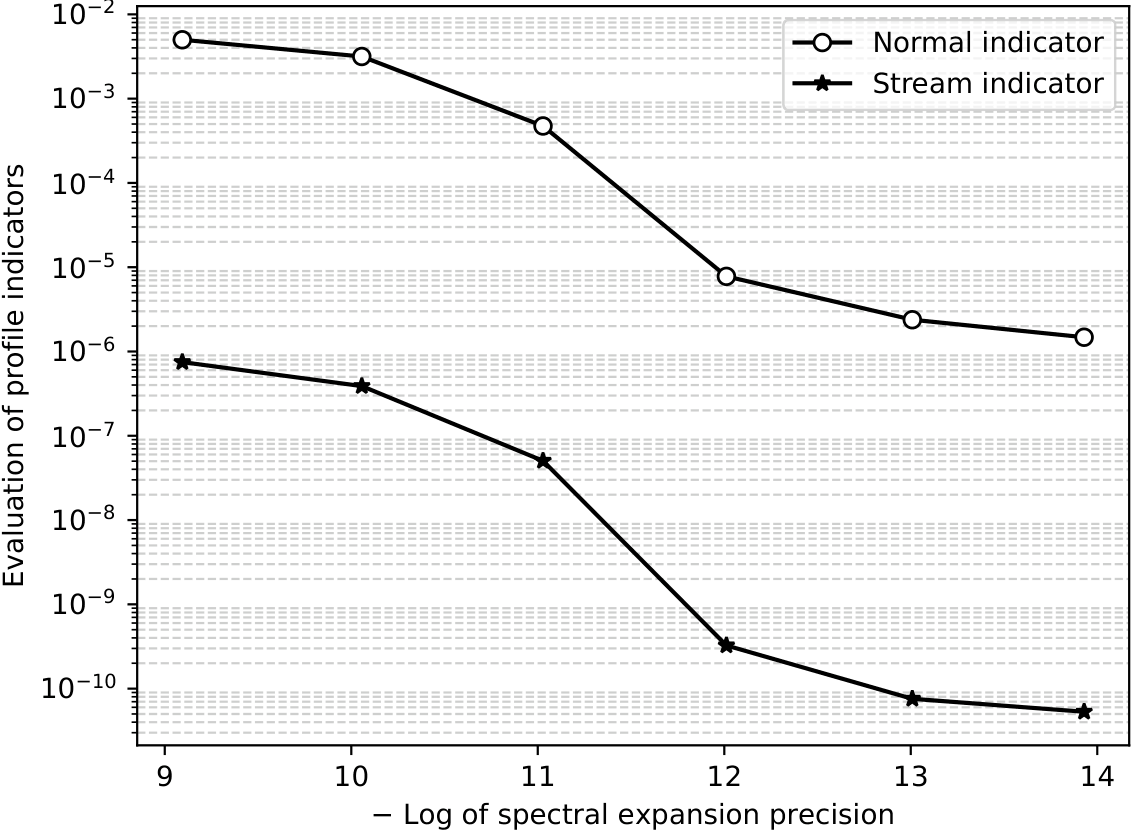}
        \subcaption{$\measureD$ as the normalized pushforward measure $\boldgamma_*\mesLebesgue / \abs{\gammadom}$}
    \end{subfigure}
    \begin{subfigure}{0.49\textwidth}
        \centering
        \includegraphics[height=6cm]{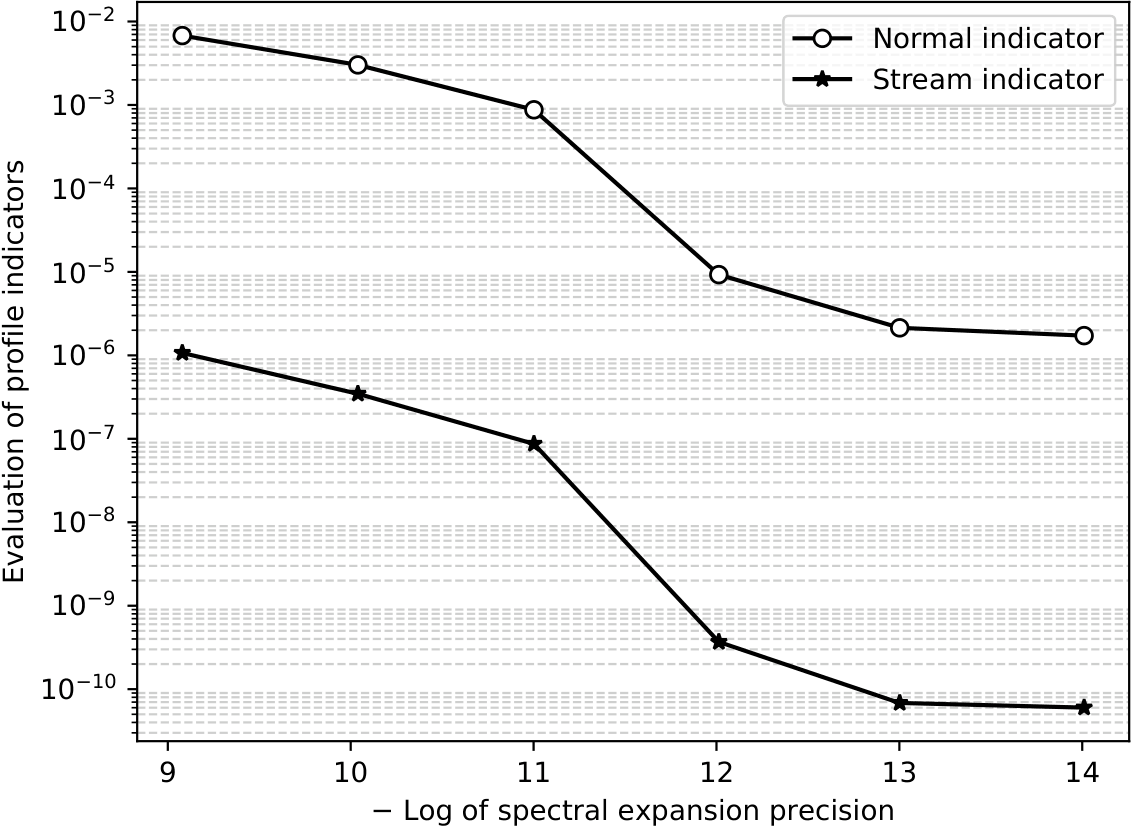}
        \subcaption{$\measureD$ as the normalized surface measure $\Hauss^{d-1} / \Hauss^{d-1}(\curveD)$}
    \end{subfigure}
    \caption{Evaluation of indicators of physical constraints around profile boundary with respect to spectral expansion convergence. The stream normal indicator of \eref{eq_stream_indicator} is shown with stars ($\star$) and the velocity normal indicator $\epsNormal$ of \eref{eq_normal_indicator} is shown with circles ($\circ$). Both of these functions were computed by fixing the GPR design framework for the NACA 0412 leading edge with the same training set and kernel hyperparameters. The horizontal axis is the negative logarithm of $\epsSpec$ from \eref{eq_epsSpec} when varying the spectral bound $\deltaSpec$ in Algorithm \ref{algo_BCGP_derivatives}. In (a) the KLE measure $\measureD$ is chosen as in \cref{Coro_1}, whereas in (b) the measure $\measureD$ is chosen as in \cref{Coro_2}.} 
    \label{fig_NACA_spectral_convergences_KL_measure}
\end{figure}

Finally, we compare in \fref{fig_NACA_spectral_convergences_KL_measure} the satisfaction of the profile boundary condition when using the PI-RBF kernel for two choices of the spectral measure $\measureD$ (see \sref{sec:BCGP}). The indicators defined in \eref{eq_stream_indicator} and \eref{eq_normal_indicator} are presented for both choices of measure. The convergence is presented on the grid $\braces{ 10^{-n}\,:\, n=9,\,10,\,\ldots, 14}$ corresponding to the spectral accuracy bound $\deltaSpec$ of the kernel truncation (see Algorithms \ref{algo_spectral_factor} and \ref{algo_BCGP_derivatives}). The discretization of the integral operator was set at $\Nint = 300$ intervals. The resulting number of spectral modes ranged from 23 at $\deltaSpec = 10^{-9}$ to 200 at $\deltaSpec = 10^{-14}$.

\section{Conclusions and perspectives}\label{sec:Conclusions}

In this work, we have developed an approach based on Gaussian process (GP) regression for simulating incompressible fluid flow fields around aerodynamic profiles by defining physics-informed GP priors. This reconstruction method is mesh-free, geometrically flexible, and is intended to provide estimates from scarce data scenarios by informing the prior distributions of the flow fields with their underlying physics.
These priors thereby satisfy a divergence-free condition for the velocity field by modelling it as a linear differential operator of a stream function which is itself described by a scalar GP. This process is in turn constrained on the boundary of an aerodynamic profile inside the domain, such that the corresponding velocity estimate satisfies a slip boundary condition on that profile. Hence the divergence-free condition is enforced in a continuous way everywhere inside the computational domain, and the slip condition is enforced all along the profile boundary alike.

The presented construction of boundary-constrained GPs is based on a spectral expansion that allows to modify any prescribed GP with a continuous kernel such that the resulting GP satisfies an homogeneous null condition on any compact set. To our knowledge, no other constraining method of a GP that is general enough in this sense has been developed so far. We have described the algorithm used for computing such constrained kernels and their derivatives with the proposed methodology. Then we have reported numerical estimates of the velocity field of incompressible flows around a cylinder profile and around the leading edge of a NACA 0412 airfoil. In the cylinder profile comparison, our results show that using boundary-constrained GPs improves the slip condition fit by three orders of magnitude. Thus this approach is competitive with respect to selecting discrete design points to inform the posterior distribution about the physical conditions on the profile boundary. Therefore, more velocity observations can be specified out of the profile boundary, optimizing a fixed observation budget. Moreover, the integration of multiple physical constraints in GP priors enhances flow reconstruction as shown in our comparison at the leading edge of the NACA airfoil. In particular, velocity reconstruction on the profile boundary is accurate under our approach, even without discrete observations, since the slip boundary condition is enforced in the prior distribution. Finally, an adequate numerical implementation allows convergence of the boundary-constrained kernel as depicted in the comparison using two choices of the spectral measure.

\subsection*{Future works}

We are interested in a design of the observations layout adapted to the dynamics of the fluid and accounting for different sources of information in order to further improve this physics-informed framework. That is, the posterior distribution could be informed with discrete observations of the stream function, the velocity and/or vorticity fields simultaneously. Along this line, we aim to enhance the design of experiments by sequentially allocating the corresponding locations in particular regions of physical interest.

A further aspect that can be improved is the choice of the base kernel within the presented framework. Possible choices may offer higher spectral precision and more accurate fit with respect to data, when using Mat\'ern structures or proper orthogonal decomposition kernels obtained from data for example \cite{Akian2022}. Furthermore, recent works on invariant-measure-informed kernels \cite{Hamzi2025} can be used to derive base kernels for our framework including physical-information from late-time statistical behavior. Invariant-measure results for nonlinear stochastic dissipative systems provide the geometry of the statistical structure for the flow dynamics. In the case of the 2D viscous Navier-Stokes equations, this structure has been proven to be quasi-Gaussian \cite{Coe2025}, which strengthens the choice of Gaussian priors for surrogate modelling \cite{Hamzi2025}. Moreover, the calibration of uncertainty quantification can be improved with more adapted techniques for hyperparameter estimation, such as cross-validation using kernel flows \cite{Owhadi2019}.

Also, we are interested in extending this framework to consider other types of profile boundary conditions, such as a no-slip condition, which are more representative of viscous flows. Ultimately, scalability could be improved by considering sparse GPs \cite{Chen2025}, which are adapted to problems involving kernel derivatives. Finally, as mentioned before, reconstruction formulas that include information from physics in the prior distributions could be directly plugged into numerical methods, such as GP regression-based Lagrangian particle simulations \cite{Owhadi2023}. We are interested in using this approach to develop physics-informed data assimilation techniques applied to fluid dynamics problems.


\section*{Funding statement}

The work of A.P--S., P.N. and O.R. was supported by SHOM (Service Hydrographique et Oc\'eanographique de la Marine) project “Machine Learning Methods in Oceanography” no-20CP07. The work of A.P.-S. and \'E.S. was also supported by ONERA.

\section*{Data Availability Statement}

The data and model implementation supporting the findings of this study are publicly available at \href{https://github.com/adrirps/Physics-Informed-BCGPs-for-Flow-Fields-Reconstruction}{https://github.com/adrirps/Physics-Informed-BCGPs-for-Flow-Fields-Reconstruction}.



\end{document}

%% file: my_macros.tex
\newcommand\abs[1]{\left\lvert #1\right\rvert}
\newcommand\norm[1]{\left\| #1\right\|}
\newcommand\inner[1]{\left\langle #1\right\rangle}
\newcommand\braces[1]{\left\{ #1\right\}}

\newcommand\brackets[1]{\left[ #1\right]}
\newcommand\parenth[1]{\left( #1\right)}

\newcommand{\Nabla}{{\boldsymbol\nabla}}
\newcommand\diver{\Nabla\cdot}

\DeclareMathOperator{\argmax}{argmax}
\newcommand{\esp}{\mathbb{E}}
\DeclareMathOperator{\Cov}{Cov}
\newcommand\GP{{\mathcal{GP}}} 
\newcommand\Normal{{\mathcal{N}}} 
\DeclareMathOperator{\Var}{Var}
\newcommand\Li{\mathcal{L}} 
\DeclareMathOperator{\Trace}{Tr} 

\DeclareMathOperator{\signo}{sgn} 
\DeclareMathOperator{\diag}{diag}

\newcommand\R{\mathbb{R}}

\newcommand\N{\mathbb{N}} 
\newcommand\Mat{\mathcal{M}} 

\newcommand{\C}{\mathcal{C}} 

\newcommand\boldgamma{{\boldsymbol{\gamma}}} 
\newcommand\gammadom{{\mathcal{P}}} 
\newcommand\curveD{\Gamma} 
\newcommand\Dcylinder{{\curveD_\text{cyl}}}
\newcommand\DNACA[1]{{ \curveD_{#1} }}

\newcommand\ycamber{ y_{\text{c}}  } 
\newcommand\ythickness{{ y_{\text{t}} }} 

\newcommand\DomG{{\Omega}} 
\newcommand\measureD{{ \nu }} 
\newcommand\mesLebesgue{{ \eta }} 

\newcommand\G{{G}} 
\newcommand\Gobstacle{{\G_0}} 
\newcommand\Gobsdist[1]{{\G_0^{(#1)}}} 
\newcommand\Ggram{{\boldsymbol G}} 
\newcommand\GgramModif{\widetilde{\boldsymbol G}} 
\newcommand\K{{\boldsymbol K}} 
\newcommand\Kobstacle{{\K_0}} 
\newcommand\Beta{{\boldsymbol \beta}} 
\newcommand\BetaBCGP{{\boldsymbol \beta_0}} 

\newcommand\RKHS{{\mathcal{H}}} 
\newcommand\HG{{\RKHS_\G}} 
\newcommand\HKobstacle{{\RKHS_0}} 

\newcommand\ZGP{{Z}} 
\newcommand\ZGPobstacle{{\ZGP_0}} 
\newcommand\velocityGPobstacle{{\curl\,\ZGPobstacle}} 
\newcommand\velocityGP{{\curl\,\ZGP}} 

\newcommand\Top{\mathcal{T}} 
\newcommand\weightMat{{\boldsymbol H}} 
\newcommand\lameigenD{\lambda}
\newcommand\LameigenMat{{\boldsymbol\Lambda}} 
\newcommand\eigenphiD{{\phi}}
\newcommand\eigenphiExt[1]{\tilde{\eigenphiD}_{#1}} 
\newcommand\EigenphiMat{{ \boldsymbol{E} }} 
\newcommand\EigenphiMatModif{\widetilde{ \boldsymbol{E} }} 
\newcommand\SpecFactor{{ \boldsymbol{S} }} 
\newcommand\normalD{{\boldsymbol n}} 
\newcommand\xiD{\xi} 

\newcommand\s{{\boldsymbol s}}
\newcommand\h{{\boldsymbol h}}
\newcommand\x{{\boldsymbol x}}
\newcommand\y{{\boldsymbol y}}
\newcommand\q{{\boldsymbol q}}
\newcommand\Q{{\boldsymbol Q}} 
\newcommand\X{{\boldsymbol X}} 
\newcommand\Y{{\boldsymbol Y}} 
\newcommand\V{{\boldsymbol V}} 

\newcommand\Hauss{{\mathcal{H}}} 

\newcommand\Xboundary[1]{\x^{#1}_\text{b}} 
\newcommand\Xboundaryset{{\X_\text{b}}} 
\newcommand\Vboundary[1]{\velocity^{#1}_\text{b}} 
\newcommand\Vboundaryset{{\boldsymbol V}_\text{b}} 
\newcommand\Xdom[1]{\x^{#1}_\text{d}} 
\newcommand\Xdomset{{\X_\text{d}}} 
\newcommand\Xset{{\X}} 
\newcommand\Xpset{{\X'}} 
\newcommand\Xtest[1]{{\x^{#1}_{\DomG}}} 
\newcommand\Xtestset{{\X_{\text{test}}}} 
\newcommand\Ntest{{N_{\text{test}}}} 
\newcommand\Vdom[1]{\velocity^{#1}_\text{d}} 
\newcommand\Vdomset{{\boldsymbol V}_\text{d}} 
\newcommand\Xobstacle[1]{\x^{#1}_\curveD} 
\newcommand\Xobstacleset{{\X_\curveD}} 
\newcommand\Yset{{\boldsymbol X}} 
\newcommand\Uset{{\boldsymbol V}} 

\newcommand\vj{u}
\newcommand\velocity{{\boldsymbol\vj}} 
\newcommand\vstream{{\psi}} 
\newcommand\vstreamapprox{{\psi^\star}} 
\newcommand\vorticity{{ \omega }} 
\newcommand\uinlet{{ \velocity_{\text{in}} }} 
\newcommand\uoutlet{{ \velocity_{\text{out}} }} 
\newcommand\uwall{{ \velocity_{\text{w}} }} 
\newcommand\uapprox{{ \velocity^\star }} 
\newcommand\vortapprox{{\vorticity^\star}} 
\newcommand\ugt{{ \velocity^\dagger }} 
\newcommand\upostcov{{\boldsymbol \Sigma^\star }} 
\newcommand\upoststd[1]{\sigma_{#1}^\star} 
\newcommand\upostStd[1]{\Sigma_{#1}^\star}

\newcommand\Nint{I} 
\newcommand\Neigen{J} 
\newcommand\Nderiv{q} 
\newcommand\Ngram{ {N_{\text{Gram}}} } 

\newcommand\Ndom{{N_{\DomG}}} 
\newcommand\Nobstacle{{N_\curveD}} 
\newcommand\Nboundary{N_\text{b}} 

\newcommand\eref[1]{Eq.~(\ref{#1})}
\newcommand\fref[1]{Figure~\ref{#1}}
\newcommand\sref[1]{Sect.~\ref{#1}}
\newcommand\tref[1]{Table~\ref{#1}}
\newcommand\cref[1]{Corollary~\ref{#1}}

\newcommand{\itr}{\mathsf{T}}
\newcommand{\demi}{\frac{1}{2}}
\newcommand\II{{\boldsymbol I}}
\newcommand\dd{{\mathrm d}}
\newcommand{\xp}{\x^\prime}
\newcommand{\curl}{{\textbf{curl}}}

\newcommand{\normal}{{\boldsymbol n}}
\newcommand{\epsDiver}{\epsilon_{\text{div}}}
\newcommand{\epsNormal}{\epsilon_n}
\newcommand{\epsSpec}{\epsilon}
\newcommand{\deltaSpec}{\delta}
\newcommand{\errRMSE}{\epsilon_{\text{RMSE}}}

\newcommand{\Reynolds}{\text{Re}}
\newcommand{\vzero}{{\boldsymbol 0}}
\newcommand{\law}{{\mathscr L}}
\newcommand{\stdv}{\sigma}
\newcommand{\lcor}{\ell}
\newcommand{\nugget}{\eta}

\newcommand{\Loss}{{\mathcal L}}

\newtheorem{theoremcounter}{Dummy}[section]
\newtheorem{remark}[theoremcounter]{Remark}
\newtheorem{proposition}[theoremcounter]{Proposition}
\newenvironment{proof_sketch}[1][Sketch of the proof]{\par\par\noindent\textbf{#1.} }{\hfill$\square$\par}
\newtheorem{corollary}[theoremcounter]{Corollary}

%% file: figures/domain_for_GPR.tex
\begin{tikzpicture}[scale=0.135]
    \draw[thick] (7,0) rectangle (60.0,20.0);
    
    \fill[gray!30] (7,1) rectangle (60,19);    

    \foreach \x in {8, 9, ..., 60} {
         \draw[fill=black, rotate around={45:(\x, 0)}] (\x, 0) rectangle (\x+0.1, 1);
    }

    \foreach \x in {8, 9, ..., 60} {
         \draw[fill=black, rotate around={45:(\x, 19)}] (\x, 19) rectangle (\x+0.1, 20);
    }
    
    \fill[white] (25.0,10.0) circle (2.5);
    \draw[thick] (25.0,10.0) circle (2.5);
    
    \draw[dotted, thick] (7.,0) -- (7.,20.0);
    \draw[dotted, thick] (60.0,0) -- (60.0,20.0);
    
    \node[below] at (5.5,-0.5) { 0 };
    \node[below] at (61,-0.5) { $L_1$ };
    \node[below] at (4,21) { $L_2$ };

    
    
    
    \draw[dotted, thick] (25.0,10.0) -- (23.2300,11.7700);
    
    \node[above] at (35, 9.5) {profile:};
    \node[above] at (35, 6.5) {$\velocity\cdot\normalD = 0$};
    

    \node[above] at (32.25, 21.0) {top wall: $\uwall$};
    \node[above] at (32.25, -5.5) {bottom wall: $\uwall$};
    
    \node[above] at (3.5, 8.0) { $\uinlet$};
    \node[above] at (64, 8.0) { $\uoutlet$};
    
    


    \foreach \x in {7, 10, 13, ..., 60}
        \fill[blue] (\x, 20) circle (0.5);
    \fill[blue] (60.3, 20) circle (0.5);
    
    \foreach \x in {7, 10, 13, ..., 60}
        \fill[blue] (\x, 0) circle (0.5);
    \fill[blue] (60.3, 0) circle (0.5);
        
    \foreach \y in { 2, 4, 6, ..., 18}
        \fill[orange] (60, \y) circle (0.5);
    \foreach \y in { 2, 4, 6, ..., 18}
        \fill[orange] (7, \y) circle (0.5);
    
    \fill[red] (12, 7) circle (0.5);
    \fill[red] (11, 16) circle (0.5);
    \fill[red] (16, 10) circle (0.5);
    \fill[red] (18, 17) circle (0.5);
    \fill[red] (19, 3) circle (0.5);
    \fill[red] (21, 6) circle (0.5);
    \fill[red] (29, 17) circle (0.5);
    \fill[red] (32, 5) circle (0.5);
    \fill[red] (38, 4) circle (0.5);
    \fill[red] (42, 6) circle (0.5);
    \fill[red] (47, 9) circle (0.5);
    \fill[red] (52, 6) circle (0.5);
    \fill[red] (53, 16) circle (0.5);
    \fill[red] (57, 11) circle (0.5);




\end{tikzpicture}